\documentclass[useAMS,usenatbib,letterpaper]{mn2e}
\usepackage{psfrag}
\usepackage{url}
\usepackage{times}
\usepackage{graphicx}
\usepackage{amssymb}

\newcommand{\scinot}[2]
{\ensuremath{#1\hspace{-.15em}\times \hspace{-.15em}10^{#2}}}
\newcommand{\scinotone}[1]
{\ensuremath{10^{#1}}}
\newcommand{\comment}[1]{}

\title[String signal denoising] 
{Wavelet domain Bayesian denoising of string signal in the cosmic
  microwave background}

\author[Hammond et al.]{
D. K. Hammond$^{1}$\thanks{E-mail: david.hammond@epfl.ch}, 
Y. Wiaux$^{1}$\thanks{E-mail: yves.wiaux@epfl.ch}, 
P. Vandergheynst$^{1}$\thanks{E-mail: pierre.vandergheynst@epfl.ch}\\
$^{1}$Institute of Electrical Engineering, 
Ecole Polytechnique F\'ed\'erale de Lausanne (EPFL), 
CH-1015 Lausanne, Switzerland\\}

\begin{document}
\date{\today}
\pagerange{\pageref{firstpage}--\pageref{lastpage}} \pubyear{2009}
\maketitle
\label{firstpage}

\begin{abstract}
  An algorithm is proposed for denoising the signal induced by cosmic
  strings in the cosmic microwave background (CMB).  A Bayesian
  approach is taken, based on modeling the string signal in the
  wavelet domain with generalized Gaussian distributions.  Good
  performance of the algorithm is demonstrated by simulated
  experiments at arcminute resolution under noise conditions including
  primary and secondary CMB anisotropies, as well as instrumental
  noise.
\end{abstract}

\begin{keywords}
  methods: data analysis --
  techniques: image processing --
  cosmic microwave background.
\end{keywords}

\section{Introduction}
\label{sec:i}

Observations of the cosmic microwave background (CMB) and of the Large
Scale Structure of the Universe (LSS) have led to the definition of a
concordance cosmological model. Recently, analysis of the temperature
data of the CMB over the whole celestial sphere from the Wilkinson
Microwave Anisotropy Probe (WMAP) satellite experiment has played a
dominant role in designing this precise picture of the Universe
\citep{bennett03,spergel03,hinshaw07,spergel07,hinshaw09,komatsu09}.
Experiments dedicated to the observation of small portions of the
celestial sphere have also provided their contribution, including the
Arcminute Cosmology Bolometer Array Receiver (ACBAR) experiment
\citep{reichardt09}, the Boomerang experiment \citep{jones06}, and the
Cosmic Background Imager (CBI) experiment \citep{readhead04}.

According to the concordance cosmological model, the cosmic structures
and the CMB originate from Gaussian adiabatic perturbations seeded in
the early phase of inflation of the Universe. However, cosmological
scenarios motivated by theories for the unification of the fundamental
interactions predict the existence of topological defects resulting
from phase transitions at the end of inflation
\citep{vilenkin94,hindmarsh95a,hindmarsh95b,turok90}. These defects
would have participated to the formation of the cosmic structures,
also imprinting the CMB. In particular, cosmic strings are a line-like
version of such defects which are also predicted in the framework of
fundamental string theory \citep{davis05}. As a consequence, the issue
of their existence is a central question in cosmology today.

Cosmic strings are parametrized by a string tension $\mu$, i.e. a mass
per unit length of string, which sets the overall amplitude of the
contribution of a string network.  Their main signature in the CMB is
characterized by temperature steps along the string positions. This
localized effect, known as the Kaiser-Stebbins (KS) effect
\citep{kaiser84,gott85}, hence implies a non-Gaussian imprint of the
string network in the CMB. The most numerous strings appear at an
angular size around $1$ degree on the celestial sphere. CMB
experiments with an angular resolution much below $1$ degree are thus
required in order to resolve the width of cosmic strings.

Experimental constraints have been obtained on a possible string
contribution in terms of upper bounds on the string tension $\mu$
\citep{perivolaropoulos93,bevis04,wyman05,wyman06,bevis07,fraisse07}.
In this context, even though observations largely fit with an origin
of the cosmic structures in terms adiabatic perturbations, room is
still available for the existence of cosmic strings.

The purpose of the present work is to develop an effective method for
mapping the string network potentially imprinted at high angular
resolution in CMB temperature data, in the perspective of forthcoming
arcminute resolution experiments. The observed CMB signal can be
modeled as a linear superposition of a statistically isotropic but
non-Gaussian string signal proportional to an unknown string tension,
with statistically isotropic Gaussian noise comprising the standard
component of the CMB induced by adiabatic perturbations as well as
instrumental noise.

We take a Bayesian approach to this denoising problem, based on
statistical models for both the string signal and noise. Our denoising
is done in the wavelet domain, using a steerable wavelet transform
well adapted for representing the strongly oriented features present
in the string signal. We show that the string signal coefficients are
well described by generalized Gaussian distributions (GGD's), which
are fit at each wavelet scale using a training simulation borrowed
from the set of realistic string simulations recently produced by
\citet{fraisse08}.

We develop a Bayesian least squares method for the denoising, where
each coefficient of the wavelet decomposition is estimated as the
expectation value of its posterior probability distribution given the
observed value. This estimation is split into two parts.  Assuming the
string tension is known, we use the GGD model to compute an estimate
of the string signal. However, the true string tension is unknown. We
deal with this by using a preliminary power spectral model (PSM) to
calculate a posterior probability for the string tension. We then
construct our overall estimator by weighting the GGD based estimators
using this posterior probability distribution for the string
tension. Finally, the string network itself can be mapped by taking
the magnitude of the gradient of the denoised signal
\citep{fraisse08}.

The denoising algorithm that we present may be considered as a modular
component of a larger data analysis.  Firstly notice that the PSM might
be replaced by any other model allowing computation of the posterior
probability distribution of the string tension, notably those which
rely on the best experimental bounds on the string tension.
Secondly, as our method produces a temperature map of the same size as
the input, it may also find use as a pre-processing step for other
methods for cosmic string detection based on explicit edge detection
\citep{jeong05,lo05,amsel07}.

The performance of our denoising algorithm is evaluated under
different conditions, with astrophysical noise components including
various contributions to the standard component of the CMB
(i.e. primary and secondary CMB anisotropies), as well as instrumental
noise.  Three quantitative measures of performance are considered,
namely the signal-to-noise ratio, correlation coefficient, and
kurtosis of the magnitude of gradient of the string signal. Our
analyses rely on $100$ test simulations of a string signal buried in
the noise. For each string tension and noise condition considered, the
test simulations are produced by linear superposition of a unique test
string simulation, also from \citep{fraisse08}, with independent noise
realizations.

In each noise condition, we find that the lowest values for the string
tension down to which our quantitative measures show effective
denoising are very close to the
lowest value where strings begin to be visible by eye. Moreover, we
acknowledge that this value is slightly larger than a detectability
threshold set on the basis of the PSM.

The remainder of this paper is organized as follows. In Section
\ref{sec:ssn}, we discuss the string signal and the noise at arcminute
resolution, and we describe our numerical simulations. In Section
\ref{sec:wss}, we describe the  steerable wavelet
formalism on the plane and study the sparsity of the string signal in
terms of the wavelet decomposition. In Section \ref{sec:bd}, we define
in detail our wavelet domain Bayesian denoising (WDBD) algorithm. In
Section \ref{sec:WDBDap}, we evaluate the performance of the
algorithm.  We finally conclude in Section \ref{sec:c}.

\section{String signal and noise}
\label{sec:ssn}

In this section, we describe the string signal, discuss current and
expected future experimental constraints, and detail the various noise
components at arcminute resolution. We also describe the numerical
simulations used for our developments.

\subsection{String signal}

In an inflationary cosmological model, the phase transitions
responsible for the formation of a cosmic string network occur after
the end of inflation, so as to produce observable defects.  From the
epoch of last scattering until today, the cosmic string network
continuously imprints the CMB. The so-called scaling solution for the
string network implies that the most numerous strings are imprinted
just after last scattering and have a typical angular size around $1$
degree, of the order of the horizon size at that time
\citep{vachaspati84,
  kibble85,albrecht85,bennet86,bennet89,albrecht89,bennet90,allen90}.
Longer strings are also imprinted in the later stages of the Universe
evolution, but in smaller number, according to the number of
corresponding horizon volumes required to fill the sky.

The main signature of a cosmic string in the CMB is described by the
KS effect according to which a temperature step is induced in the CMB
along the string position. The relative amplitude of this step reads
as\begin{equation} \frac{\delta
    T}{T}=\left(8\pi\gamma\beta\right)\rho,\label{2-1}\end{equation}
where $\beta=v/c$, $\gamma=(1-\beta^{2})^{-1/2}$ is the relativistic
gamma factor, and $\rho$ is a dimensionless parameter uniquely
associated with the string tension $\mu$ through \begin{equation}
  \rho=\frac{G\mu}{c^{2}},\label{2-2}\end{equation} where $G$ stands
for the gravitational constant and $c$ for the speed of light. In the
following we call $\rho$ the string tension.

Analytical models relying on the KS effect and the scaling property
were defined to simulate the string signal imprinted in the
CMB. However, in order to produce precise CMB maps accounting for the
full non-linear evolution of the string network, one needs to resort
to numerical simulations. On small angular scales, realistic
simulations can be produced by stacking CMB maps induced in different
redshift ranges between last scattering and today. The simulations we
use in this work have been produced by this technique
\citep{bouchet88,fraisse08}.

\newcommand{\mathbi}[1]{\textbf{\em #1}}
\newcommand{\vecx}{\mathbi{x}}
\newcommand{\veck}{\mathbi{k}}
\newcommand{\vecp}{\mathbi{p}}

The string signal is understood as a realization of a statistically
isotropic but non-Gaussian process on the celestial sphere with an
overall amplitude rescaled by the string tension $\rho$, and
characterized by a nearly scale-free angular power spectrum:
$C_{l}^{s}(\rho)=\rho^{2}C_{l}^{s}$, where the positive integer index
$l$ stands for the angular frequency index on the sphere.  An
analytical expression of this spectrum was provided for $l$ larger
than a few hundreds by \citet{fraisse08}, on the basis of their
simulations. We consider CMB experiments with a small field of view
corresponding to an angular opening $\tau\in[0,2\pi)$ on the celestial
sphere. In this context, the small portion of the celestial sphere
accessible is identified to a planar patch of size $\tau\times\tau$,
and we may consider planar signals in terms of Cartesian coordinates
$\vecx=(x,y)$. The spatial frequencies may be denoted as
$\veck=(k_{x},k_{y})$ with a radial component
$k=(k_{x}^{2}+k_{y}^{2})^{1/2}$.  In this Euclidean limit, the radial
component identifies with the angular frequency on the celestial
sphere, below some band limit set by the resolution of the experiment
under consideration: $l=k<B$.  Analogously, the planar power spectrum,
depending only on the radial component $k$ for a statistically
isotropic signal, identifies with the angular power spectrum of the
original signal on the sphere. In particular, for $k$ larger than a few
hundreds, the nearly scale-free planar power spectrum of the string
signal $s(\vecx)$ reads as 
\begin{equation}
  P^{s}\left(k,\rho\right)=\rho^{2}P^{s}\left(k\right),\label{2-3}
\end{equation}
with $P^{s}(k)=C_{l}^{s}$ for $l=k$.

In this context, the observed CMB signal can be understood as a linear
superposition of the string signal and statistically isotropic noise
of astrophysical and instrumental origin.  As will be discussed in
detail below, this noise is modeled as Gaussian with some angular
power spectrum $C_{l}^{n}$. In the Euclidean limit, the corresponding
planar power spectrum for the noise $n(\vecx)$ may be written as
$P^{n}(k)=C_{l}^{n}$ for $l=k$. The observed noisy signal $f(\vecx)$
is given by:

\begin{equation} \label{2-4}
f\left(\vecx\right)=s\left(\vecx\right)+n\left(\vecx\right).
\end{equation}
Let us notice that we consider zero mean signals, identifying perturbations
around statistical means.

\subsection{Experimental constraints}

Current CMB experiments achieve an angular resolution on the celestial
sphere of the order of $10$ arcminutes, corresponding to a limit
angular frequency not far above $B\simeq10^{3}$. At such resolutions,
the standard component of the CMB primarily contains the Gaussian
perturbations induced by adiabatic perturbations at last scattering,
i.e. when the Universe became essentially transparent to
radiation. These Gaussian anisotropies are referred to as the primary
CMB anisotropies. In this context, any possible string signal is
confined to amplitudes largely dominated by these primary
anisotropies.  The constraints mainly come from a best fit analysis of
the angular power spectrum of the overall CMB signal in the WMAP
temperature data
\citep{perivolaropoulos93,bevis04,wyman05,wyman06,bevis07,fraisse07}.
The tightest of these constraints \citep{fraisse07} gives the
following upper bound at $68$ per cent confidence level:

\newcommand{\rhoexp}{\rho_{\mbox{{\scriptsize exp}}}}
\newcommand{\rhomax}{\rho_{\mbox{{\scriptsize max}}}}

\begin{equation}\label{2-5}
\rho\leq\rhoexp=\scinot{2.1}{-7}.
\end{equation}

Algorithms have also been designed for the explicit identification
of cosmic strings through the observation of the KS effect on CMB
temperature data. The results of the analysis of the full-sky WMAP
data typically provide constraints on the string tension two orders
of magnitude wider than the best fit analysis of the CMB angular power
spectrum, i.e. roughly $\rho<\scinotone{-5}$ \citep{jeong05,lo05}.
The limited angular resolution
of the WMAP data relative to a typical string width is actually more
harmful for the explicit local detection of cosmic strings than for
the estimation of a global parameter such as the string tension through
the analysis of an angular power spectrum.
Corresponding bounds have also been studied in the perspective of
experiments providing higher resolution observation of the CMB on
small portions of the sky \citep{amsel07}. 

\subsection{Noise at arcminute resolution}

\begin{figure}
\begin{center}
\psfrag{1e+04}[][]{$10^4$}
\psfrag{1e+03}[][]{$10^3$}
\psfrag{1e+02}[][]{$10^2$}
\psfrag{1e+00}[][]{$1$}
\psfrag{1e-02}[][]{$10^{-2}$}
\includegraphics[width=8cm]{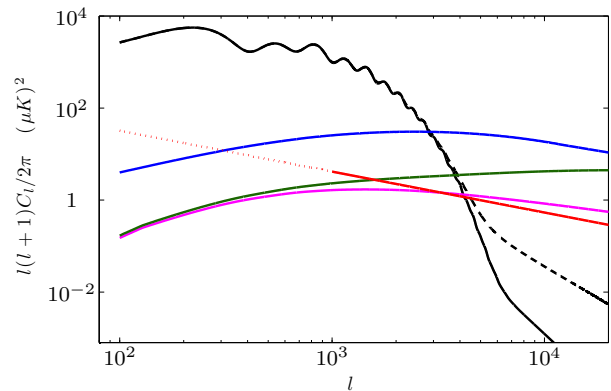}
\end{center}
\caption{\label{fig:spectra} 
  Angular power spectra of string signal and noise (borrowed from
  \citet{fraisse08}) as a function of angular frequency in the range
  $l\in[10^2,\scinot{2}{4}]$ in $\log_{10}-\log_{10}$ axes
  scaling. The spectrum of the string signal is represented for a
  string tension $\rho=\scinot{2}{-7}$ $ $in terms of its analytical
  expression valid at high angular frequencies (red straight
  line). The noise spectra (ordered by decreasing amplitude at
  low angular frequencies) are: the primary CMB anisotropies (black solid
  line) and the gravitational lensing correction (black dashed line),
  the thermal SZ effect in the Rayleigh-Jeans limit (blue solid line),
  the non-linear kinetic SZ effect (green solid line), and the
  Ostriker-Vishniac effect (magenta solid line).}
\end{figure}

Forthcoming experiments will provide access to higher angular
resolution. The Planck Surveyor satellite experiment will provide
full-sky CMB data at a resolution of $5$ arcminutes, i.e.  with
$B\simeq\scinot{2}{3}$ \citep{bouchet04}
\footnote{See also Planck Bluebook at
  http://www.rssd.esa.int/Planck.}.  The Atacama Cosmology Telescope
(ACT) \citep{kosowsky06}, or the South Pole Telescope (SPT)
\citep{ruhl04} will map the CMB on small portions of the celestial
sphere at a resolution around $1$ arcminute, i.e. with
$B\simeq10^{4}$. A slightly higher resolution might be reached by the
radio interferometer Arcminute Microkelvin Imager (AMI)
\citep{jones02,barker06,zwart08}. We consider the issue of the
explicit mapping of the string network in the context of such high
angular resolution experiments. At these resolutions, the so-called
secondary CMB anisotropies, induced by interaction of CMB photons with
the evolving universe after last scattering, will dominate the primary
anisotropies and must be accounted for in the standard component of
the CMB.

For the sake of our analyses, we consider that the standard
cosmological parameters (i.e. excluding the string tension) are fixed
at their values in the context of the concordance cosmological model,
while the string tension remains undetermined. This approximation is
supported by the already tight experimental bounds (\ref{2-5}) on
the string tension. In other words, we assume that even if the true
string tension is non-zero it is to be small, and the true values of
the standard cosmological parameters are close to their present
concordance values. In this context, the angular power spectrum of
both the primary and secondary anisotropies may be
computed on the basis of the assumed concordance values for the
standard cosmological parameters.

The statistically isotropic Gaussian primary anisotropies exhibit
exponential damping at high angular frequencies. This contrasts with
the slow decay of the nearly scale-free angular power spectrum of the
string signal, which thus dominates over the primary anisotropies at
high enough angular frequencies.

The secondary anisotropies include gravitational effects such as the
Integrated Sachs-Wolfe (ISW) effect, the Rees-Sciama (RS) effect, and
gravitational lensing, as well as re-scattering effects such as the
thermal and kinetic Sunyaev-Zel'dovich (SZ) effects.  The SZ effects
dominate these secondary anisotropies
\citep{sunyaev80,komatsu02,fraisse08}.  The ISW and RS effects
associated with the time evolution of the standard gravitational
potentials can be neglected at these angular frequencies.  One may
thus restrict the secondary anisotropies considered in the noise to
the linear (Ostriker-Vishniac) and non-linear kinetic SZ effects, as
well as thermal SZ effect.  The SZ effects are actually non-Gaussian,
spatially dependent, and the kinetic and thermal effects are
correlated.  As a simplifying assumption, we treat these two effects
as two independent statistically isotropic Gaussian noise components.
The effect of gravitational lensing is very small relative to the SZ
effects, but we still take it into account as a correction to the
angular power spectrum of the primary anisotropies.

At arcminute resolution, the thermal and kinetic SZ effects have
standard deviations around $10\,\mu\textnormal{K}$ and
$5\,\mu\textnormal{K}$ respectively. They also have a slow decay at
high angular frequencies and will dominate the string signal for
string tension values below the current experimental bound. Arcminute
CMB experiments are in fact primarily dedicated to the detection of
these secondary anisotropies. Unlike the other effects considered, which have the same
black body spectrum as the primary anisotropies, the thermal SZ effect
on the CMB temperature depends on the frequency of observation. Its
amplitude decreases from the Rayleigh-Jeans limit (null frequency) and
around $217\mbox{\, GHz}$ where it is expected to vanish, before
increasing again at higher frequencies. Figure \ref{fig:spectra}
represents the angular power spectra as a function of the angular
frequency $l$, for a string signal with string tension
$\rho=\scinot{2}{-7}$, the primary CMB anisotropies and the
correction due to gravitational lensing, the Ostriker-Vishniac and
non-linear kinetic SZ effect, and the thermal SZ effect in the
Rayleigh-Jeans limit. These spectra are explicitly borrowed from
\citet{fraisse08}, again assuming concordance values for the standard
cosmological parameters.

Instrumental noise also obviously affects signal
acquisition. Corresponding expected amplitudes for future experiments
should be lower than the amplitude of secondary anisotropies, but
still with a standard deviation very roughly around
$1\mu\textnormal{K}$ per pixel \citep{kosowsky06}. We will model
instrumental noise as Gaussian white noise, i.e. with a flat power
spectrum.

In this context, the performance of the denoising algorithm to be
defined will be studied in the following limits.  As a first approach,
we consider the secondary anisotropies as a statistically isotropic
Gaussian noise with power spectrum given by the Rayleigh-Jeans limit,
that is added to the primary anisotropies.  One can also assume an
observation frequency around $217\mbox{\, GHz}$ taking advantage of
the frequency dependence of the thermal SZ effect, and include in the
noise secondary anisotropies in absence of this effect. This is
equivalent to including only the kinetic SZ effect and gravitational
lensing in the secondary anisotropies.  Notice that the future ACT
will have one of its acquisition frequencies at $215\mbox{\, GHz}$
\citep{kosowsky06}. In these two cases, instrumental noise is
considered to be negligible and simply discarded. These two different
noise conditions are respectively denoted as SA$+$tSZ (secondary
anisotropies with thermal SZ effect) and SA$-$tSZ (secondary
anisotropies without thermal SZ effect) in the following. Analyzing
these limits can reveal to what extent the kinetic and thermal SZ
effect hamper the  denoising of the string signal,
as a function of the string tension.

Notice that component separation techniques relying on the
non-Gaussianity of the thermal SZ effect have been designed for its
extraction from the CMB temperature data, on the basis of
multi-frequency observations
\citep{hobson98,delabrouille03,maisinger04,pires06,bobin08}.  Other
component separation techniques relying on the non-Gaussianity of the
kinetic SZ effect and on its correlation with the thermal SZ effect
have also been proposed for its extraction from the CMB temperature
data \citep{forni04}. In that regard, a global component separation
technique might be envisaged in order to simultaneously extract all
non-Gaussian components of the CMB temperature data, including the
string signal.

In the context of our string signal denoising approach, the
performance of a denoising algorithm can also be examined in the limit
where the noise only includes primary anisotropies and instrumental
noise, assuming secondary anisotropies have been correctly separated.
The case without instrumental noise is denoted as PA$-$IN (primary
anisotropies without instrumental noise) and will be studied in order
to understand the behaviour of the denoising algorithm in ideal noise
conditions. The case with instrumental noise with a standard deviation
of $1 \mu K$, denoted as PA$+$IN (primary anisotropies with
instrumental noise), is also considered.

For the sake of our analyses, foreground emissions such as Galactic
dust or point sources \citep{kosowsky06} are disregarded.

\subsection{Numerical simulations}

\begin{figure*}
\begin{center}
\includegraphics[width=5.5cm]{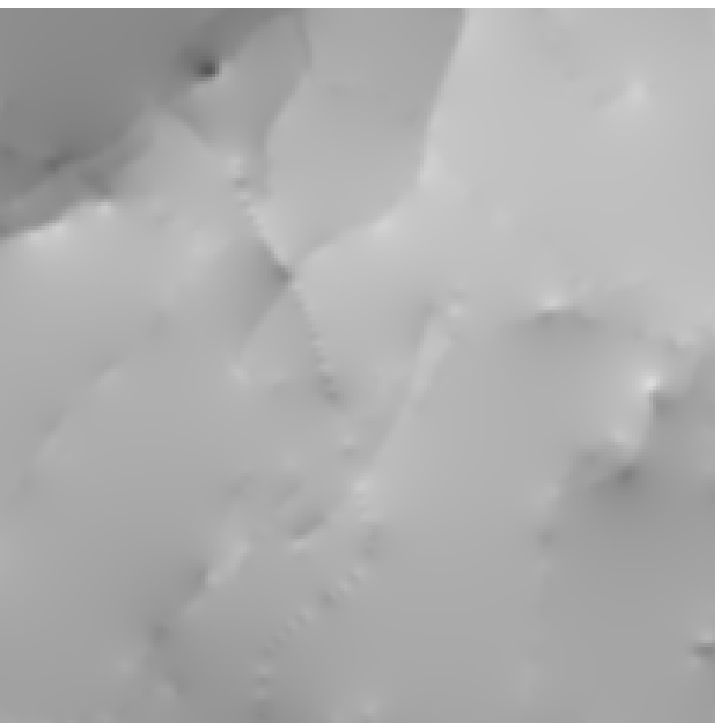}\hspace{2mm}
\includegraphics[width=5.5cm]{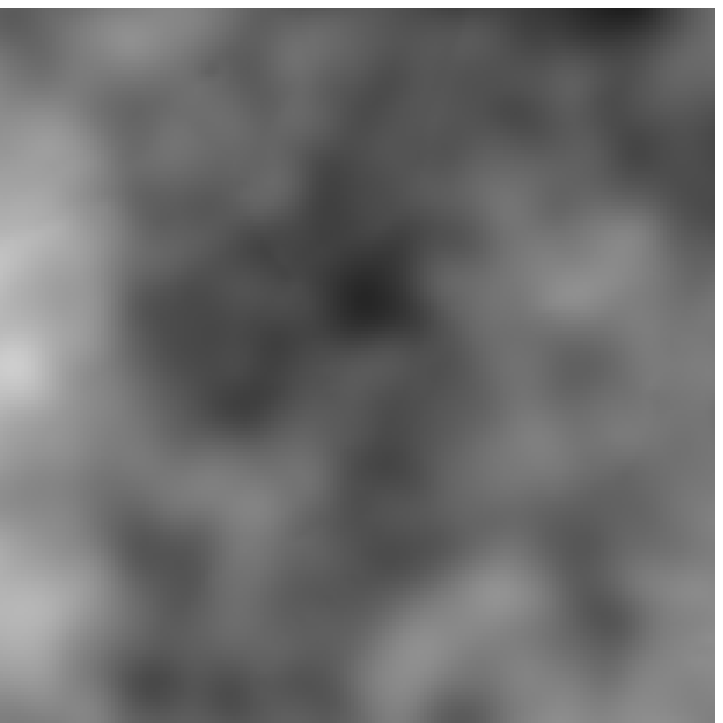}\hspace{2mm}
\includegraphics[width=5.5cm]{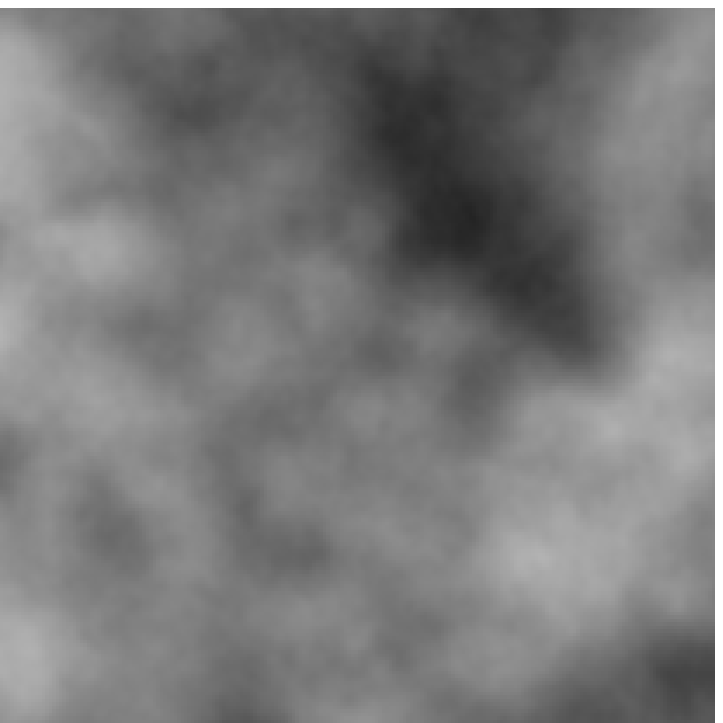}\vspace{2mm}

\includegraphics[width=5.5cm]{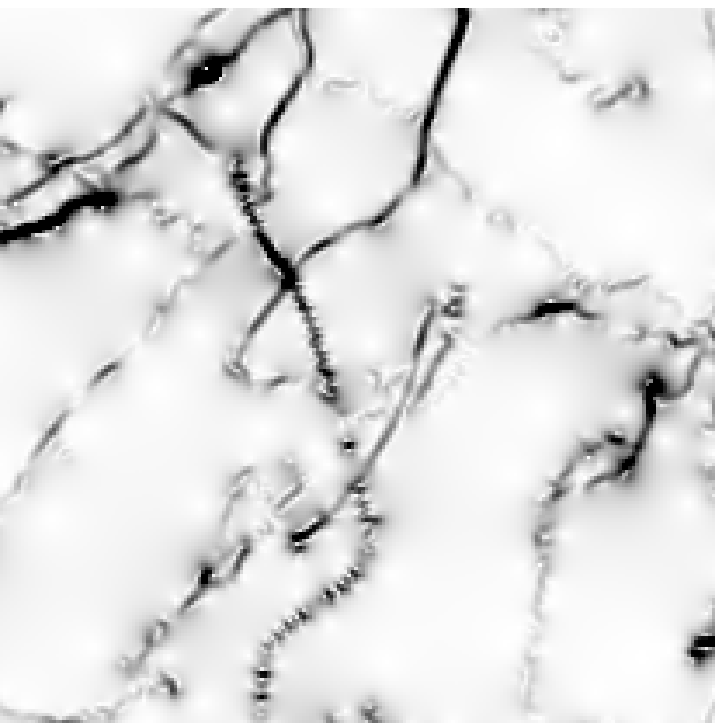}\hspace{2mm}
\includegraphics[width=5.5cm]{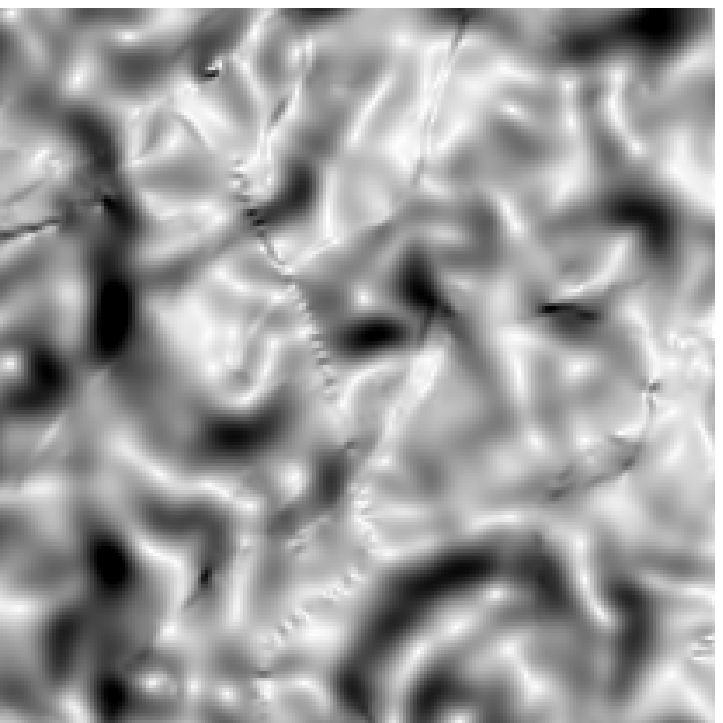}\hspace{2mm}
\includegraphics[width=5.5cm]{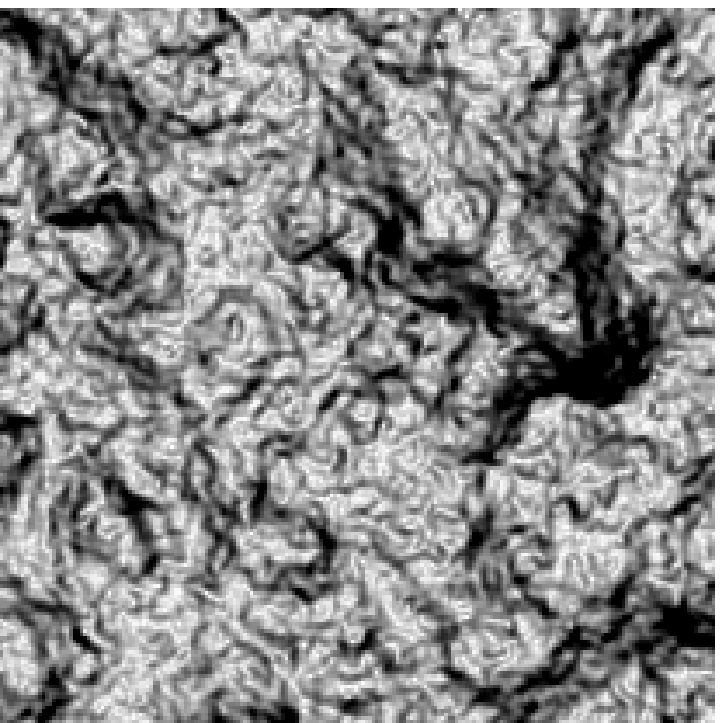}
\end{center}
\caption{\label{fig:maps-gradients}
  Simulated maps of the string signal with noise at $1$ arcminute
  resolution, on a field of view of $\tau'=1.4^{\circ}$.  The top left
  panel represents the test simulation of the string signal.  The top
  middle and top right panels represent the combinations of this
  string signal simulation for a string tension $\rho=\scinot{2}{-7}$
  in the noise conditions PA$-$IN and SA$-$tSZ respectively. The
  bottom panels represent corresponding maps of the magnitude of
  gradient.}
\end{figure*}

Our denoising approach is based on explicitly describing the
statistical properties of the string signal on small angular
scales. We need precise simulations of the string signal on planar
patches for both training and validation of our method. We use $2$
simulations of the string signal borrowed from the full set of $84$
simulations produced by \citep{fraisse08}. The first simulation of the
string signal is used as training data for fitting the prior
probability distributions for the coefficients of the wavelet
decomposition of the signal $s$, while the second is reserved for
testing the algorithm. In the four noise conditions considered
(PA$-$IN, PA$+$IN, SA$-$tSZ, or SA$+$tSZ), this test string signal
simulations is combined with $100$ independent realizations of the
noise in order to produce multiple test simulations.

The simulations are defined on planar patches of size $\tau\times\tau$
for a field of view defined by an angular opening $\tau=7.2^{\circ}$.
The finite size of the patch induces a discretization of the spatial 
frequencies below the band limit $B$:
$\veck=(2\pi/\tau)\vecp=50\vecp$ with $\vecp=(p_{x},p_{y})$
and for integer values $p_{x}$ and $p_{y}$ with $-L\leq p_{x},p_{y}<L$
with $L=\tau B/2\pi=B/50$. The original maps are sampled on grids with
$2L\times2L$ uniformly sampled points $\vecx_{i}$ with $1\leq
i\leq4L^{2}$ for $L=512$. The corresponding pixels thus have an
angular size around $0.42'$. The corresponding band limit on $k_{x}$
and $k_{y}$ thus reads $B\simeq\scinot{2.5}{4}$.

The astrophysical and instrumental components of the noise are modeled
as statistically isotropic Gaussian noise on a planar patch with the
appropriate power spectra. We consider instrumental noise with a
standard deviation of $1\mu K$.  For each noise component, a
simulation may easily be produced by taking the Fourier transform of
Gaussian white noise, renormalizing each Fourier frequency value by
the square root of the corresponding power spectrum, and inverting the
Fourier transform \citep{rocha05}. In each noise condition considered,
an overall noise simulation is obtained by simple superposition of the
required independent components simulated. The power spectrum of the
noise $P^{n}(k)$ is the sum of the individual spectra.

We also include the effect of the experimental beam of a typical
arcminute experiment in the training string signal simulation, as well
as in all test simulations for each noise condition considered. We
simply model this effect by convolution of the string signal and
astrophysical noise components with a Gaussian kernel with a full
width at half maximum (FWHM) of $1$ arcminute. This corresponds to a
Gaussian tapering of angular frequencies with a FWHM of
$\scinot{2}{4}$, which effectively limits the angular frequencies not
far above $B\simeq\scinotone{4}$.  Hence the power spectrum
$ $$P^{s}(k,\rho)$ of the string signal in relation (\ref{2-3}) and the power
spectrum of the astrophysical noise components are multiplied by the
square modulus of the Fourier transform of the experimental beam. The
corresponding power spectra of the string signal and of the noise in
each noise condition considered are respectively denoted as
$\tilde{P}^{s}(k,\rho)$ and $\tilde{P}^{n}(k)$.

For illustration, Figure \ref{fig:maps-gradients} represents simulated
maps of the string signal and noise at the resolution considered, as
well as corresponding maps of the magnitude of gradient. For
visualization purposes, we show only one fifth of the total field of
view, corresponding to an angular opening $\tau'=1.4^{\circ}$. At a
tension $\rho=\scinot{2}{-7}$, with noise only including the
primary CMB anisotropies, the strings are not visible by eye in the
original map itself, while part of the network appears in the map of
the magnitude of gradient. This illustrates the natural enhancement of
high frequency features such as temperature steps by the gradient
operator. At the same string tension, the presence of the secondary
anisotropies adds noise at the highest angular frequencies and the
strings are not visible by eye anymore in either the original map
or the map of the magnitude of gradient, already when the thermal SZ
effect is discarded.

\section{Wavelets and Signal Sparsity}\label{sec:wss}

In this section, we firstly describe the steerable wavelet transform
and reformulate the denoising problem in the wavelet domain.  We then
detail the probability distributions that we use to describe the
marginal statistics of the wavelet coefficients of string signal and
the noise.

\subsection{Steerable wavelets} \label{sec:sw}

\begin{figure}
\begin{center}
\includegraphics[width=8cm,height=5cm]{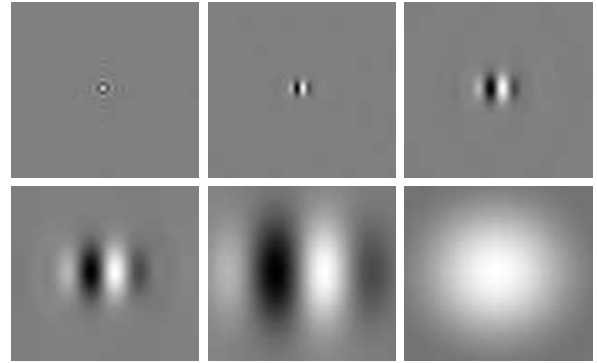}
\end{center}
\caption{\label{fig:wavelet} 
  Steerable Pyramid wavelets with $N=6$ basis orientations, centered
  at the origin of the plane.  From top left to bottom right, the
  highpass filter $\gamma_{h}$, the wavelets $\gamma_{q,j}$ with
  orientation $\chi_{N}=\pi$ (i.e.  $q=N$) for the spatial scales
  $j=1$ to $j=4$, and the lowpass filter $\gamma_{l}$. }
\end{figure}

Wavelet transforms have become widely used in data analysis and image
processing in recent years, and have found numerous applications in
astrophysics \citep{hobson99,barreiro01,starck06a}. In general, a
particular transform will be useful for signal modeling and denoising
if the properties of the signal of interest are easier to describe, or
more distinct from the noise process, in the transform domain than in
the original domain. The cosmic string signal is characterized by
localized, oriented edge-like discontinuities. This motivates the use
of a transform that is well adapted for representing localized
oriented features. 

Standard orthogonal wavelets are well localized, but are not well
suited for arbitrarily oriented features as they have strong bias for
horizontal and vertical orientations due to their tensor product
construction. Instead, in this paper we use a steerable wavelet
transform. The transform is parametrized by a number of orientations
$N$ and spatial scales $J$.  The output of the transform is given by
the convolution of the original signal $f$ with a set of filters at
different scales. These filters are formed by scaling and rotating a
single, ``mother'' wavelet $\gamma(\vecx)$. For the discrete numerical
transform, the rotation is sampled at $N$ equally spaced angles
$\chi_q = q\pi/N$ for integer $q$ with $1\leq q \leq N$. The scalings
are sampled dyadically, i.e. as $2^{-j}$ for integer $j$ with $1\leq j
\leq J$.  The transform output at spatial scale $j$ and orientation
$q$ is given by $f\star \gamma_{q,j}$ where $\gamma_{q,j}(x)$ is given
by rotating $\gamma$ by $\chi_q$ and scaling by $2^{-j}$.

In order to ensure the invertibility of the transform, it is also
necessary to include residual highpass and lowpass bands, generated by
filters $\gamma_h$ and $\gamma_l$ respectively. The output of the
complete steerable wavelet transform then includes a highpass band,
$J$ sets of $N$ oriented bandpass bands, and the lowpass band.
Invertibility of the transform is important for our work, as we are
interested in reconstructing the string map which resides in the image
domain.

\newcommand{\vw}{\mathbi{W}} 
\newcommand{\mR}{R} 

We adopt the notation $\vw^f$ to denote the full vector of wavelet
coefficients for a given input signal $f$, where we have implicitly
vectorized and concatenated the subbands corresponding to different
scales and orientations. We will write $W^f_I$ to specify individual
coefficients, where $I$ is a multi-index specifying the scale,
orientation and spatial location of the coefficient. We will denote by
$\mR$ the inverse wavelet transform operator, so that
\begin{equation}\label{eq:wavelet-inverse}
f(\vecx) = \left[\mR \vw^f\right] (\vecx)
\end{equation}

In this work, we use a particular implementation known as the
Steerable Pyramid\footnote{See also steerable pyramid implementation
  available for download at
  \url{http://www.cns.nyu.edu/~eero/STEERPYR/}.} \citep{simoncelli92}.
We use the transform with $N=6$ orientations and $J=4$ spatial
scales. The corresponding wavelet filters are shown in Figure
\ref{fig:wavelet} for orientation $\chi_N=\pi$.  In particular, the
filters that we employ have odd symmetry, which is especially appropriate
for representing the edge-like discontinuities present in the string
signal.

\subsection{Problem reformulation in wavelet domain}

\begin{figure*}
\psfrag{-1.3e+06}{\scinot{-1.3}{6}}
\psfrag{1.3e+06}{\scinot{1.3}{6}}
\psfrag{-1.4e+07}{\scinot{-1.4}{7}}
\psfrag{1.4e+07}{\scinot{1.4}{7}}
\psfrag{-9.5e+07}{\scinot{-9.5}{7}}
\psfrag{9.5e+07}{\scinot{9.5}{7}}
\psfrag{-5.1e+08}{\scinot{-5.1}{8}}
\psfrag{5.1e+08}{\scinot{5.1}{8}}
\psfrag{-3.7e+05}{\scinot{-3.7}{5}}
\psfrag{3.7e+05}{\scinot{3.7}{5}}
\psfrag{-2.1e+10}{\scinot{-2.1}{10}}
\psfrag{2.1e+10}{\scinot{2.1}{10}}
\psfrag{-1.3e+06}{\scinot{-1.3}{6}}
\psfrag{1.3e+06}{\scinot{1.3}{6}}
\psfrag{-1.4e+07}{\scinot{-1.4}{7}}
\psfrag{1.4e+07}{\scinot{1.4}{7}}
\psfrag{-9.5e+07}{\scinot{-9.5}{7}}
\psfrag{9.5e+07}{\scinot{9.5}{7}}
\psfrag{-5.1e+08}{\scinot{-5.1}{8}}
\psfrag{5.1e+08}{\scinot{5.1}{8}}
\psfrag{-3.7e+05}{\scinot{-3.7}{5}}
\psfrag{3.7e+05}{\scinot{3.7}{5}}
\psfrag{-2.1e+10}{\scinot{-2.1}{10}}
\psfrag{2.1e+10}{\scinot{2.1}{10}}
\psfrag{-2}{-2}
\psfrag{-3}{-3}
\psfrag{-4}{-4}

  \begin{center}
    \includegraphics[width=5cm]{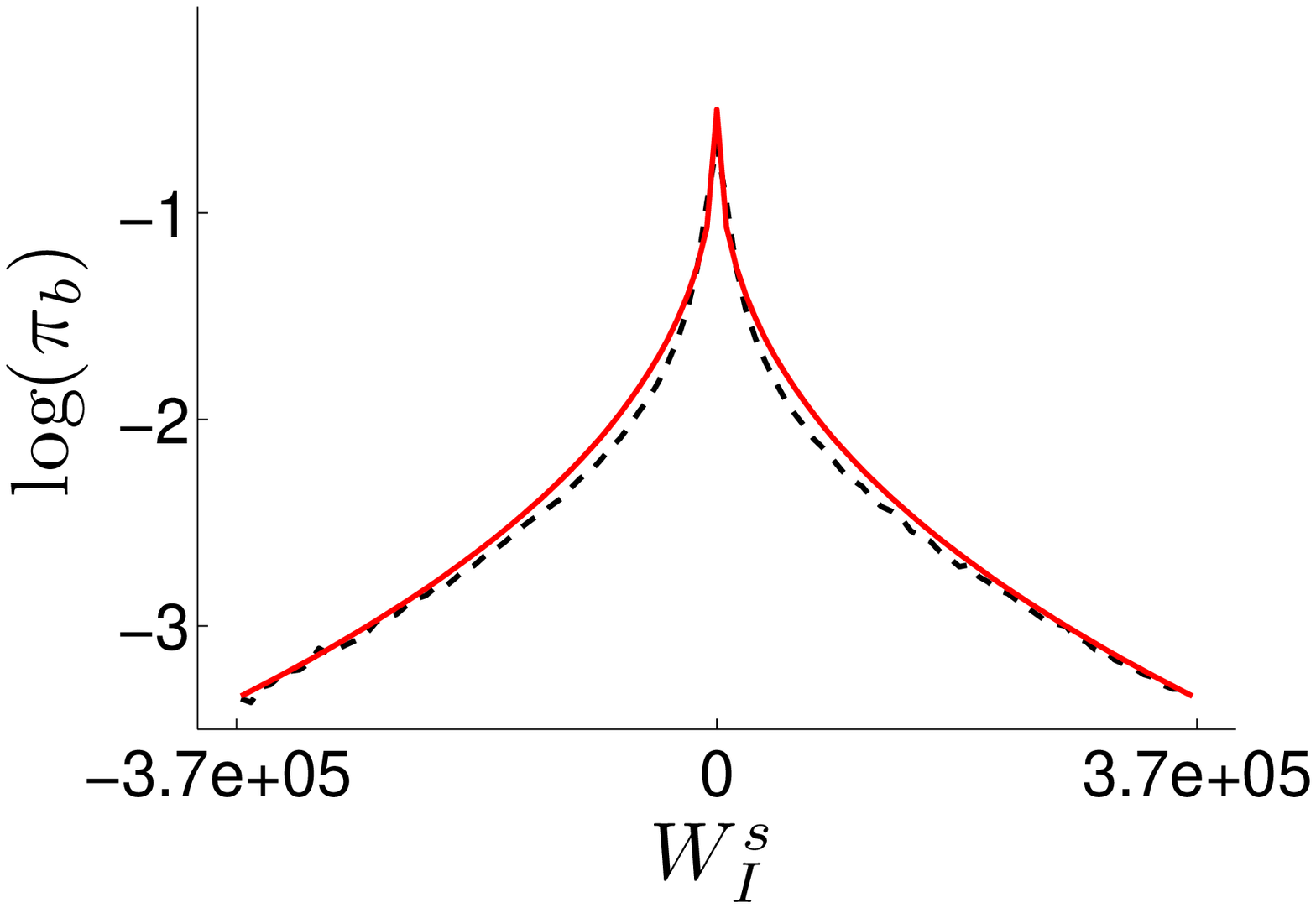}
    \hspace{1cm}
    \includegraphics[width=5cm]{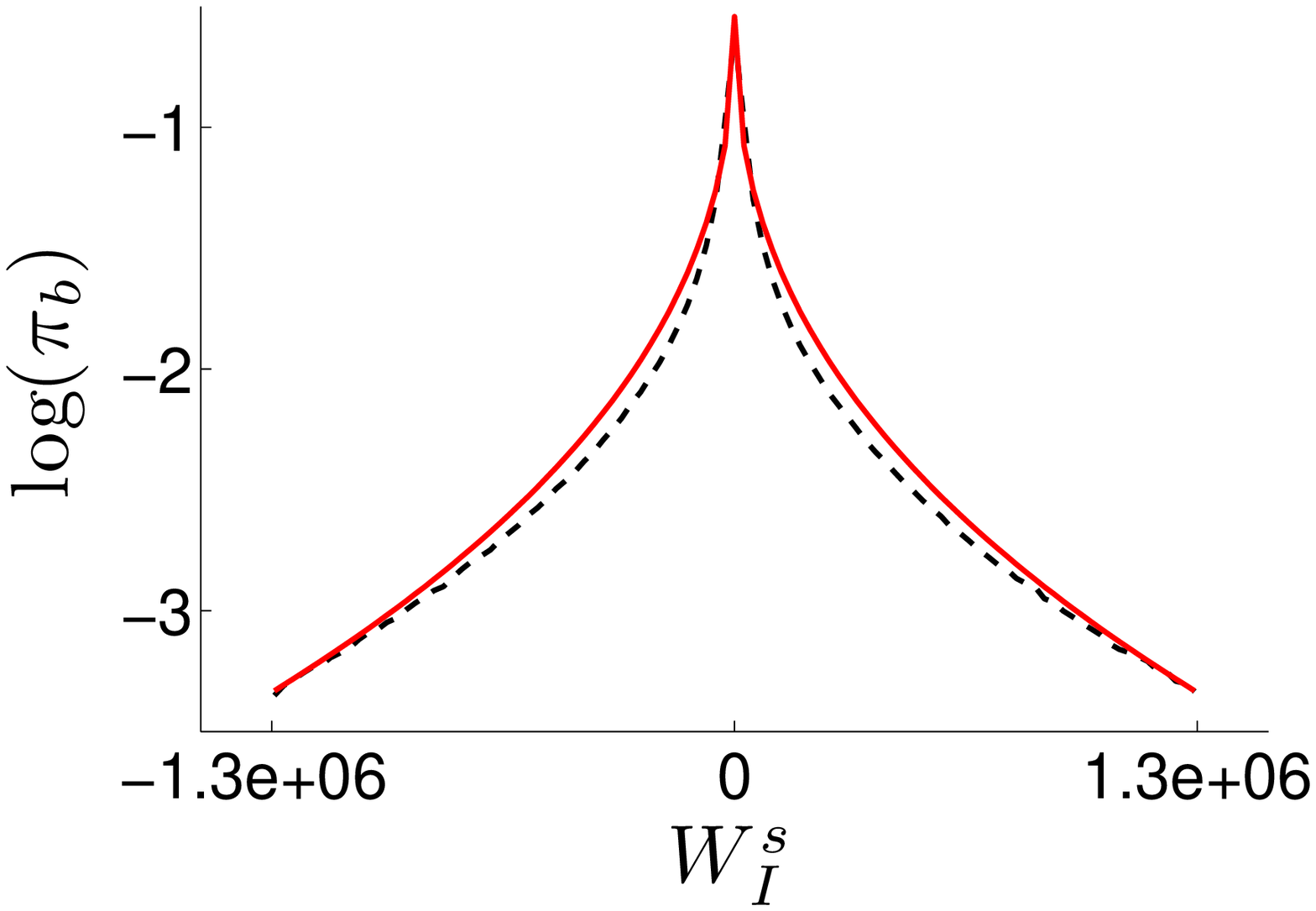}
    \hspace{1cm}
    \includegraphics[width=5cm]{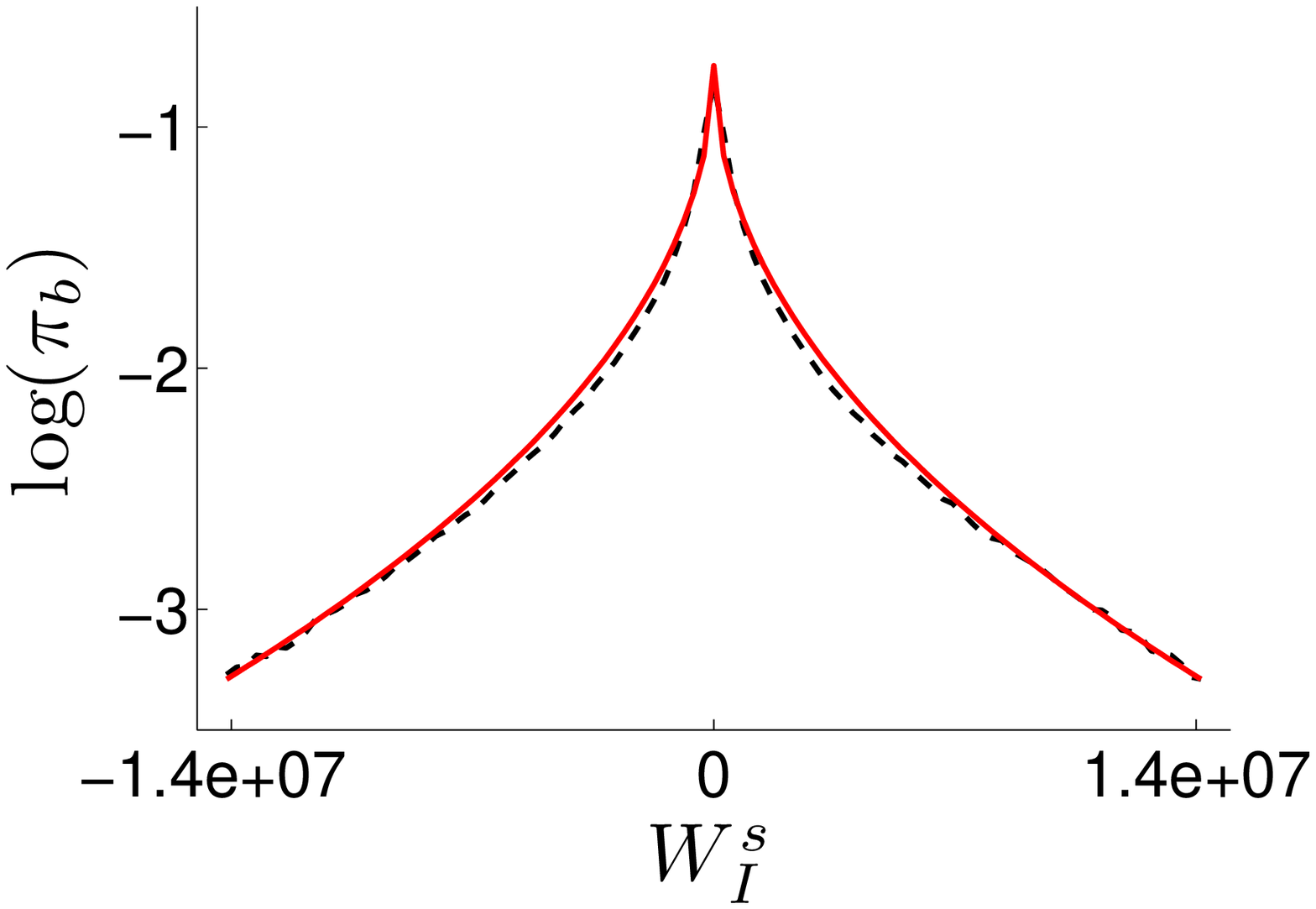}
    \vspace{1cm}
    \includegraphics[width=5cm]{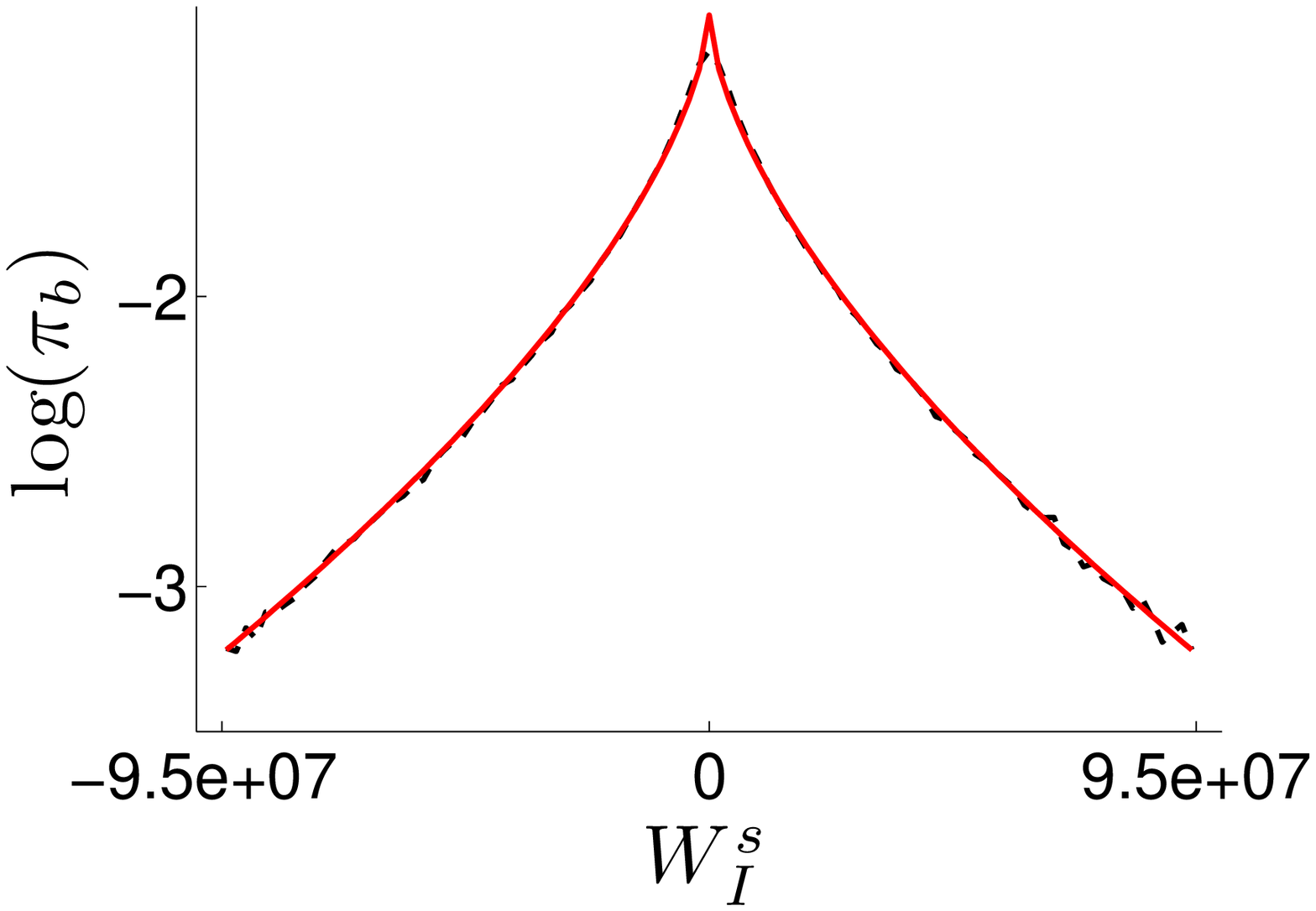}
    \hspace{1cm}
    \includegraphics[width=5cm]{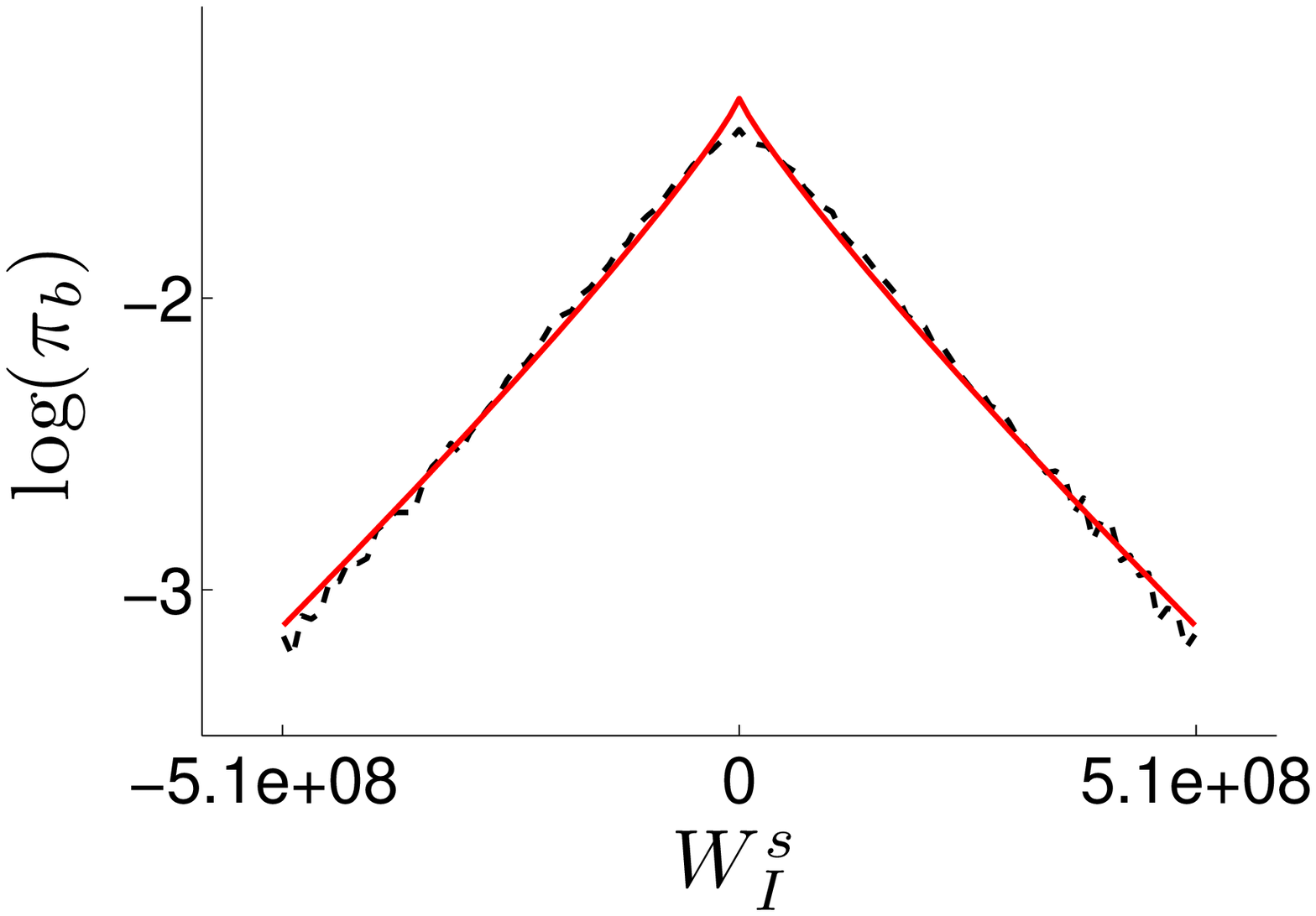}
    \hspace{1cm}
    \includegraphics[width=5cm]{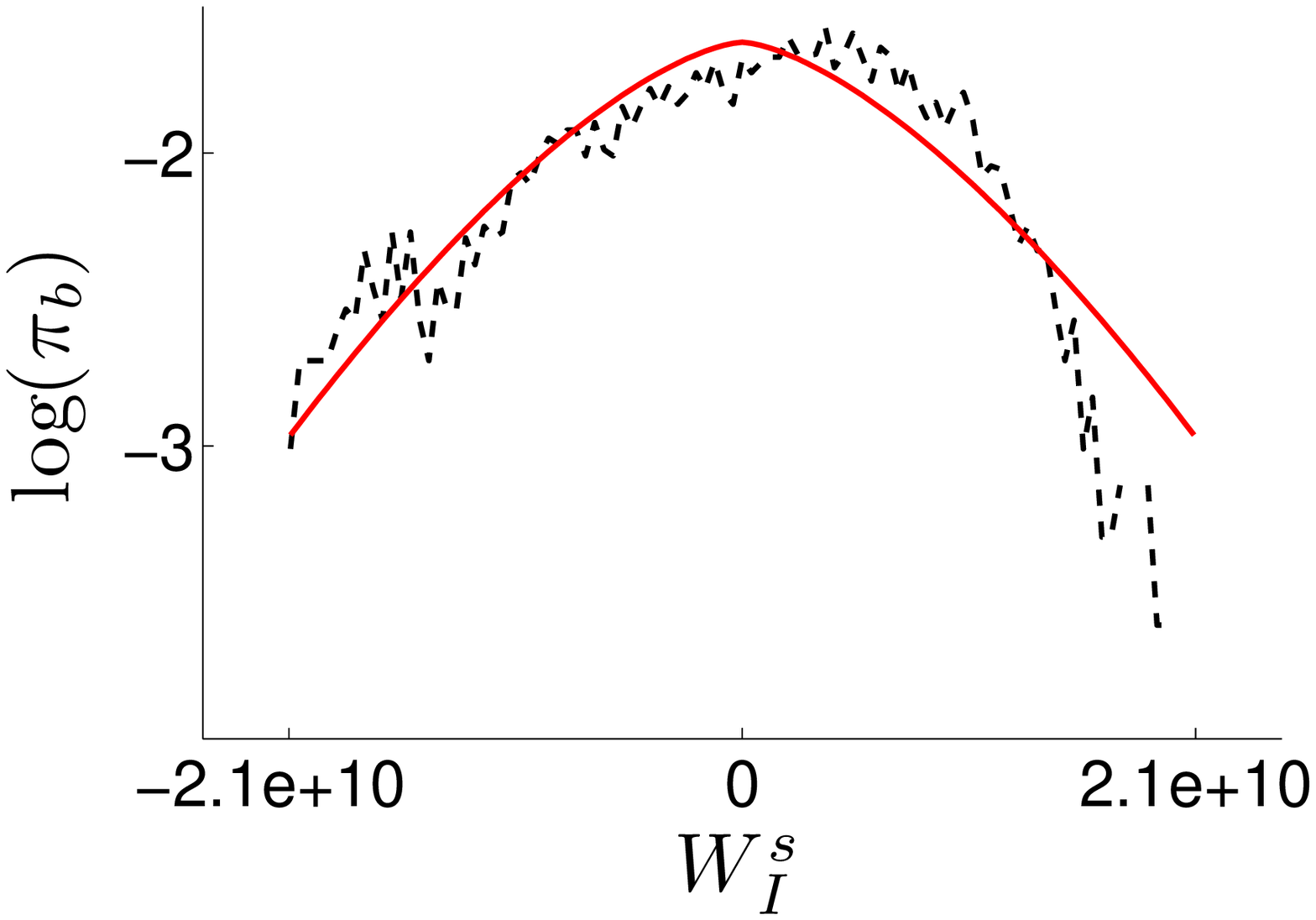}
  \end{center}
  \vspace{-1cm}
  \caption{ \label{fig:histograms} Logarithm of the modeled prior
    GGD's $\pi_{b}$ for the wavelet coefficients $W_{I}^{s}$ of a
    string signal (red solid curves), computed using $N=6$
    orientations and $J=4$ spatial scales, superimposed on the
    histograms of the corresponding coefficients from the training
    simulation (black dashed curves) for $b=\{h,j,l\}$ and $1\leq
    j\leq J=4$.  From top left to bottom right, the coefficients
    associated with the highpass filter $\gamma_{h}$, with the
    wavelets $\gamma_{q,j}$ for the spatial scales $j=1$ to $j=4$, and
    with the lowpass filter $\gamma_{l}$. }
\end{figure*}

\begin{table*} 
  \begin{center}
    \begin{tabular*}{0.9\textwidth}{@{\extracolsep{\fill}}cccccccc}
      \hline 
      \noalign{\vskip\doublerulesep}
      $b$ & $u_{b}$ & $v_{b}$ & $\sigma_{b}^{s}(\rho=1)$  & $\kappa_{b}^{s}$ & $\sigma_{b}^{n}$ (PA$-$IN)$$ & $\sigma_{b}^{n}$(SA$-$tSZ)$$ & $\sigma_{b}^{n}$(SA$+$tSZ)\tabularnewline[\doublerulesep]
 \hline
 \noalign{\vskip\doublerulesep}
 $h$ & $\scinot{3.8}{3}$ & $0.41$ & $\scinot{1.2}{5}$ & $47$ & $\scinot{4.7}{-4}$ & $\scinot{8.6}{-2}$ & $0.16 \comment{\scinot{1.6}{-1}}$\tabularnewline
 \noalign{\vskip\doublerulesep}
 $j=1$ & $\scinot{1.6}{4}$ & $0.42$ & $\scinot{4.1}{5}$ & $42$ & $\scinot{4.5}{-3}$ & $0.26 \comment{\scinot{2.6}{-1}}$ & $0.56 \comment{\scinot{5.6}{-1}}$\tabularnewline
 \noalign{\vskip\doublerulesep}
 $j=2$ & $\scinot{3.8}{5}$ & $0.49$ & $\scinot{4.5}{6}$ & $26$ & $0.55 \comment{\scinot{5.5}{-1}}$ & $2.3 \comment{\scinot{2.3}{0}}$ & $5.4 \comment{\scinot{5.4}{0}}$\tabularnewline
 \noalign{\vskip\doublerulesep}
 $j=3$ & $\scinot{7.2}{6}$ & $0.63$ & $\scinot{3.1}{7}$ & $14$ & $32$ \comment{$\scinot{3.2}{1}$} & $35$ \comment{$\scinot{3.5}{1}$} &$44$ \comment{$\scinot{4.4}{1}$}\tabularnewline
 \noalign{\vskip\doublerulesep}
 $j=4$ & $\scinot{9.9}{7}$ & $0.88$ & $\scinot{1.8}{8}$ & $7.3$ & $\scinot{4.9}{2}$ & $\scinot{4.9}{2}$ & $\scinot{5.1}{2}$\tabularnewline
 \noalign{\vskip\doublerulesep}
 $l$ & $\scinot{9.8}{9}$ & $1.5$ & $\scinot{8.3}{9}$ & $3.7$ & $\scinot{2.5}{4}$ & $\scinot{2.5}{4}$ & $\scinot{2.5}{4}$\tabularnewline
 \hline
\end{tabular*}
\end{center}
\caption{\label{tab:statmodel}
  Parameters for the modeled prior GGD's $\pi_{b}$
  and noise Gaussian distributions $g_{b}$ for the spatial scales
  $b=\{h,j,l\}$ and $1\leq j\leq J=4$. The first column identifies
  the spatial scale $b$. The next four columns identify the parameters
  $u_{b}$ and $v_{b}$, and the corresponding standard deviations
  $\sigma_{b}^{s}$ at $\rho=1$ and kurtoses $\kappa_{b}^{s}$ for
  the prior GGD's. The last three columns identify the standard deviations
  $\sigma_{b}^{n}$ for the noise distributions in the noise conditions
  PA$-$IN, SA$-$tSZ, and SA$+$tSZ. All values are given with two significant
  figures.}
\end{table*}

\label{sec:ss}
By linearity of the wavelet transform, the coefficients of the
observed signal in relation (\ref{2-4}) are a sum of the wavelet coefficients
for the string signal and the Gaussian CMB noise, i.e.
\begin{equation}\label{4-1}
  W_{I}^{f}=W_{I}^{s}+W_{I}^{n}.
\end{equation}
The overall denoising algorithm will proceed by computing the wavelet
decomposition of the observed signal, estimating the coefficients
corresponding to the string signal, and finally inverting the wavelet
transform. Our Bayesian estimator requires knowledge of probability
distributions describing the behaviour of both the string signal and
the noise.  As we shall see later, part of our denoising procedure
will assume independence of the coefficients for different $I$,
allowing it to use a model for the marginal probability distribution of the
coefficients.

Notice that by statistical isotropy of both the signal and noise
processes, the probability distributions of the wavelet coefficients
for different spatial scales $j$ do not depend on position or
orientation. For notational convenience we introduce a generalized
spatial scale $b\in\{l,j,h\}$ for $1\leq j\leq J$. The above
comment implies that the signal and noise distributions depend only on
$b$. The distribution for the string coefficients will depend on the
string tension $\rho$. We write this as the conditional probability
$\pi_b(W_I^s|\rho)$. The noise coefficient distribution will be
denoted $g_b(W^n_I)$.

\subsection{String signal distribution}
\label{sec:ssd}
The morphology of the string signal should give rise to sparse
distribution of its wavelet coefficients, i.e. many coefficients are
close to zero with a small number of large magnitude coefficients near
the temperature steps. We observe this behaviour in our training
simulation. The sparse wavelet coefficients can be successfully
modeled by a class of probability distributions known as the
Generalized Gaussian Distributions (GGD).

We thus use a GGD to  model the prior distributions $\pi_{b}$:
\newcommand{\dvs}{x} 
\begin{equation}\label{4-2}
\pi_{b}\left(\dvs\vert\rho\right)=
\frac{v_{b}}{2\left(\rho u_{b}\right)\Gamma\left(v_{b}^{-1}\right)}
\exp\left[-\Big\vert\frac{\dvs}{\rho
    u_{b}}\Big\vert^{v_{b}}\right],
\end{equation}
where $\Gamma$ is the Gamma function, and where $v_b$ and $\rho u_b$
are respectively called shape parameters and scale parameters. These
distributions all have zero statistical means as the signal itself is
defined in relation (\ref{2-4}) as a zero mean perturbation.

Let us acknowledge that GGD's have been used previously to model
wavelet coefficients for various image processing applications
including denoising \citep{simoncelli96,moulin99}, deconvolution
\citep{belge00}, and coding \citep{antonini92,mallat98}.

The shape parameters $v_{b}$ can be considered as a continuous measure
of the sparsity of the underlying distribution. Setting $v_b=2$
recovers the Gaussian distribution, which is non-sparse. Letting $v_b$
approach $0$ yields very peaked probability distributions with heavy
tails relative to Gaussian distributions, i.e. very sparse
distributions. These parameters determine the kurtoses
$\kappa_{b}^{s}$, i.e. the ratio of the fourth central moment to the
square of the variance (second central moment), by
\begin{equation}\label{eq:ggd-kurtosis}
  \kappa_{b}^{s}=\frac{\Gamma\left(5v_{b}^{-1}\right)
\Gamma\left(v_{b}^{-1}\right)}
{\left[\Gamma\left(3v_{b}^{-1}\right)\right]^{2}}.
\end{equation}

The scale parameters $\rho u_{b}$ are linearly proportional to the
standard deviations $\sigma_{b}^{s}$ of the distributions. The
corresponding variances reflect the power spectrum (\ref{2-3}) of the
string signal in the range of spatial frequencies $k$ probed by the filter at
scale $b$, and thus also scale as $\rho^{2}$:
\begin{equation}\label{eq:ggd-variance}
\left(\sigma_{b}^{s}\right)^{2}=
\rho^{2}\frac{\Gamma\left(3v_{b}^{-1}\right)}
{\Gamma\left(v_{b}^{-1}\right)}u_{b}^{2}.
\end{equation}

The parameters $v_{b}$ and $u_{b}$ are estimated by a moment method
from the wavelet decomposition of the training simulation of the
string signal for a given string tension $\rho$.  For each spatial scale, the sample variance and kurtosis
are calculated, then equations (\ref{eq:ggd-kurtosis}) and
(\ref{eq:ggd-variance}) are solved numerically to obtain $v_{b}$ and
$u_{b}$.

Figure \ref{fig:histograms} shows the modeled prior GGD's $\pi_{b}$
for the coefficients of the string signal with the steerable wavelet
$\gamma$ with $N=6$ orientations and $J=4$ spatial scales (see
Figure \ref{fig:wavelet}). The GGD's are superimposed on the
histograms of the corresponding coefficients from the training
simulation.  As the distributions for coefficients of different
orientations at the same spatial scale will be identical by
statistical isotropy, the corresponding histograms are produced by
aggregating the coefficients over all $6$ orientations. Qualitatively,
we see that the prior distributions $\pi_{b}$ are well modeled by
GGD's, which justifies our choice of parameters $N$ and $J$.

The estimated values of the parameters $u_{b}$ and $v_{b}$, and
corresponding standard deviations $\sigma_{b}^{s}$ and kurtoses
$\kappa_{b}^{s}$ are reported in the columns two to five of Table
\ref{tab:statmodel}. Notice that the shape parameters measured for
the highpass band ($b=h$) and for the four bandpass bands ($j=1$ to $j=4$)
are significantly lower than $2$, corresponding to very sparse
distributions.  The larger value for the shape parameter for the
lowpass band justifies our choice of $J=4$ for the maximal spatial
scale. At the scales accounted for by the lowpass filter, the signal
coefficients are not significantly non-Gaussian and will not be very
sparsely distributed. The reconstruction of temperature steps
therefore does not strongly rely on those scales.

\subsection{Noise distribution}

The Gaussian probability distributions $g_{b}$ for the noise
coefficients $W_{I}^{n}$ are defined as:
\newcommand{\dvn}{x} 
\begin{equation}\label{4-5}
  g_{b}\left(\dvn\right)=\frac{1}{\sigma_{b}^{n}\sqrt{2\pi}}
  \exp\left[-\frac{1}{2}\left(\frac{\dvn}{\sigma_{b}^{n}}\right)^{2}\right].
\end{equation}

These distributions are all zero mean, as the noise itself is defined
in relation (\ref{2-4}) as a zero mean perturbation.
For each of the noise conditions PA$-$IN,
PA$+$IN, SA$-$tSZ, or SA$+$tSZ, the variances $(\sigma_{b}^{n})^{2}$
can be inferred from the power spectrum $\tilde{P}^{n}(k)$ of the
noise at $1$ arcminute resolution in the range of spatial frequencies
$k$ probed by the wavelets at the different spatial scales:
\begin{equation}\label{eq:noise-variance-from-ps}
  \left(\sigma_{b}^{n}\right)^{2}=
  \frac{1}{\tau^{2}}\sum_{\{p_{x,}p_{y}\}=-L}^{L-1}
  \vert\widehat{\gamma}_{G}\left(\veck\right)\vert^{2}
  \tilde{P}^{n}\left(k\right).
\end{equation}
In this relation, the multi-index value $G$ reads as $G=(q,j)$ with
$1\leq q\leq N$ and $1\leq j\leq J$ for the oriented wavelet
coefficients, $G=l$ for the lowpass coefficients, and $G=h$ for the
highpass coefficients. Notice that due to the rotational invariance of
the noise power spectrum, the variances calculated in
equation (\ref{eq:noise-variance-from-ps}) do not depend on the orientation
$q$. The values of the standard deviations $\sigma_{b}^{n}$ for the
noise conditions PA$-$IN, SA$-$tSZ, and SA$+$tSZ are listed in the last
three columns of Table \ref{tab:statmodel}.

\section{Bayesian denoising}

\label{sec:bd}
In this section, we define in detail our wavelet domain Bayesian
denoising (WDBD) algorithm. We define an overall Bayesian least
squares estimator as an average of estimation functions evaluated at
each value of the unknown string tension, weighted by the posterior
probability distribution for the string tension.  We then discuss
Wiener filtering as a standard alternative to our WDBD algorithm.

\subsection{Bayesian least squares}
\newcommand{\estm}{\overline}

In a Bayesian approach the signal coefficients $\vw^{s}$ are estimated
from their posterior probability distribution given the coefficients
of the observed signal $\vw^{f}$: $p(\vw^{s}\vert\vw^{f})$.  Under our
general assumption that the standard cosmological parameters are fixed
at their concordance values while the string tension remains
undetermined this posterior probability distribution reads as
\begin{equation}\label{eq:integrate_over_rho1}
p\left(\vw^{s}\vert\vw^{f}\right)=
\int\textnormal{d}\rho\, p\left(\rho\vert\vw^{f}\right)\, 
p\left(\vw^{s}\vert\vw^{f},\rho\right),
\end{equation}
where $p(\rho\vert\vw^{f})$ is the posterior probability distribution
function for $\rho$ given $\vw^{f}$ and $p\left(\vw^{s}\vert\vw^{f},\rho\right)$
is the posterior probability distribution function for $\vw^{s}$
given $\vw^{f}$ and $\rho$.

Several possible methods for selecting an appropriate estimate given
the posterior probability distribution are possible. Maximizing this
distribution leads to the maximum \emph{a posteriori} (MAP) estimate.
Other approaches may consist of minimizing some expected cost
function.  We employ the well known Bayesian least squares estimate
which minimizes a quadratic cost function. This estimate is given by
the expectation value of the posterior probability distribution.
Using relation (\ref{eq:integrate_over_rho1}) and the linearity of the
expectation value, our estimator may be written as
\begin{equation}\label{eq:global-estimate}
\estm{\vw^{s}}=
E\left[\vw^{s}\vert\vw^{f}\right]=
\int\textnormal{d}\rho\, p\left(\rho\vert\vw^{f}\right)\,
\estm{\vw^{s}}\left(\rho\right),
\end{equation}
with
\begin{equation}\label{eq:local-estimate}
\estm{\vw^{s}}\left(\rho\right) =
E\left[\vw^{s}\vert\vw^{f},\rho\right] \\
=\int\textnormal{d}\vw^{s}\,\vw^{s}\, 
p\left(\vw^{s}\vert\vw^{f},\rho\right).
\end{equation}
This estimation of the signal coefficients is thus given by the mean
of the estimations for different string tensions, weighted by
posterior probability distribution for $\rho$ given the observed
signal.

\subsection{Estimation functions}

\begin{figure}
\begin{center}
\includegraphics[width=8cm]{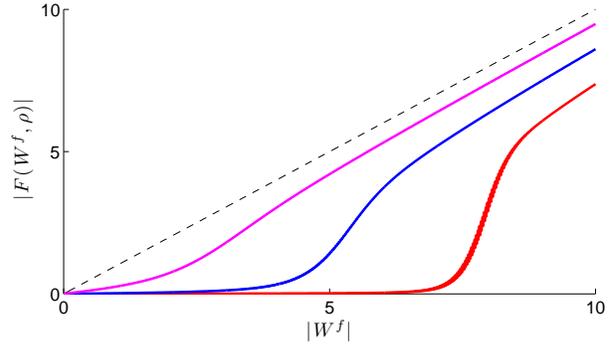}
\end{center}
\caption{\label{fig:estimation-function} Bayesian least squares
  estimation functions $F(\cdot,\rho)$ depending on observed
  coefficient $W_I^{f}$ at a non-specified spatial scale and for
  various string tensions $\rho$. We consider a shape parameter
  $v=0.5$ and various scale parameters $\rho u$ identifying various
  standard deviations $\sigma^{s}$ of the coefficients of the string
  signal, all for a unit standard deviation of the noise
  $\sigma^{n}=1$.  The black dashed curve shows the limit
  $\rho\rightarrow\infty$, where the estimation function is the
  identity function.  The upper solid curve (magenta) relates to $\rho
  u=0.1$, i.e. $\sigma^{s}\simeq1.1$, the middle solid curve (blue)
  relates to $\rho u=\scinot{1.5}{-2}$, i.e. $\sigma^{s}=0.16
  \comment{\scinot{1.6}{-1}}$, while the lower solid curve (red)
  relates to $\rho u=\scinot{5}{-3}$,
  i.e. $\sigma^{s}=\scinot{5.5}{-2}$.}
\end{figure}

We concentrate firstly on computation of
$\estm{\vw^{s}}\left(\rho\right)$ from equation
(\ref{eq:local-estimate}). 
Notice that the coefficients of the wavelet decomposition of the signal
and noise are correlated at different orientations, spatial scales and
positions. Formally one should construct probability distributions
accounting for these correlations. However, this would require
computing expectations in a space with dimension equal to the number
of coefficients of the wavelet decomposition.
In this perspective, approaches accounting for correlations of the
string signal developed in the framework of maximum entropy methods
(MEM) \citep{gull99,maisinger04} might be considered. However such
methods assume entropic prior models and are therefore not directly
compatible with our prior model in terms of GGD's for the coefficients
of the string signal.

\renewcommand{\sc}{x} 
\newcommand{\dc}{y} 
\newcommand{\nc}{n} 

Accordingly, we employ the simplifying assumption that the wavelet
coefficients for both the signal and noise for different values of $I$
are independent, after conditioning on the string tension $\rho$.
Under this assumption, the integral in expression
(\ref{eq:local-estimate}) may be factorized, and each coefficient
$\estm{W_I^s}$ of the estimate depends only on the corresponding
observed value $W_{I}^{f}$.  To simplify the notation in the
following, we write $\sc=W_I^s$ and $\dc=W_I^f$ to refer to the
individual pure string coefficients and observed signal
coefficients. We shall see that the resulting estimator will also
depend on the spatial scale $b$.

Each coefficient $\sc$ may then be estimated as a function of the
corresponding observed value $\dc$. By Bayes' theorem we have that
$p(\sc|\dc,\rho)\propto p(\dc|\sc,\rho) p(\sc|\rho)$.  The probability
$p(\sc|\rho)$ is exactly the marginal probability for each
coefficient, which we have modeled as the GGD $\pi_b$. Conditioned on
the signal $\sc$, the probability of observing $\dc$ is equal to the
probability of the noise coefficient being equal to exactly
$\dc-\sc$. Thus $p(\dc|\sc,\rho)$ is equal to $g_b(\dc-\sc)$. We thus
have the posterior probability distribution
\begin{equation}
p(x|y,\rho) =  \comment{\frac{1}{C}}C^{-1}\mbox{ } g_b(y-x) \pi_b(x|\rho) 
\end{equation}
and our Bayesian estimator at spatial scale $b$ is
\begin{equation}
  E[\sc|\dc,\rho]=
  C^{-1} \comment{\frac{1}{C}} \int \sc g_b(\dc-\sc) \pi_b(\sc|\rho) dx
\end{equation}
with normalization $C=\int g_b(\dc-\sc) \pi_b(\sc|\rho) d\sc$.  This
expression depends only on the observed coefficient $\dc$, the tension
$\rho$ and the scale $b$. This defines the estimation function
$F_b(\dc,\rho) = E[\sc|\dc,\rho]$. Returning to our original notation,
the estimated string coefficients are given by evaluation of the
estimation function at each scale, i.e.
\begin{equation}\label{5-8}
\estm{W_{I}^{s}}\left(\rho\right)=
F_{b}\left(W_{I}^{f},\rho\right).
\end{equation}

In practice, these estimation functions are computed by numerical
integration and tabulated for the different spatial scales $b$ and
the required range of string tensions.  Figure
\ref{fig:estimation-function} shows generic shapes of
 estimation functions $F(\cdot,\rho)$ at a non-specified
spatial scale and for various string tensions $\rho$. For the sake of
illustration, we consider a shape parameter $v=0.5$ and various scale
parameters $\rho u$ identifying various standard deviations
$\sigma^{s}$ of the coefficients of the string signal, all for a unit
standard deviation of the noise $\sigma^{n}=1$. Notice that the
estimation functions are odd, and always shrink the magnitude of their
input, i.e.  $\vert F(W_I^{f},\rho)\vert<\vert W_I^{f}\vert$.
Qualitatively, they behave as a smooth thresholding operation on the
observed coefficient $\vert W_I^{f}\vert$, sending small magnitude
coefficients closer to zero while preserving large magnitude
coefficients. For small string tensions, the noise dominates the
signal and the effective thresholding is more severe, while for large
string tensions the noise becomes negligible and $F_b(\cdot,\rho)$
reduces to the identity.

In the particular case of Gaussian signal coefficients ($v=2$), the
Bayesian least squares estimation is equivalent to simple Wiener filtering
of the coefficients. Also notice that in the case of Laplacian signal
coefficients ($v=1$), the estimation function for MAP estimation
would reduce to the well known soft-thresholding operation \citep{moulin99}.
By definition, this specific instance of thresholding operation sends
to zero coefficients with an absolute value below some threshold,
and reduces the absolute value of coefficients above the threshold
by the value of the threshold itself.

\subsection{Posterior string tension distribution}

By Bayes' theorem, the posterior probability distribution function
for $\rho$ given the  observed signal $p(\rho\vert\vw^{f})$
is simply obtained from the likelihood $\mathcal{L}(\vw^{f}\vert\rho)$
and the prior probability distribution function $p(\rho)$ on $\rho$.
For complete consistency, the likelihood should be calculated using
the framework of the model established for the coefficients, based
on relation (\ref{4-1}) and on the prior GGD's $\pi_{b}$. However,
while this model by construction accounts for the non-Gaussianity,
i.e. sparsity, of the string signal, it ignores the correlation between
coefficients.

We have observed that a likelihood yielding a more precise
localization of the string tension value can actually be obtained
using a power spectral model. Such a model
assumes both the string signal and the noise arise from statistically
isotropic Gaussian random processes, such that their Fourier
coefficients are independent Gaussian variables.  This assumption
relies on the idea that the characteristic temperature steps of the
string signal are smoothed by projection on the non-local imaginary
exponentials defining the Fourier basis. Under this model, as the
string signal and noise are independent, the observed signal $f$ has a
power spectrum
\begin{equation}\label{5-9}
\tilde{P}\left(k,\rho\right)=
\tilde{P}^{n}\left(k\right)+\tilde{P}^{s}\left(k,\rho\right),
\end{equation}

In this setting the likelihood can be computed most easily in terms
of the Fourier transform $\widehat{f}$ of the  observed signal. Accounting
for the complex value of the Fourier coefficients as well as for the
symmetry $\widehat{f}(-\veck)=\widehat{f}^{*}(\veck)$ that holds
for real signals $f(\vecx)$, this likelihood reads as: 
\begin{equation}\label{5-10}
\mathcal{L}\left(\widehat{f}\vert\rho\right)=
\prod_{\{p_{x,}p_{y}\}=-L}^{L-1}
\frac{1}{\sqrt{\pi\tilde{P}\left(k,\rho\right)}}
\exp\left[-\frac{1}{2} 
\frac{\vert\widehat{f}\left(\veck\right)\vert^{2}}
{\tilde{P}\left(k,\rho\right)}
\right],
\end{equation}
where $\vert \cdot \vert$ stands for the modulus of a complex variable.
The posterior probability distribution function for $\rho$ given
the  observed signal thus reads as 
\begin{equation}\label{5-11}
p\left(\rho\vert\widehat{f}\right)=
D^{-1}\mbox{ }p\left(\rho\right)\mathcal{L}\left(\widehat{f}\vert\rho\right)
\end{equation}
with normalization $D=\int p(\rho)\mathcal{L}(\hat{f}|\rho) d\rho$.
We take the prior $p(\rho)$ to be flat in an interval
$\rho\in[0,\rhomax]$, with an upper bound $\rhomax$ large enough
relative to the upper bound associated with the best experimental
constraints (\ref{2-5}): $\rhomax>\rhoexp$.  In practice,
$\mathcal{L}(\widehat{f}\vert\rho)$ decays so rapidly for large $\rho$
that the resulting posterior is not sensitive to the value $\rhomax$
provided that it is greater than the effective support of
$\mathcal{L}(\widehat{f}\vert\rho)$.

We use this PSM posterior $p(\rho\vert\widehat{f})$ in the place of
$p(\rho|\vw^f)$ in equation (\ref{eq:global-estimate}).  Each
component of the string coefficient $\vw^s$ is thus estimated as
\begin{equation}\label{eq-int_F}
\estm{W^s_I} = \int p(\rho|\widehat{f}) F_b(W^f_I,\rho) d\rho.
\end{equation}
In practice, this integral is computed numerically by sampling 20
values of $\rho$ chosen to cover the effective support of
$p(\rho|\widehat{f})$.

The estimated string signal in the original image domain is then given
by inverting the wavelet transform, i.e.
\begin{equation}\label{eq:image-domain-string-estimate}
\estm{s}(\vecx) = \left[\mR \estm\vw^s\right] (\vecx)
\end{equation}

\subsection{Alternative Wiener filtering}

In order to obtain a more precise estimation of the posterior
probability distribution function for $\rho$, we have explicitly set
up a PSM assuming that the string signal arises from
a statistically isotropic Gaussian random process, such that its
Fourier coefficients are independent Gaussian variables, just as for
the noise. 

At each string tension allowed by $p(\rho\vert\widehat{f})$, one may
now consider estimating the string signal $s$ from the observed signal
$f$ simply using this Gaussianity assumption. In this case the
Bayesian least squares estimate for a string tension $\rho$ reduces to
Wiener filtering in the Fourier domain, so that:

\begin{equation}\label{5-14}
\estm{\widehat{s}}\left(\veck,\rho\right)=
\frac{\tilde{P}^{s}\left(k,\rho\right)}
 {\tilde{P}^{n}\left(k\right)+\tilde{P}^{s}\left(k,\rho\right)}
\widehat{f}\left(\veck\right).
\end{equation}

Analogously to relation (\ref{eq-int_F}), the estimate of the string signal in
the Fourier domain is
\begin{equation}
\estm{\widehat{s}}(\veck) = \int p(\rho|\widehat{f}) 
\estm{\widehat{s}}(\veck,\rho) d\rho,
\end{equation}
and the estimate in the image domain is recovered by inverting the
Fourier transform.

Let us acknowledge the fact that this alternative Wiener filtering
based procedure relies only on the knowledge of the power spectra of
both the signal and noise, while our WDBD approach relies on a
training simulation for an explicit modeling of the prior GGD's for
the coefficients of a wavelet decomposition of the string
signal. However, from the theoretical point of view it is clear that
the Wiener filtering approach, which disregards the non-Gaussianity of
the signal to be recovered, will be less effective at identifying this
signal than our WDBD procedure, which explicitly accounts for the
corresponding sparsity. While the Gaussianity assumption is useful for
estimating a single global parameter such as the string tension on the
basis of a PSM, it is not optimal for the explicit
reconstruction of the sparse features of the string network
itself. This fact is illustrated in our analysis  the algorithm
performance in the next section.

\section{Algorithm performance} \label{sec:WDBDap}

In this section, we firstly define the WDBD performance criteria to be
the signal-to-noise ratio, correlation coefficient, and kurtosis of
the map of the magnitude of gradient of the string signal. We then
study the algorithm performance in each noise condition, in comparison
with Wiener filtering. We also examine a detectability threshold on
the string tension based on on the PSM, and compare it with an eye
visibility threshold for the WDBD algorithm.

\subsection{WDBD performance criteria}

As already emphasized, denoising may be used as a pre-processing step
for other methods for cosmic string detection based on explicit edge
detection \citep{jeong05,lo05,amsel07}. The relative performance
of such methods before and after denoising might be an effective criterion
for evaluating the denoising performance itself. Here we evaluate
the performance of WDBD independently of any further processing.

The overall denoising is effective for string mapping if the magnitude
of gradient of the denoised signal closely resembles the magnitude of
gradient of the true string signal.  A simple qualitative measure of
the denoising performance is given by whether the string network is
visible in the magnitude of gradient of the denoised signal. We define
the eye visibility threshold as the minimum string tension around
which the overall denoising and mapping by the magnitude of the gradient
begins to exhibit string features visible by eye. We will augment this
qualitative assessment of the denoising performance with three
quantitative measures, namely the signal-to-noise ratio, the
correlation coefficient and the kurtosis of the magnitude of the
gradient of the denoised string signal. The first two of these are
computed with respect to the original known signal, while the kurtosis
is computed only using the denoised signal. The kurtosis is known to
be a good statistic for discriminating between models with and without
cosmic strings \citep{moessner94}.

\newcommand{\vecnabla}{\mathbf{\nabla}}

The signal-to-noise ratio is defined in terms of the magnitude of
gradient $\vert\vecnabla s\vert(\vecx)$ of the original string
signal $s(\vecx)$ in relation (\ref{2-4}), and of the magnitude of gradient
$\vert\vecnabla \overline{s}\vert(\vecx)$ of the denoised signal
$\overline{s}(\vecx)$ in relation (\ref{eq:image-domain-string-estimate}) as 
\begin{equation}\label{6-1}
\textnormal{SNR}^{(\vert\vecnabla s\vert,\vert\vecnabla \overline{s}\vert)}=
-20\log_{10}\frac{\sigma^{(\vert\vecnabla s\vert-\vert\vecnabla \overline{s}\vert)}}{\sigma^{\vert\vecnabla s\vert}},
\end{equation}
where
$\sigma^{(\vert\vecnabla s\vert-\vert\vecnabla \overline{s}\vert)}$
and $\sigma^{\vert\vecnabla s\vert}$ respectively stand for the
standard deviations of the discrepancy signal
$\vert\vecnabla s\vert-\vert\vecnabla \overline{s}\vert$ and of
the original signal $\vert\vecnabla s\vert$. The standard deviations
are estimated from the sample variances on the basis of the signal
realizations concerned. With this definition, the
$\textnormal{SNR}^{(\vert\vecnabla s\vert,\vert\vecnabla \overline{s}\vert)}\in\mathbb{R}$
is measured in decibels (dB). Large negative and positive values are
respectively associated with large and small discrepancy signals
relative to the original signal. An exact recovery of the string
network would provide an infinite signal-to-noise ratio.  We will
consider that the denoising is effective in terms of signal-to-noise
ratio for the values of $\rho$ where this statistic is larger after
denoising than before, and positive.

The correlation coefficient is defined in terms of the magnitude of
gradient of the original and denoised string signals as
\begin{equation}\label{6-2}
r^{(\vert\vecnabla s\vert,\vert\vecnabla \overline{s}\vert)}=
\frac{\textnormal{cov}^{(\vert\vecnabla s\vert,\vert\vecnabla \overline{s}\vert)}}
{\sigma^{\vert\vecnabla s\vert}\sigma^{\vert\vecnabla \overline{s}\vert}},
\end{equation}
where
$\textnormal{cov}^{(\vert\vecnabla s\vert,\vert\vecnabla \overline{s}\vert)}$
stands for the covariance between $\vert\vecnabla s\vert$ and
$\vert\vecnabla \overline{s}\vert$.  This signal covariance is also
estimated from the sample covariance on the basis of the signal
realizations concerned. An exact recovery of the string network would
provide a unit correlation coefficient.  The null value corresponds to
a reconstruction completely decorrelated from the original signal. We
will consider that the denoising is effective in terms of correlation
coefficient for the values of $\rho$ where this statistic is larger
after denoising than before, and positive.

Analogously, kurtoses are estimated from the sample kurtoses on the
basis of the signal realizations concerned. The estimated kurtosis of
the magnitude of gradient of pure Gaussian noise is distributed around
a mean value across all test simulations
$\kappa^{\vert\vecnabla n\vert}\simeq3$, even though the magnitude
of gradient itself is not Gaussian. At the arcminute resolution
considered, the estimated kurtosis of the magnitude of gradient of a
pure string signal is much higher than the value associated with pure
noise, with a value around $\kappa^{\vert\vecnabla s\vert}\simeq32$
in the training simulation. The estimated kurtosis of the magnitude of
gradient of a string signal with noise before denoising naturally lies
in the interval
$[\kappa^{\vert\vecnabla n\vert},\kappa^{\vert\vecnabla s\vert}]$,
for any value of the string tension $\rho$. Its mean value
$\kappa^{\vert\vecnabla f\vert}(\rho)$ obviously increases from
$\kappa^{\vert\vecnabla n\vert}$ for $\rho=0$ to
$\kappa^{\vert\vecnabla s\vert}$ for $\rho\rightarrow\infty$.  An
ideal denoising procedure should recover exactly the original string
signal. The mean value of the estimated kurtosis would then be raised
to $\kappa^{\vert\vecnabla s\vert}$ after denoising. In practice,
the estimated kurtosis of the magnitude of gradient after denoising is
distributed around some mean value
$\kappa^{\vert\vecnabla \overline{s}\vert}(\rho)$ as a function the
string tension $\rho$. The comparison
$\kappa^{\vert\vecnabla \overline{s}\vert}(\rho)$ after denoising
with $\kappa^{\vert\vecnabla f\vert}(\rho)$ before denoising
measures the denoising performance as a function of the string
tension. We will simply consider that the denoising is effective in
terms of kurtosis for the values of $\rho$ where this statistic is
significantly larger after denoising than before.

Our denoising experiments for each noise condition considered are
performed for string tensions equi-spaced in logarithmic scaling in
the range $\log_{10}\rho\in[-10,-05]$, corresponding to ratio values
for $\rho$ of $1.0$, $1.6$, $2.5$, $4.0$, and $6.3$ in each order of
magnitude.  For each noise condition and string tension considered, we
perform $100$ denoising simulations at 1 arcminute resolution. We
consider that the quantitative measures described above indicate
effective denoising performance for a given string tension when they
show effective performance significant over the entire ensemble of
denoising simulations.

\subsection{Noise conditions PA$-$IN and PA$+$IN}

\begin{figure*}
\begin{center}
\includegraphics[width=5.5cm]{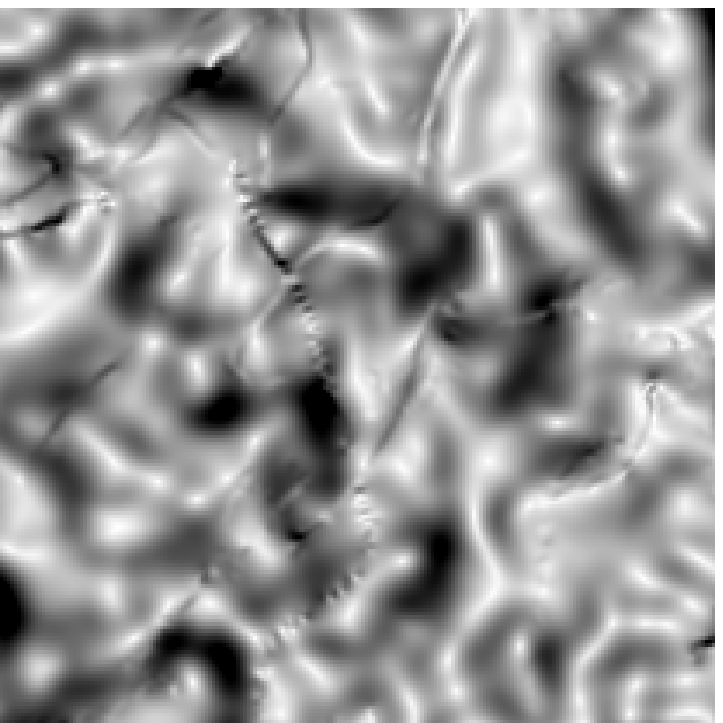}\hspace{2mm}
\includegraphics[width=5.5cm]{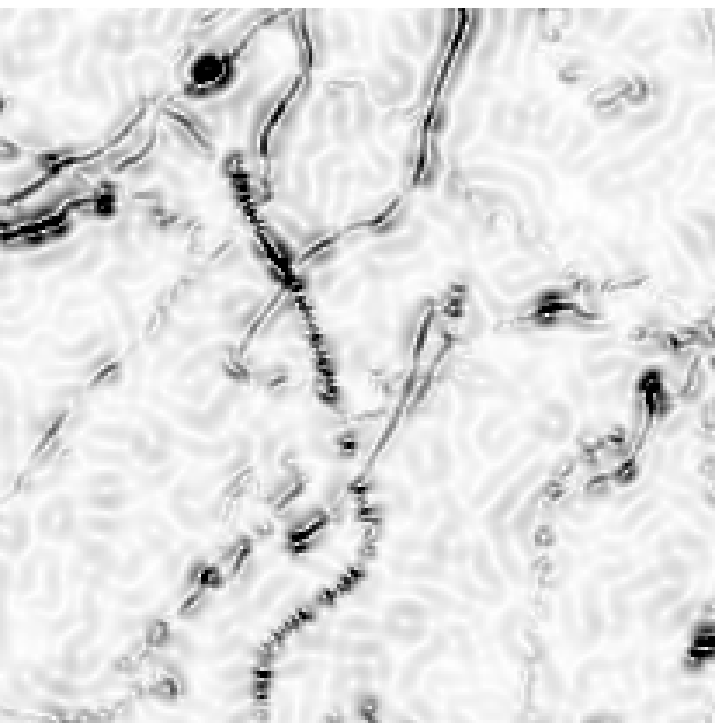}\hspace{2mm}
\includegraphics[width=5.5cm]{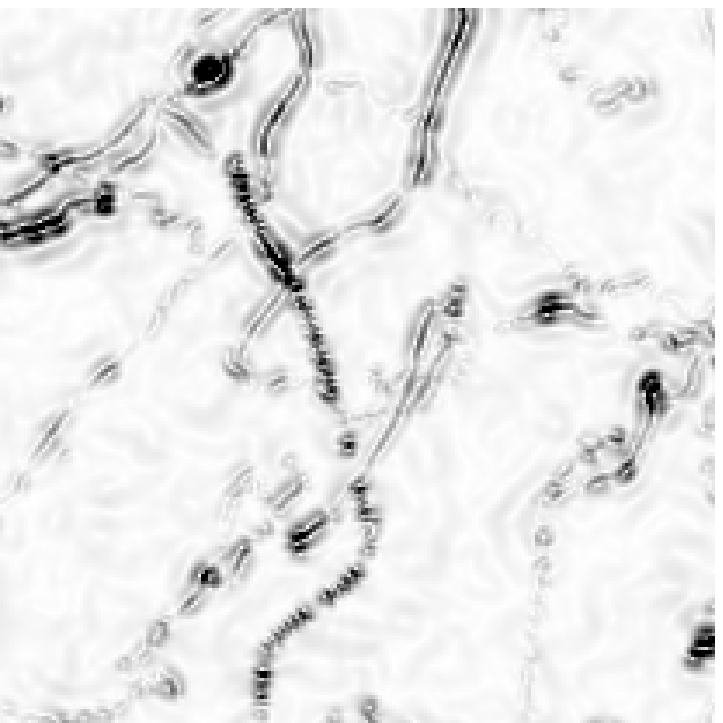}\vspace{2mm}

\includegraphics[width=5.5cm]{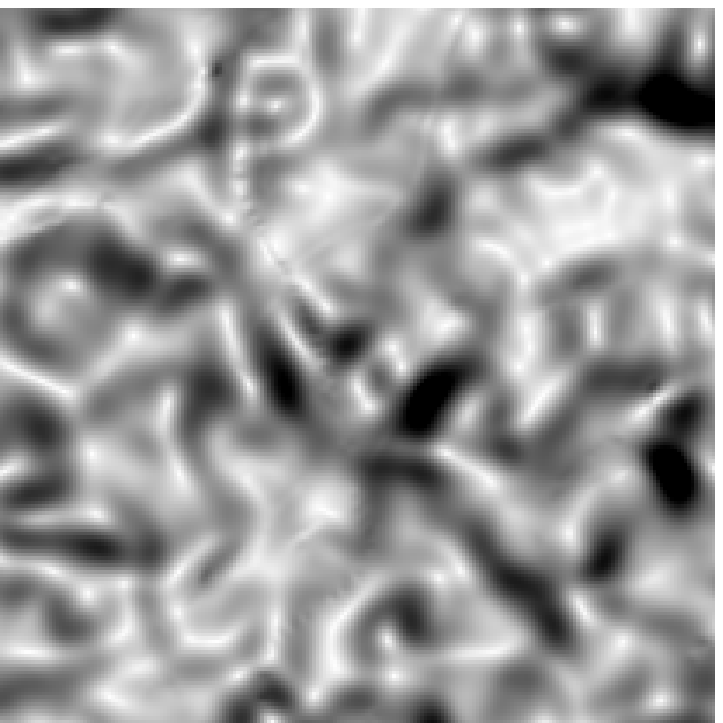}\hspace{2mm}
\includegraphics[width=5.5cm]{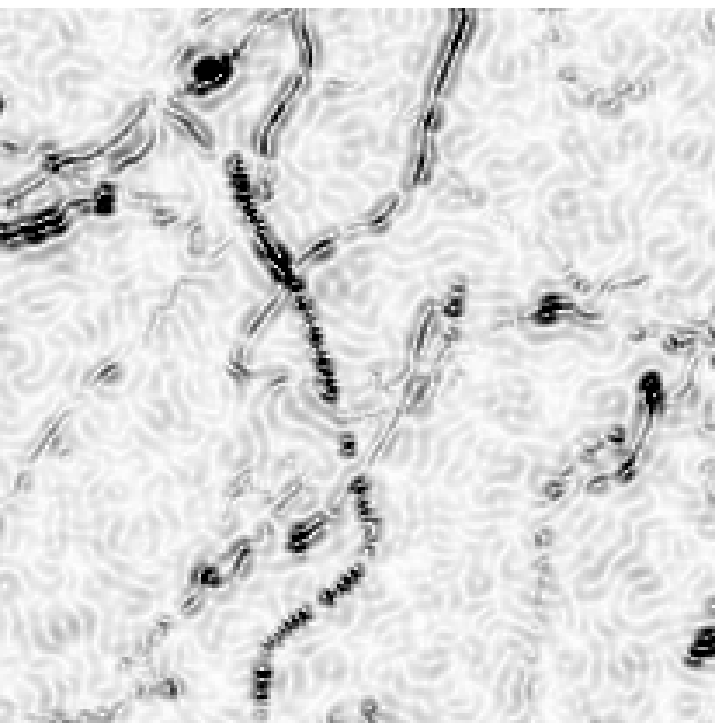}\hspace{2mm}
\includegraphics[width=5.5cm]{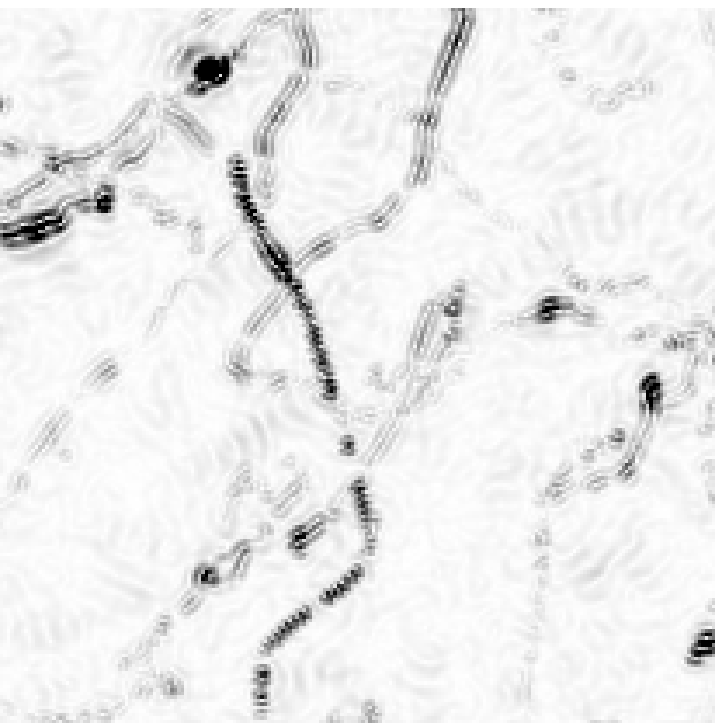}\vspace{2mm}

\includegraphics[width=5.5cm]{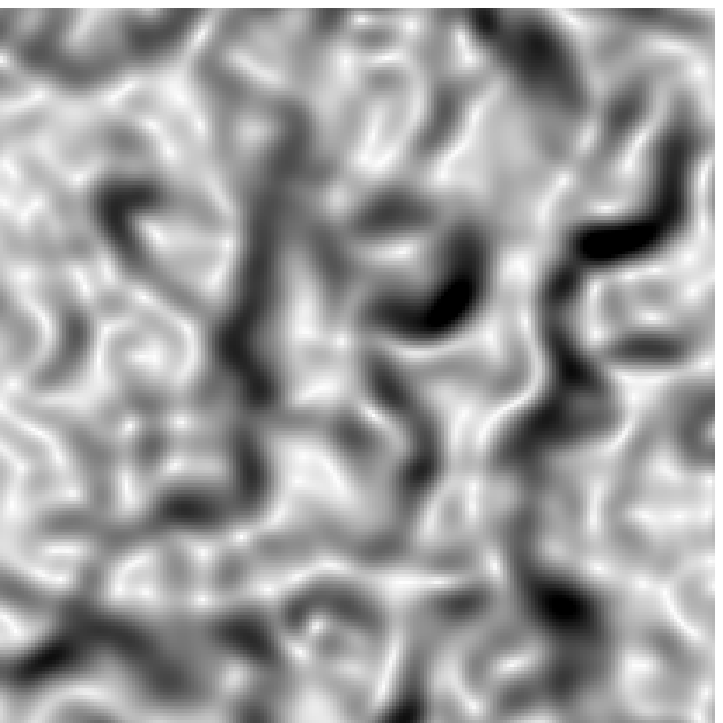}\hspace{2mm}
\includegraphics[width=5.5cm]{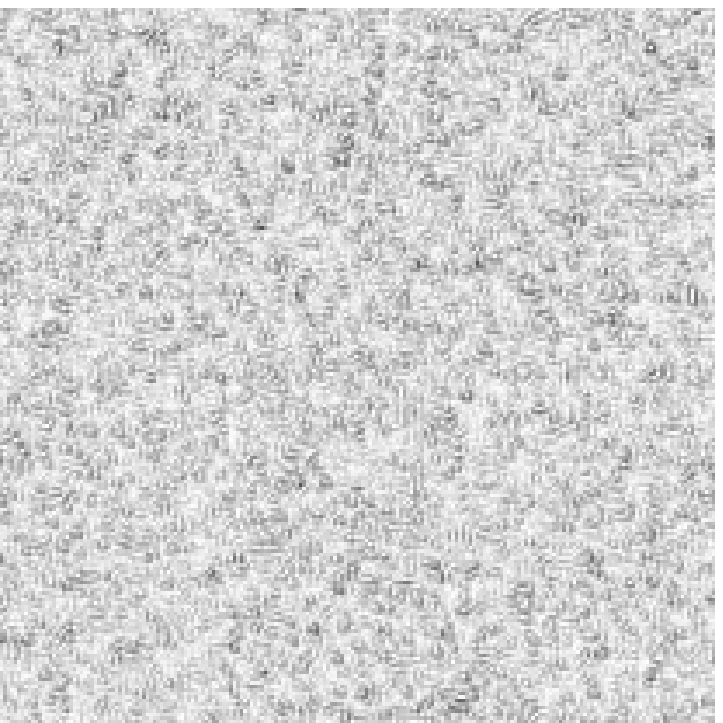}\hspace{2mm}
\includegraphics[width=5.5cm]{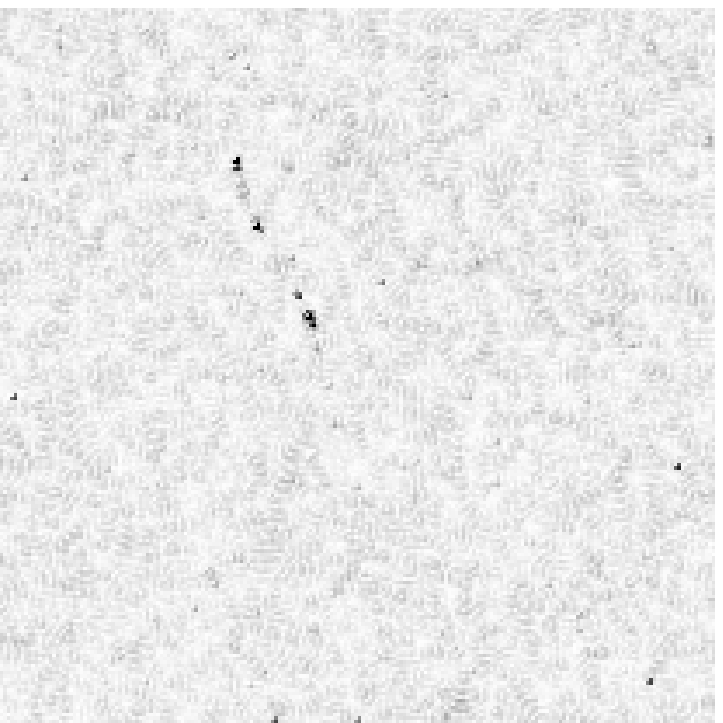}\end{center}

\caption{\label{fig:gradient-beforeafter}
  Magnitude of the gradient of the string signal before denoising
  (left panels), after Wiener filtering (middle panels) and WDBD
  (right panels), in the noise conditions PA$-$IN and for various
  string tensions at $1$ arcminute resolution on a field of view of
  $\tau'=1.4^{\circ}$. From top to bottom, the string tensions
  considered are $\rho=\scinot{2.5}{-7}$, $\rho=\scinot{6.3}{-8}$, and
  $\rho=\scinot{6.3}{-10}$.}
\end{figure*}
\begin{figure*}
\begin{center}
\newcommand{\tsc}{0.9}
\psfrag{2.2e-07}[][][\tsc]{\scinot{2.2}{-7}}
\psfrag{2.5e-07}[][][\tsc]{\scinot{2.5}{-7}}
\psfrag{2.8e-07}[][][\tsc]{\scinot{2.8}{-7}}
\psfrag{5.6e-08}[][][\tsc]{\scinot{5.6}{-8}}
\psfrag{6.3e-08}[][][\tsc]{\scinot{6.3}{-8}}
\psfrag{7.0e-08}[][][\tsc]{\scinot{7.0}{-8}}
\psfrag{5.6e-10}[][][\tsc]{\scinot{5.6}{-10}}
\psfrag{6.3e-10}[][][\tsc]{\scinot{6.3}{-10}}
\psfrag{7.0e-10}[][][\tsc]{\scinot{7.0}{-10}}
\psfrag{0.05}[][][\tsc]{0.05}
\psfrag{0.04}[][][\tsc]{0.04}
\psfrag{0.03}[][][\tsc]{0.03}
\psfrag{0.02}[][][\tsc]{0.02}
\psfrag{0.01}[][][\tsc]{0.01}
\psfrag{0}[][][\tsc]{0\hspace{2em}}
\includegraphics[width=5cm]{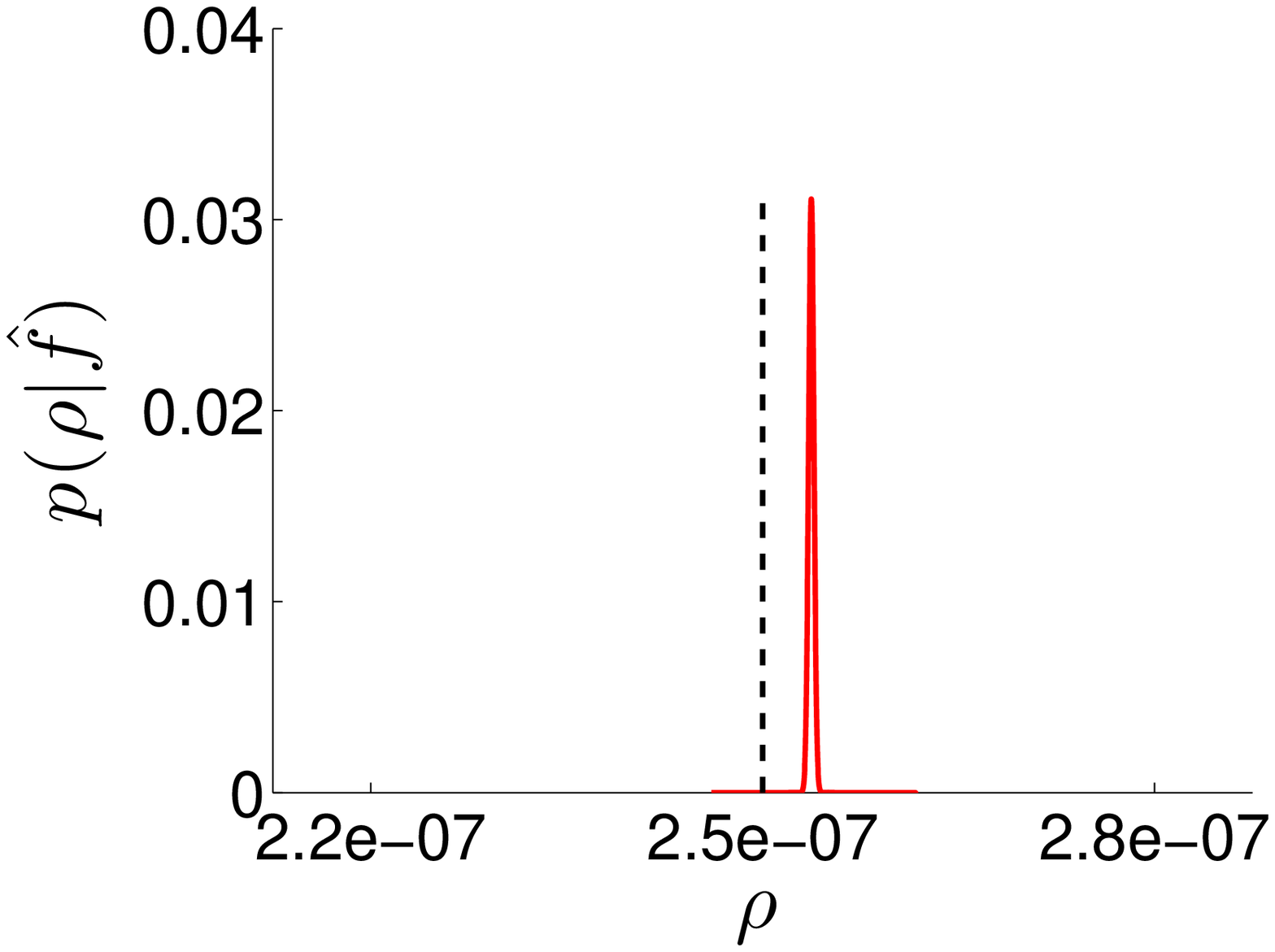}\hspace{1cm}
\includegraphics[width=5cm]{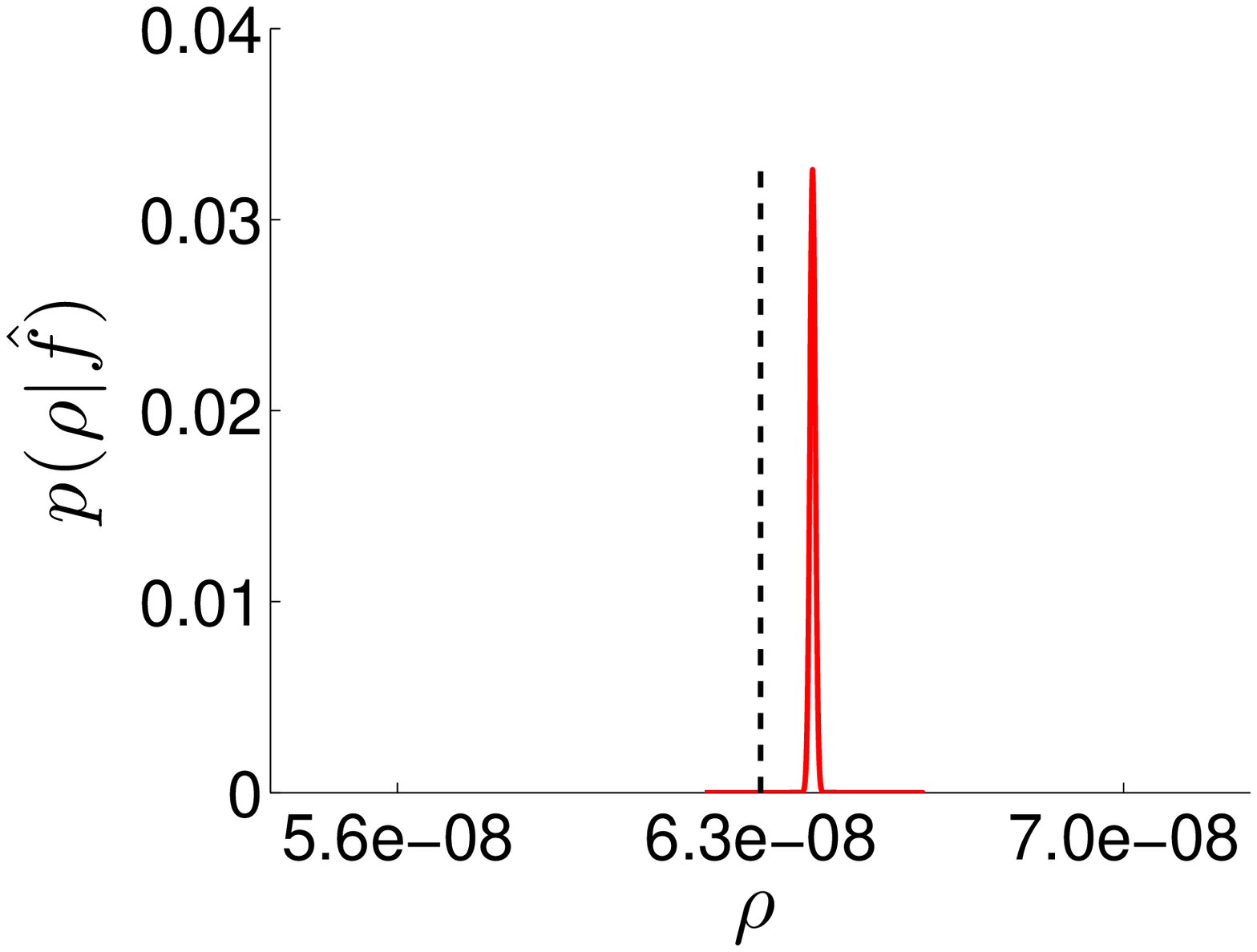}\hspace{1cm}
\includegraphics[width=5cm]{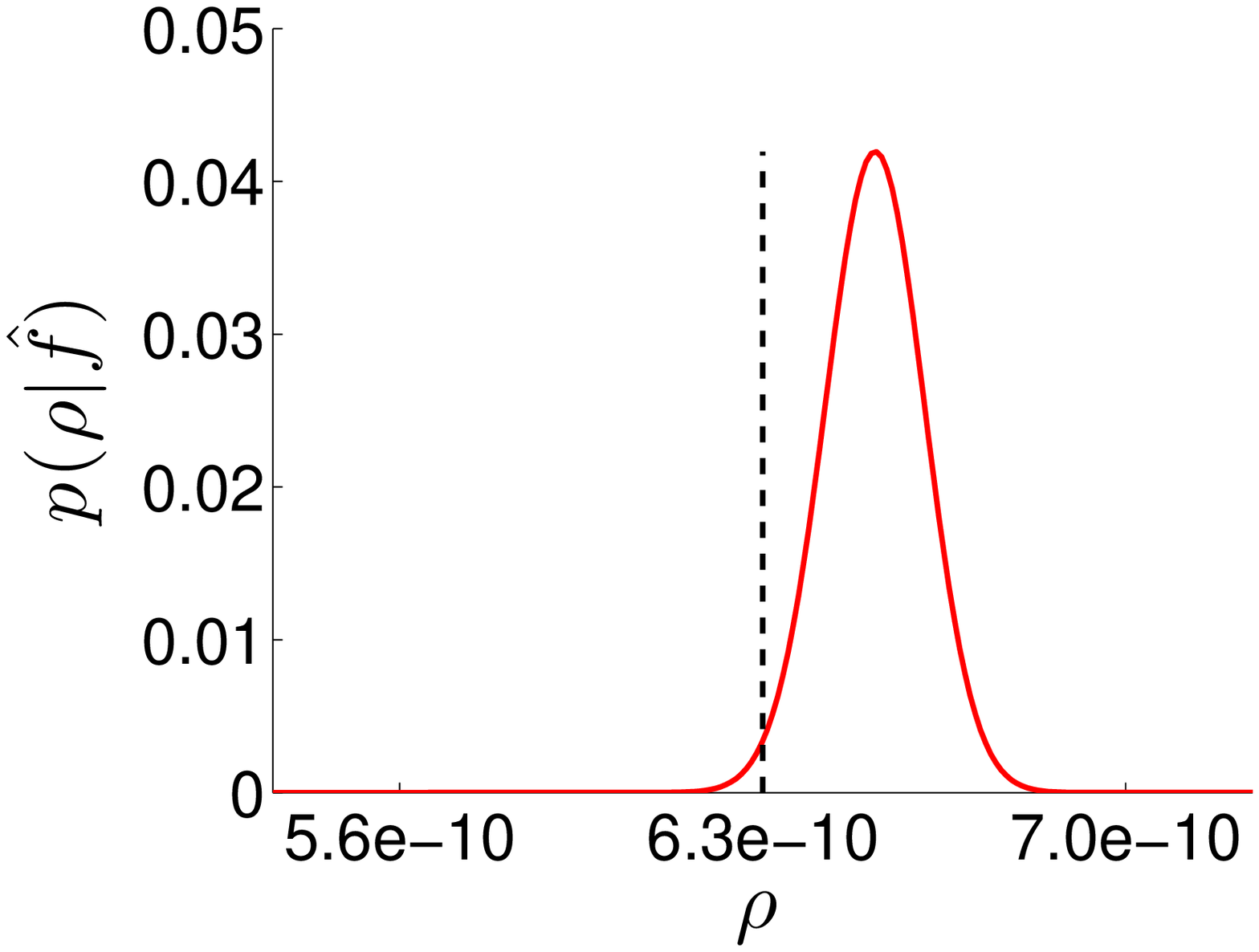}
\end{center}
\caption{\label{fig:Posterior-pdf} 
  Posterior probability distributions (red solid curves) for the
  string tension as computed from simulated signals observed for
  $\rho=\scinot{2.5}{-7}$ (left panel), $\rho=\scinot{6.3}{-8}$
  (middle panel), and $\rho=\scinot{6.3}{-10}$ (right panel), in the
  noise conditions PA$-$IN at $1$ arcminute resolution. The black
  dashed vertical lines represent the exact values of the string
  tension relative to the test string simulation.}
\end{figure*}
\begin{figure*}
\begin{center}
\newcommand{\tsc}{0.9}
\psfrag{20}[][][\tsc]{20}
\psfrag{0}[][][\tsc]{0\hspace{0.5em}}
\psfrag{-20}[][][\tsc]{-20}
\psfrag{-40}[][][\tsc]{-40}
\psfrag{-60}[][][\tsc]{-60}
\psfrag{-80}[][][\tsc]{}
\psfrag{10}[][][\tsc]{10}
\psfrag{-10}[][][\tsc]{-10}
\psfrag{-9}[][][\tsc]{-9}
\psfrag{-8}[][][\tsc]{-8}
\psfrag{-7}[][][\tsc]{-7}
\psfrag{-6}[][][\tsc]{-6}
\psfrag{-5}[][][\tsc]{-5}
\psfrag{-0.5}[][][\tsc]{-0.5}
\psfrag{0.5}[][][\tsc]{0.5}
\psfrag{1}[][][\tsc]{1\hspace{0em}}
\psfrag{1.5}[][][\tsc]{1.5}
\psfrag{50}[][][\tsc]{50}
\psfrag{100}[][][\tsc]{100}
\psfrag{150}[][][\tsc]{150}
\psfrag{200}[][][\tsc]{200}
\psfrag{snrdb}[][]{SNR (dB)}
\psfrag{correlationcoefficient}[][]{Correlation Coefficient}
\psfrag{kurtosis}[][]{Kurtosis}
\psfrag{log10rho}[][]{$\log_{10}(\rho)$}

\includegraphics[width=5cm]{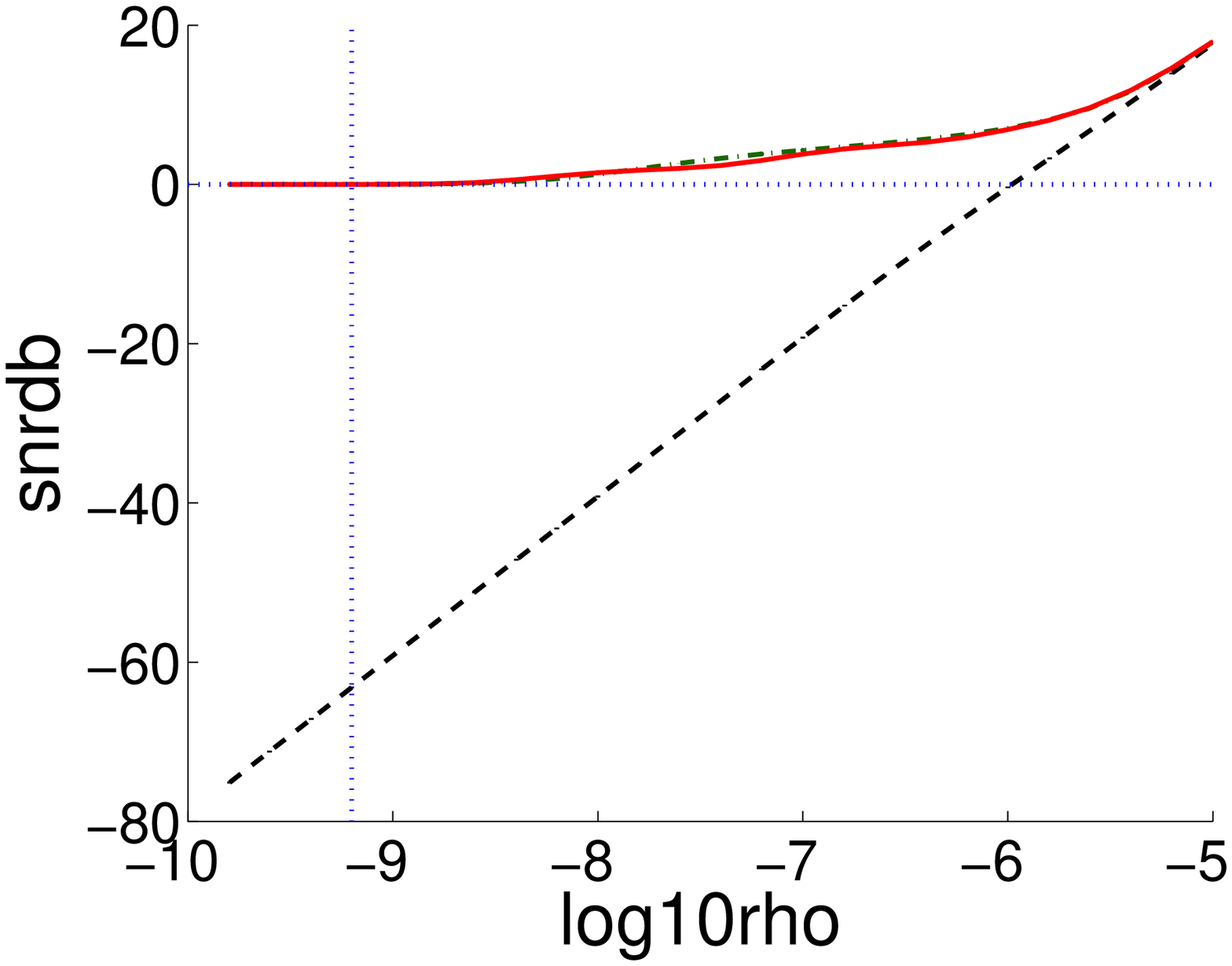}\hspace{1cm}
\includegraphics[width=5cm]{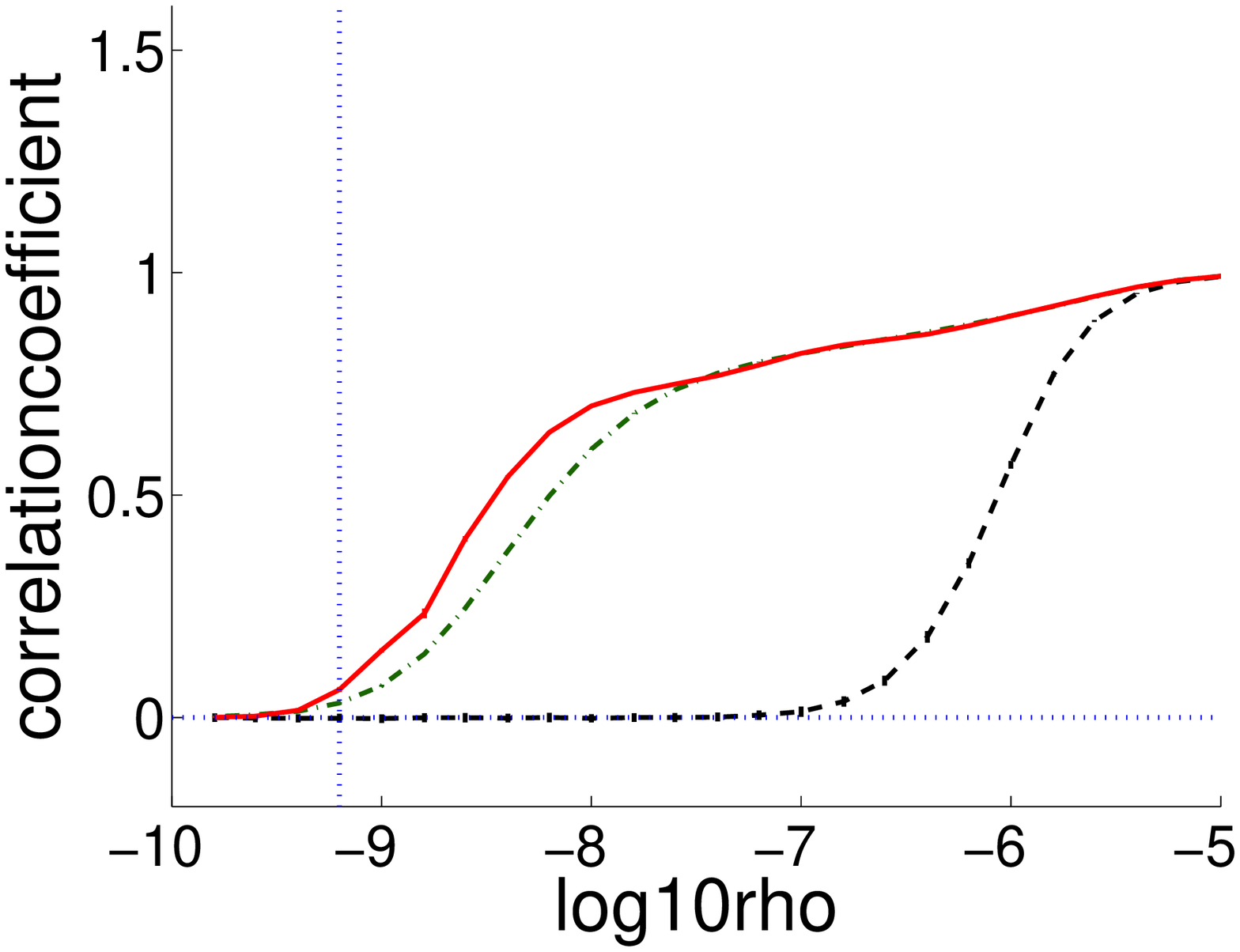}\hspace{1cm}
\includegraphics[width=5cm]{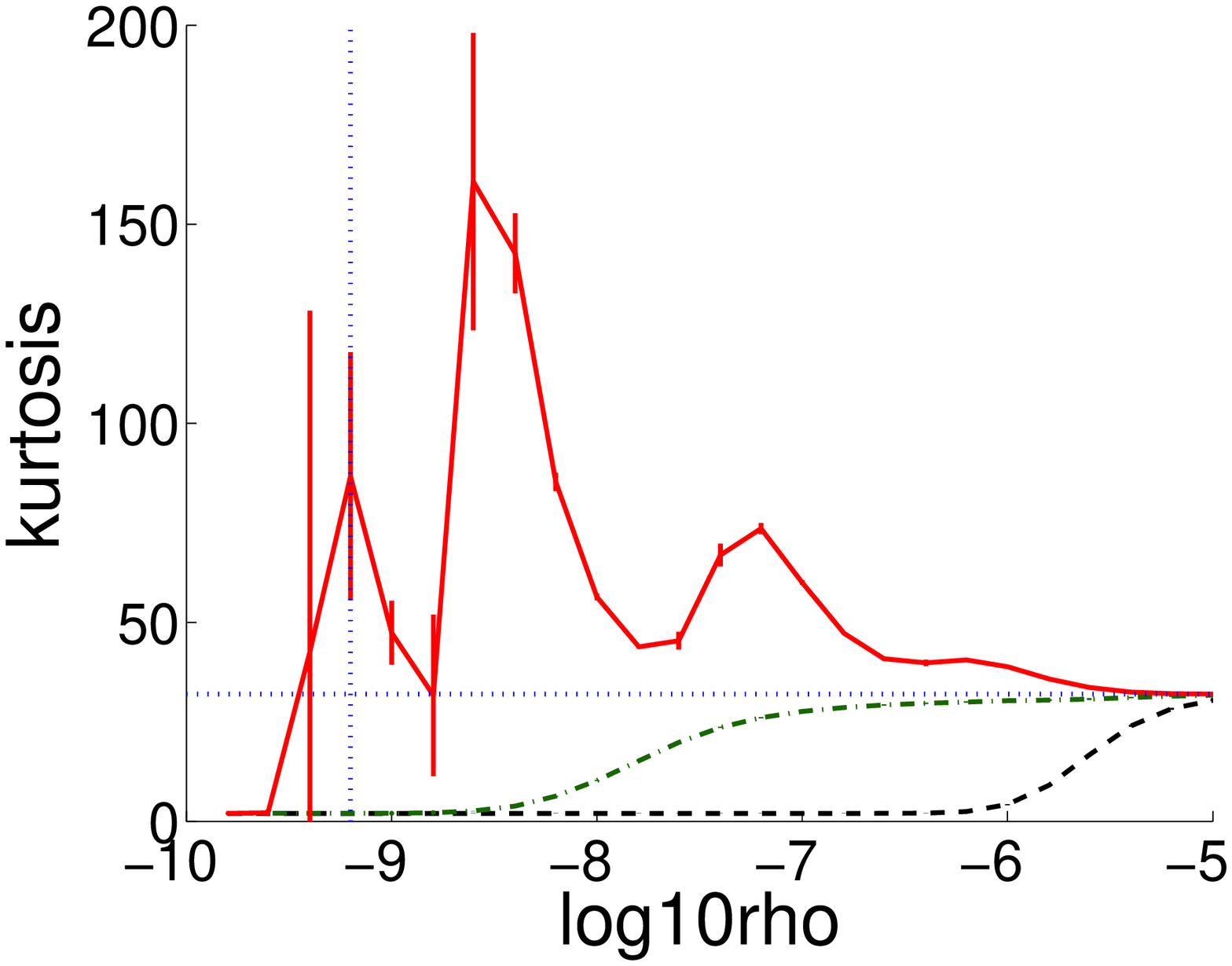}
\end{center}

\caption{\label{fig:kurtosis}
  Signal-to-noise ratio (left panel) in decibels (dB), correlation
  coefficient (middle panel), and kurtosis (right panel) of the
  magnitude of gradient as functions of the string tension in
  logarithmic scaling in the range $\log_{10}\rho\in[-10,-05]$ in the
  noise conditions PA$-$IN and at $1$ arcminute resolution. The black
  dashed curves represent values before denoising, while the red solid
  curves and green dot-dashed curves represent values after WDBD and
  Wiener filtering respectively. The vertical lines on the curves
  represent the variability at one standard deviation of the estimated
  statistic across the $100$ test simulations considered (these lines
  are not visible where smaller than the width of the curves).  The
  blue dotted vertical lines represent the eye visibility threshold
  $\rho=\scinot{6.3}{-10}$. The blue dotted horizontal lines identify
  either the limit of zero signal-to-ratio, zero correlation
  coefficient, or the kurtosis of the magnitude of gradient of a pure
  string signal: $\kappa^{\vert\vecnabla s\vert}\simeq32$.}

\end{figure*}

For the PA$-$IN condition, the magnitude of gradient of the string
signal before and after WDBD and Wiener filtering is represented in
Figure \ref{fig:gradient-beforeafter} for various string tensions from
a single simulation. Only one fifth of the total field of view of the
simulations is shown, corresponding to an angular opening
$\tau'=1.4^{\circ}$.

The visibility of the individual strings of the network is clearly
enhanced by the denoising. For a value of the string tension around
the experimental upper bound $\rho=\scinot{2.5}{-7}$, part of the
network is visible by eye before denoising.  At tension
$\rho=\scinot{6.3}{-8}$, a very reduced number of strings is visible
by eye before denoising.  For both of these string tensions, part of
the network is visible by eye through WDBD and Wiener filtering, but
the resulting map is clearly more noisy in the second case.  The value
$\rho=\scinot{6.3}{-10}$ is the lower bound on the string tension
where a very reduced number of strings is visible by eye through WDBD,
while no strings are visible by eye before denoising.  In this limit,
only string loops are actually recovered, together with some spurious
point sources. Wiener filtering only provides noise at that level.

The posterior probability distributions for the string tension are
reported in Figure \ref{fig:Posterior-pdf} as computed from the
signals observed at the three string tensions of interest in Figure
\ref{fig:gradient-beforeafter}.  The graphs highlight the high
precision of the localization of $\rho$ by the PSM described in
Section \ref{sec:bd}.  The slight offset observed is not related to a
bias of the procedure itself but is simply due to an effective
difference between the power spectrum of the test string simulation
and the analytical expression of the power spectrum $P^{s}(k)$ used in
relations (\ref{5-10}) and (\ref{5-11}). This difference may be
associated with a cosmic variance including the contribution of a
string signal.

The signal-to-noise ratio, correlation coefficient, and kurtosis of
the magnitude of gradient of the string before and after WDBD and
Wiener filtering are represented in Figure \ref{fig:kurtosis} as
functions of the string tension.  In the range of string tensions
where the denoising procedure provides visibility of strings by eye,
it appears clearly that WDBD and Wiener filtering both significantly
increase the signal-to-noise ratio and correlation coefficient to
strictly positive values.  At low string tensions the correlation
coefficient is also significantly higher for WDBD than for Wiener
filtering. This represents a first quantitative measure of the
superiority of our approach. The kurtosis of the magnitude of gradient
is also significantly increased from its value before denoising
towards higher values through WDBD. The peak obtained at low string
tensions, with kurtosis values above the expected value around
$\kappa^{\vert\vecnabla s\vert}\simeq32$, reflects the fact that the
denoising recovers a thresholded version of the string signal in that
range, only keeping localized loops in the limit identified by the
visibility by eye (see Figure \ref{fig:gradient-beforeafter}).  At low
string tensions, Wiener filtering essentially fails to increase the
kurtosis values towards the expected value. We interpret this failure
as a quantitative measure of the fact that Wiener filtering fails to
remove a substantial part of the noise, in contrast with WDBD. This
represents a second quantitative measure of the superiority of our
approach. Let us emphasize that the lowest string tensions where each
of our quantitative measures begin to show effective denoising
performance for the WDBD algorithm are very close to the eye
visibility threshold. 

The degradation of the denoising performance due to instrumental noise
is probed in the noise condition PA$+$IN, with an instrumental noise
level of $1\mu K$.  For this case we omit a complete analysis of all
our quantitative measures.  We simply notice that such a small level
of instrumental noise already significantly affects the denoising
performance by raising the eye visibility threshold by more than one
order of magnitude. The only reason why an effective reconstruction of
strings may be achieved down to so small string tensions in the noise
conditions PA$-$IN is simply that, at high spatial frequencies, the
string signal with a nearly scale-free power spectrum largely
dominates the primary anisotropies with an exponentially damped power
spectrum.  This advantage is lost as soon as high frequency noise is
added, in particular instrumental noise.

\subsection{Noise conditions SA$-$tSZ and SA$+$tSZ}

\begin{figure*}
\begin{center}
\includegraphics[width=5.5cm]{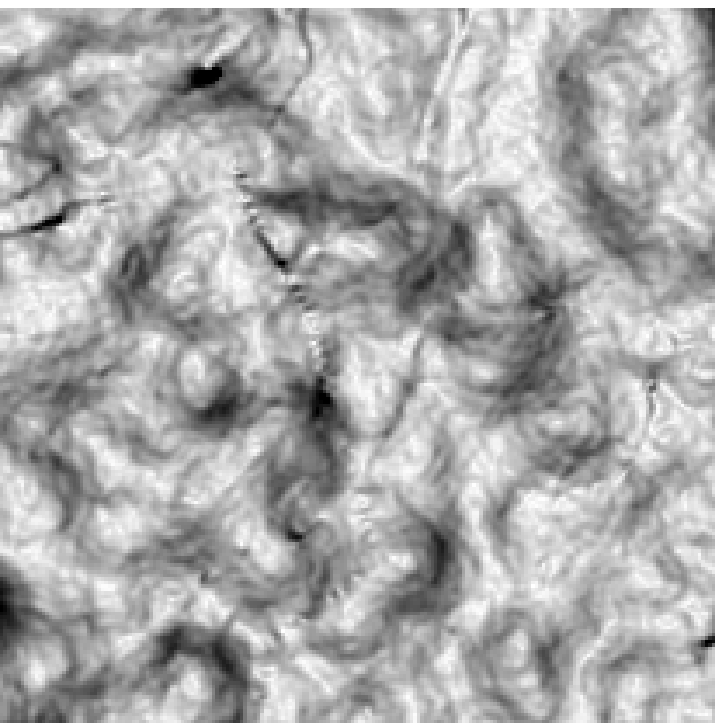}\hspace{2mm}
\includegraphics[width=5.5cm]{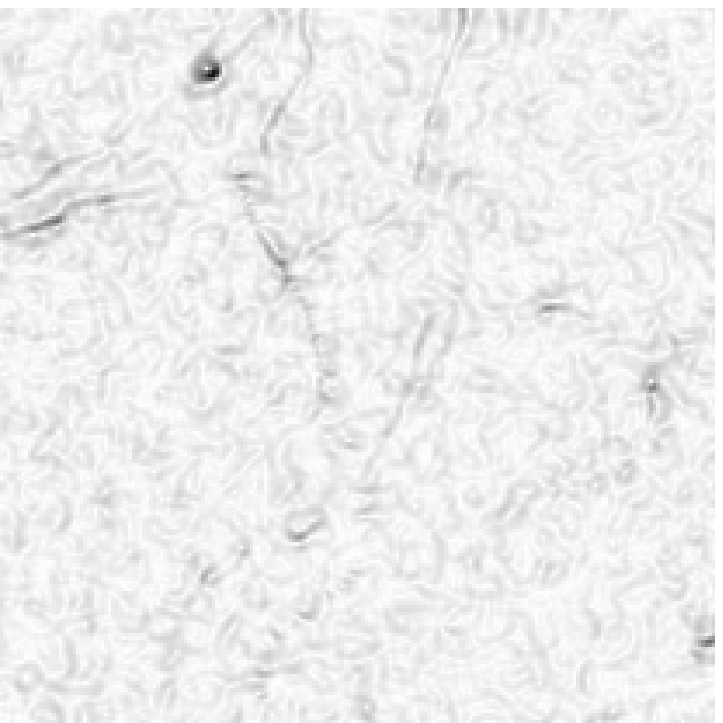}\hspace{2mm}
\includegraphics[width=5.5cm]{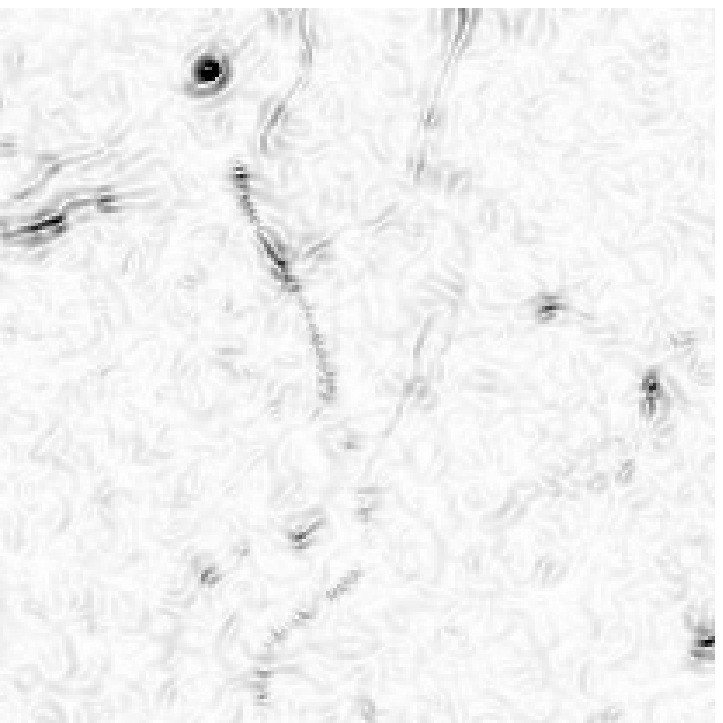}\vspace{2mm}

\includegraphics[width=5.5cm]{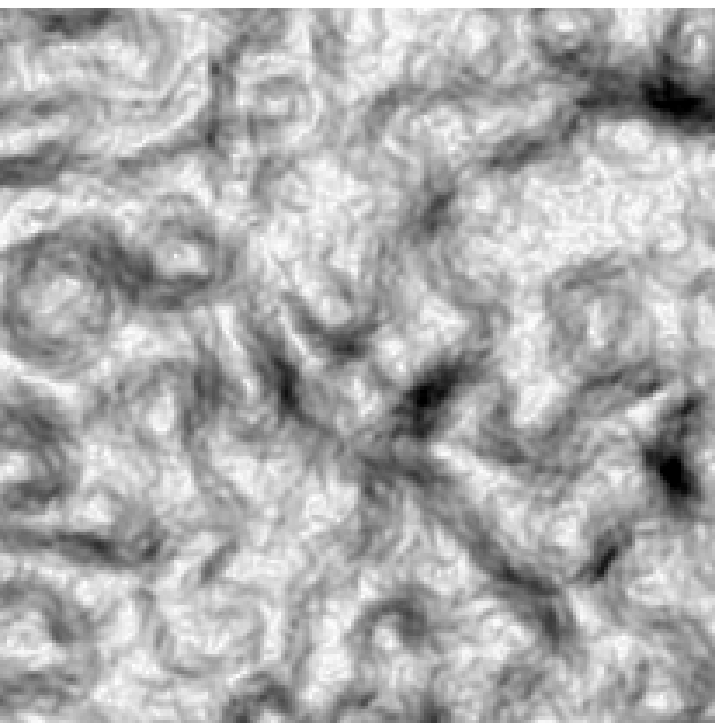}\hspace{2mm}
\includegraphics[width=5.5cm]{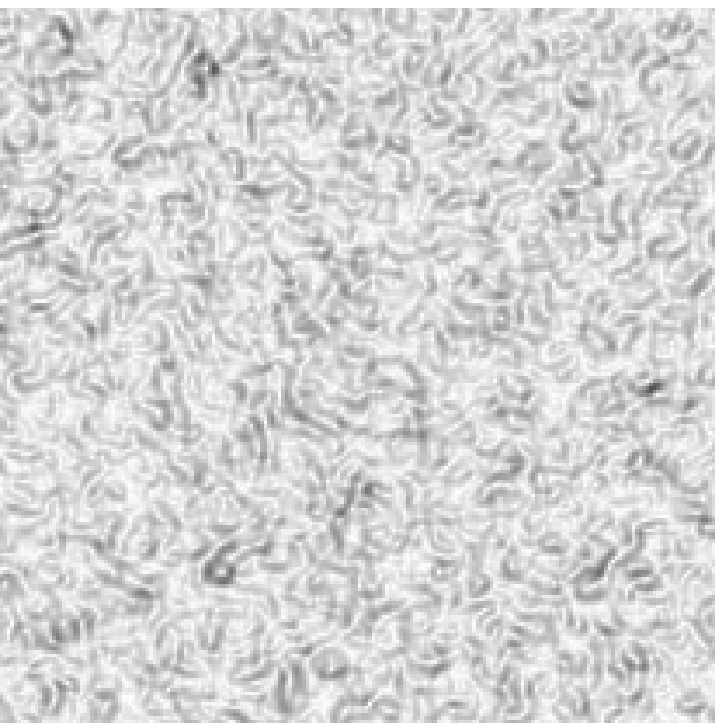}\hspace{2mm}
\includegraphics[width=5.5cm]{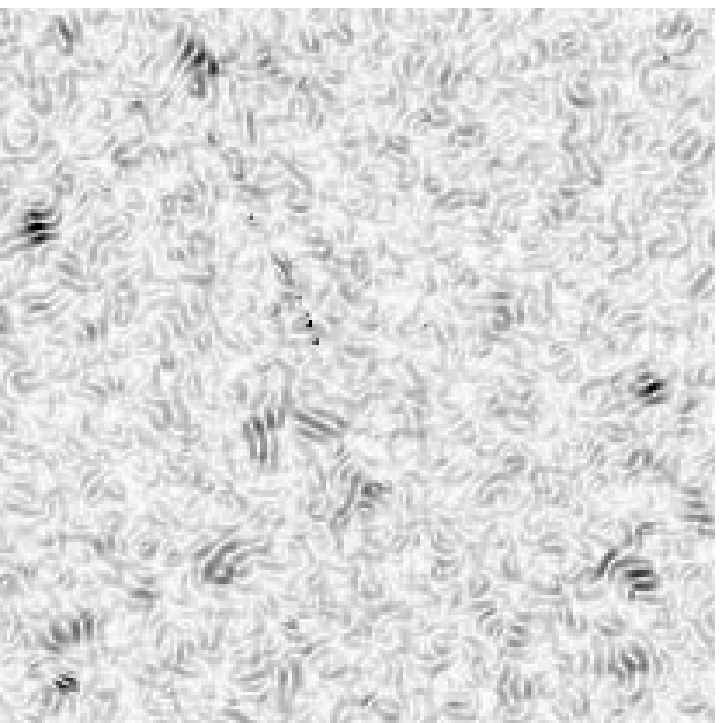}\vspace{2mm}

\includegraphics[width=5.5cm]{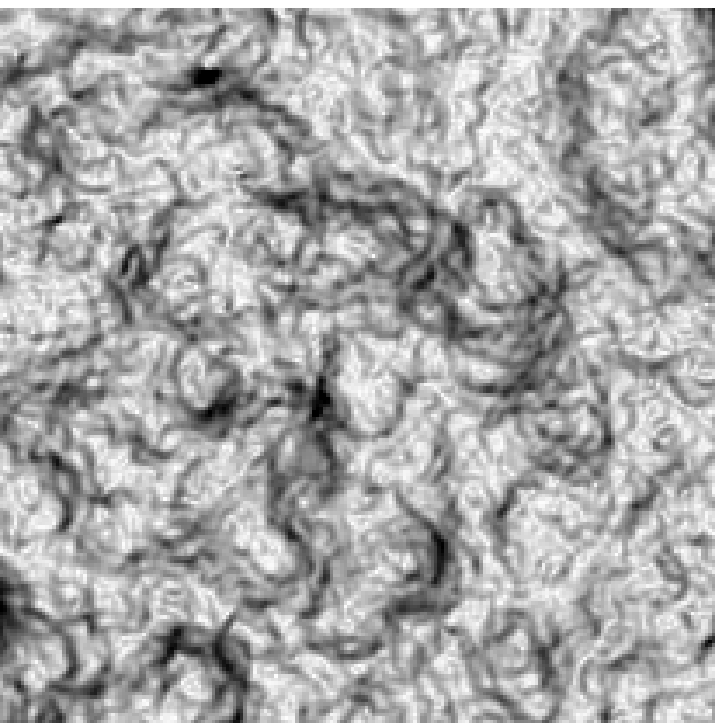}\hspace{2mm}
\includegraphics[width=5.5cm]{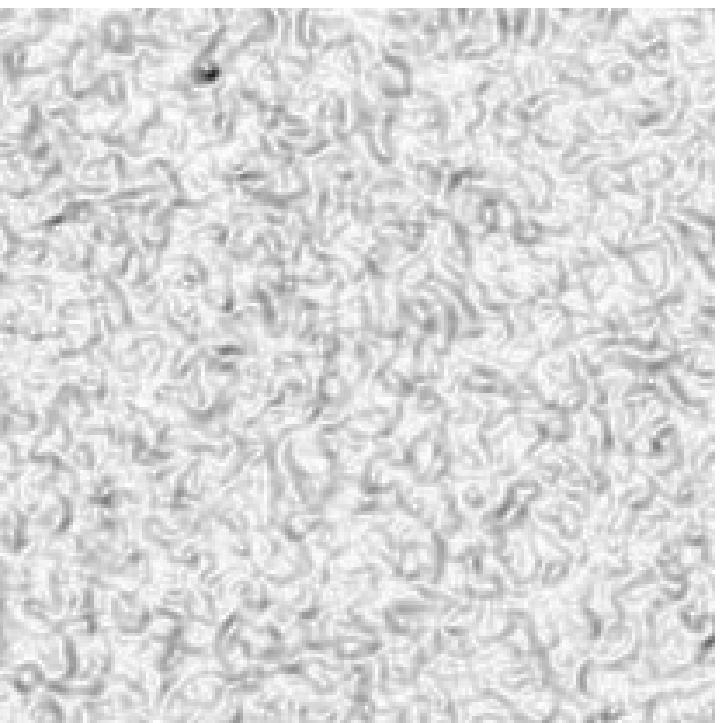}\hspace{2mm}
\includegraphics[width=5.5cm]{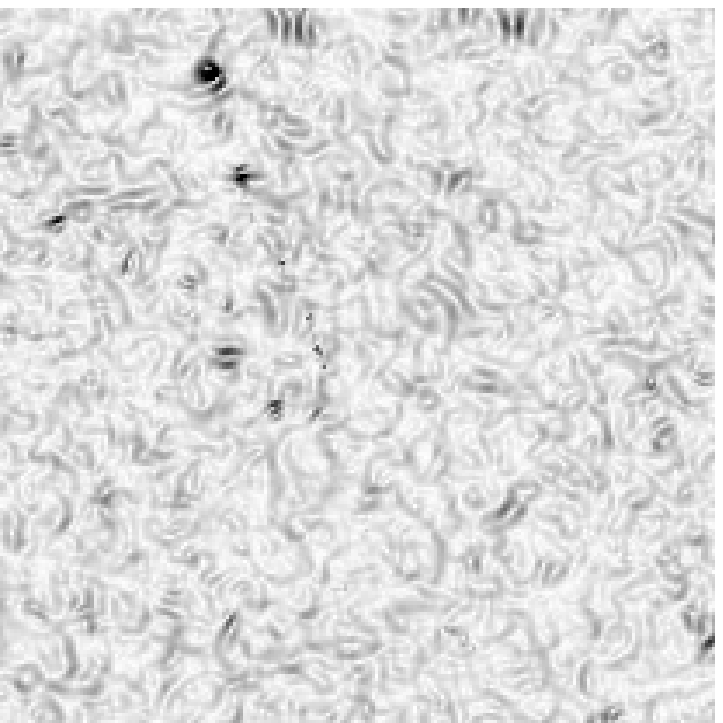}
\end{center}

\caption{\label{fig:gradient-beforeafter-secondary}
  Magnitude of the gradient of the string signal before denoising
  (left panels), after Wiener filtering (middle panels) and WDBD
  (right panels), for various string tensions at $1$ arcminute
  resolution on a field of view of $\tau'=1.4^{\circ}$.  The two top
  panel rows relate to the noise conditions SA$-$tSZ for string
  tensions $\rho=\scinot{4.0}{-7}$ and $\rho=\scinot{1.0}{-7}$
  respectively. The bottom panel row relates to the noise conditions
  SA$+$tSZ for a string tension $\rho=\scinot{2.5}{-7}$.}
\end{figure*}
\begin{figure*}
\begin{center}
\newcommand{\tsc}{0.9}
\psfrag{3.6e-07}[][][\tsc]{\scinot{3.6}{-7}}
\psfrag{4.0e-07}[][][\tsc]{\scinot{4.0}{-7}}
\psfrag{4.4e-07}[][][\tsc]{\scinot{4.4}{-7}}
\psfrag{8.9e-08}[][][\tsc]{\scinot{8.9}{-8}}
\psfrag{1.0e-07}[][][\tsc]{\scinot{1.0}{-7}}
\psfrag{1.1e-07}[][][\tsc]{\scinot{1.1}{-7}}
\psfrag{2.2e-07}[][][\tsc]{\scinot{2.2}{-7}}
\psfrag{2.5e-07}[][][\tsc]{\scinot{2.5}{-7}}
\psfrag{2.8e-07}[][][\tsc]{\scinot{2.8}{-7}}
\psfrag{0.05}[][][\tsc]{0.05}
\psfrag{0.04}[][][\tsc]{0.04}
\psfrag{0.03}[][][\tsc]{0.03}
\psfrag{0.025}[][][\tsc]{0.025}
\psfrag{0.02}[][][\tsc]{0.02}
\psfrag{0.015}[][][\tsc]{0.015}
\psfrag{0.01}[][][\tsc]{0.01}
\psfrag{0.005}[][][\tsc]{0.005}

\psfrag{0}[][][\tsc]{0\hspace{2em}}
\includegraphics[width=5cm]{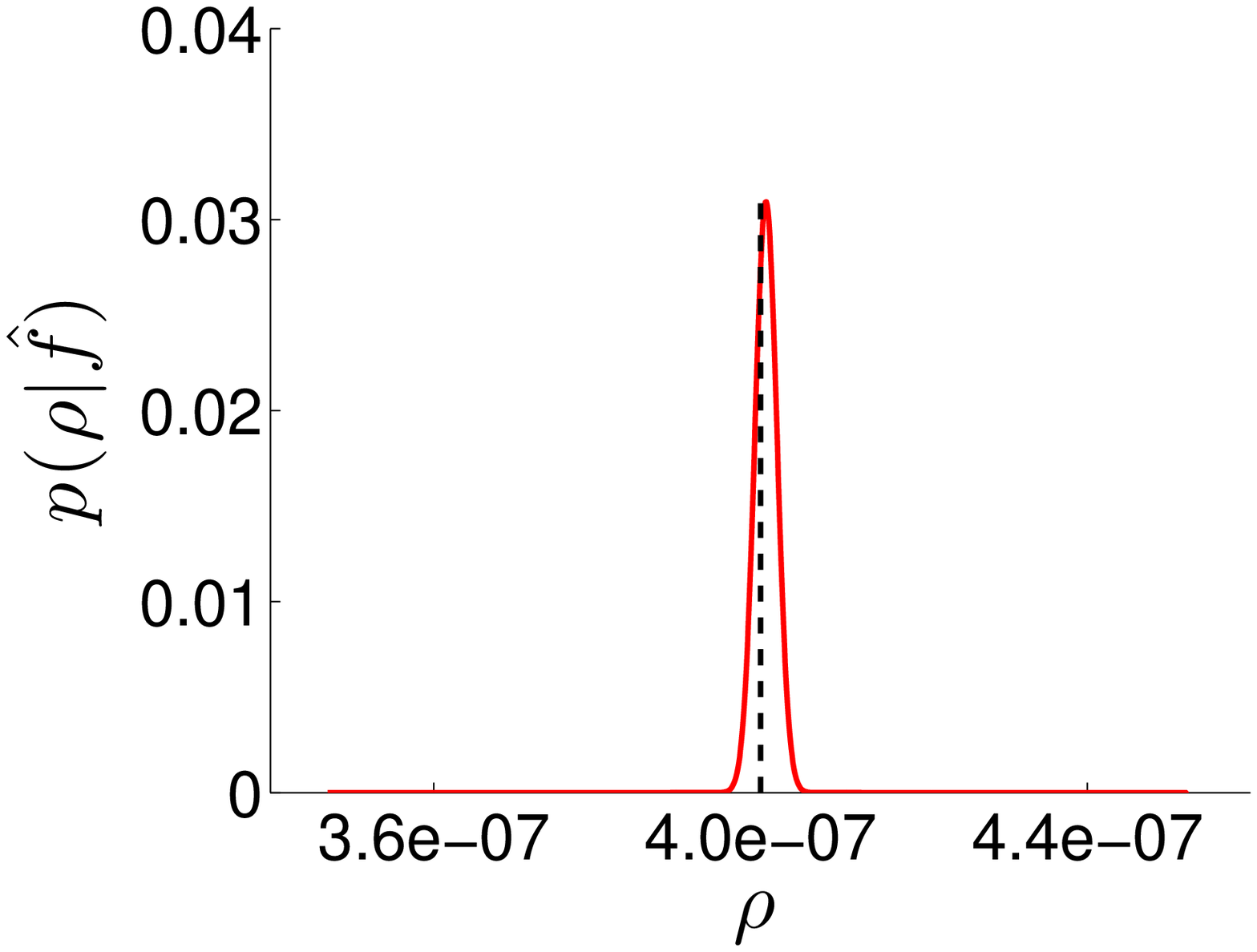}\hspace{1cm}
\includegraphics[width=5cm]{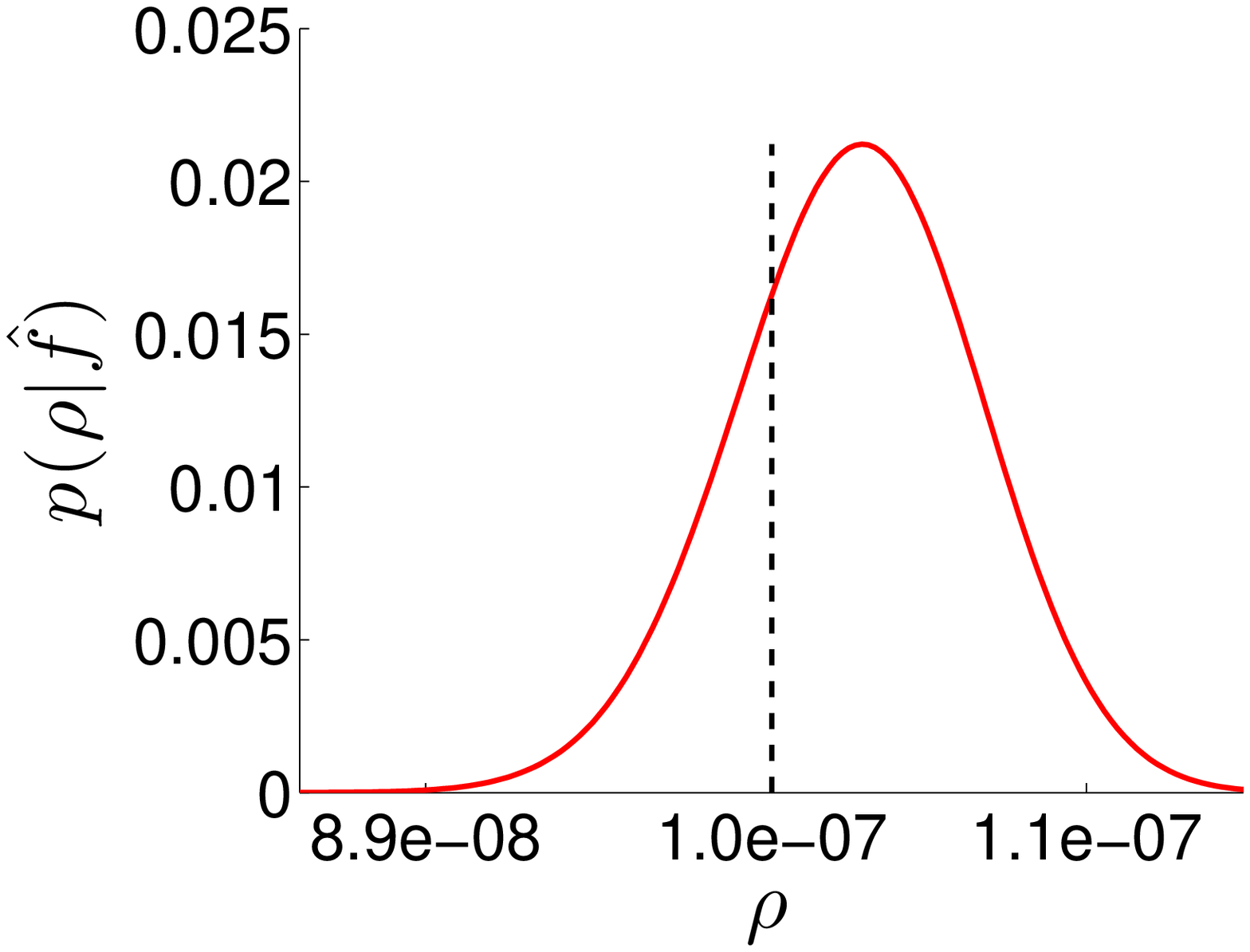}\hspace{1cm}
\includegraphics[width=5cm]{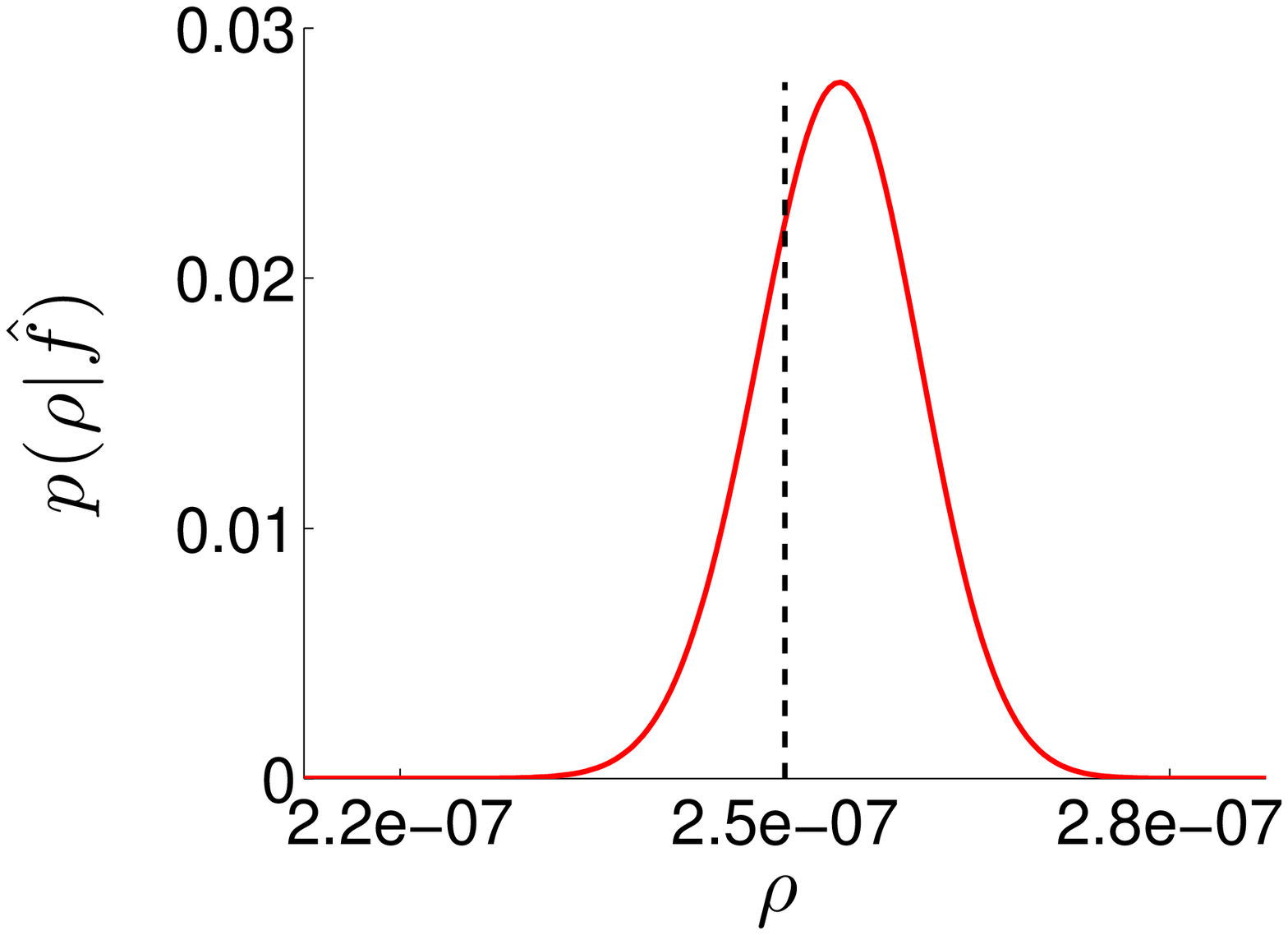}\end{center}

\caption{\label{fig:Posterior-pdf-secondary}
  Posterior probability distributions (red solid curves) for the
  string tension as computed from a simulated signal observed at $1$
  arcminute resolution. The left and middle panels relate to the noise
  conditions SA$-$tSZ for string tensions $\rho=\scinot{4.0}{-7}$ and
  $\rho=\scinot{1.0}{-7}$ respectively.  The right panel relates to
  the noise conditions SA$+$tSZ for a string tension
  $\rho=\scinot{2.5}{-7}$. The black dashed vertical lines represent
  the exact values of the string tension relative to the test string
  simulation.}

\end{figure*}
\begin{figure*}
\begin{center}
\newcommand{\tsc}{0.9}
\psfrag{20}[][][\tsc]{20}
\psfrag{0}[][][\tsc]{0\hspace{0.5em}}
\psfrag{-20}[][][\tsc]{-20}
\psfrag{-40}[][][\tsc]{-40}
\psfrag{-60}[][][\tsc]{-60}
\psfrag{-80}[][][\tsc]{}
\psfrag{10}[][][\tsc]{10}
\psfrag{-10}[][][\tsc]{-10}
\psfrag{-9}[][][\tsc]{-9}
\psfrag{-8}[][][\tsc]{-8}
\psfrag{-7}[][][\tsc]{-7}
\psfrag{-6}[][][\tsc]{-6}
\psfrag{-5}[][][\tsc]{-5}
\psfrag{-0.5}[][][\tsc]{-0.5}
\psfrag{0.5}[][][\tsc]{0.5}
\psfrag{1}[][][\tsc]{1\hspace{0em}}
\psfrag{1.5}[][][\tsc]{1.5}
\psfrag{50}[][][\tsc]{50}
\psfrag{100}[][][\tsc]{100}
\psfrag{150}[][][\tsc]{150}
\psfrag{200}[][][\tsc]{200}
\psfrag{snrdb}[][]{SNR (dB)}
\psfrag{correlationcoefficient}[][]{Correlation Coefficient}
\psfrag{kurtosis}[][]{Kurtosis}
\psfrag{log10rho}[][]{$\log_{10}(\rho)$}
\includegraphics[width=5cm]{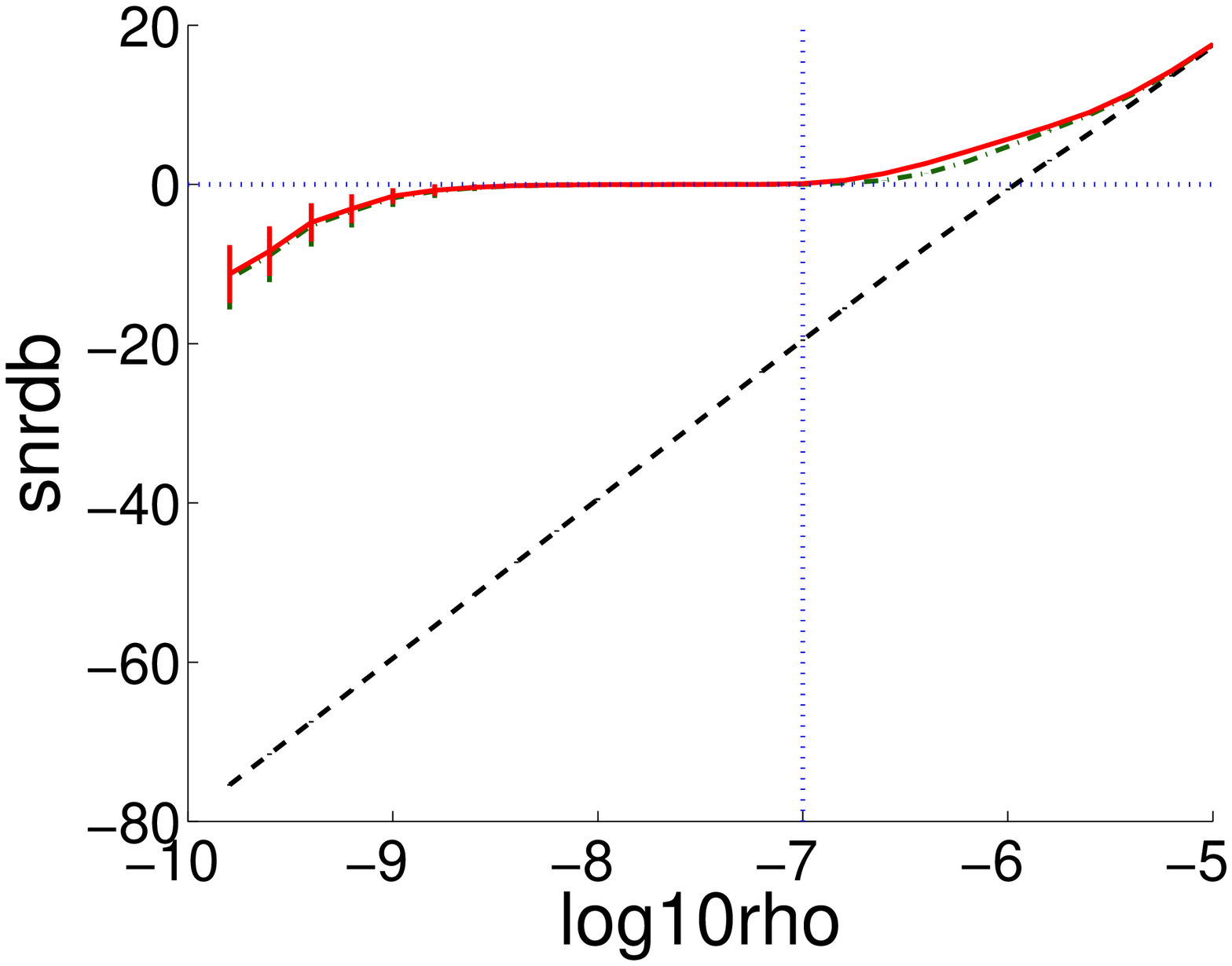}\hspace{1cm}
\includegraphics[width=5cm]{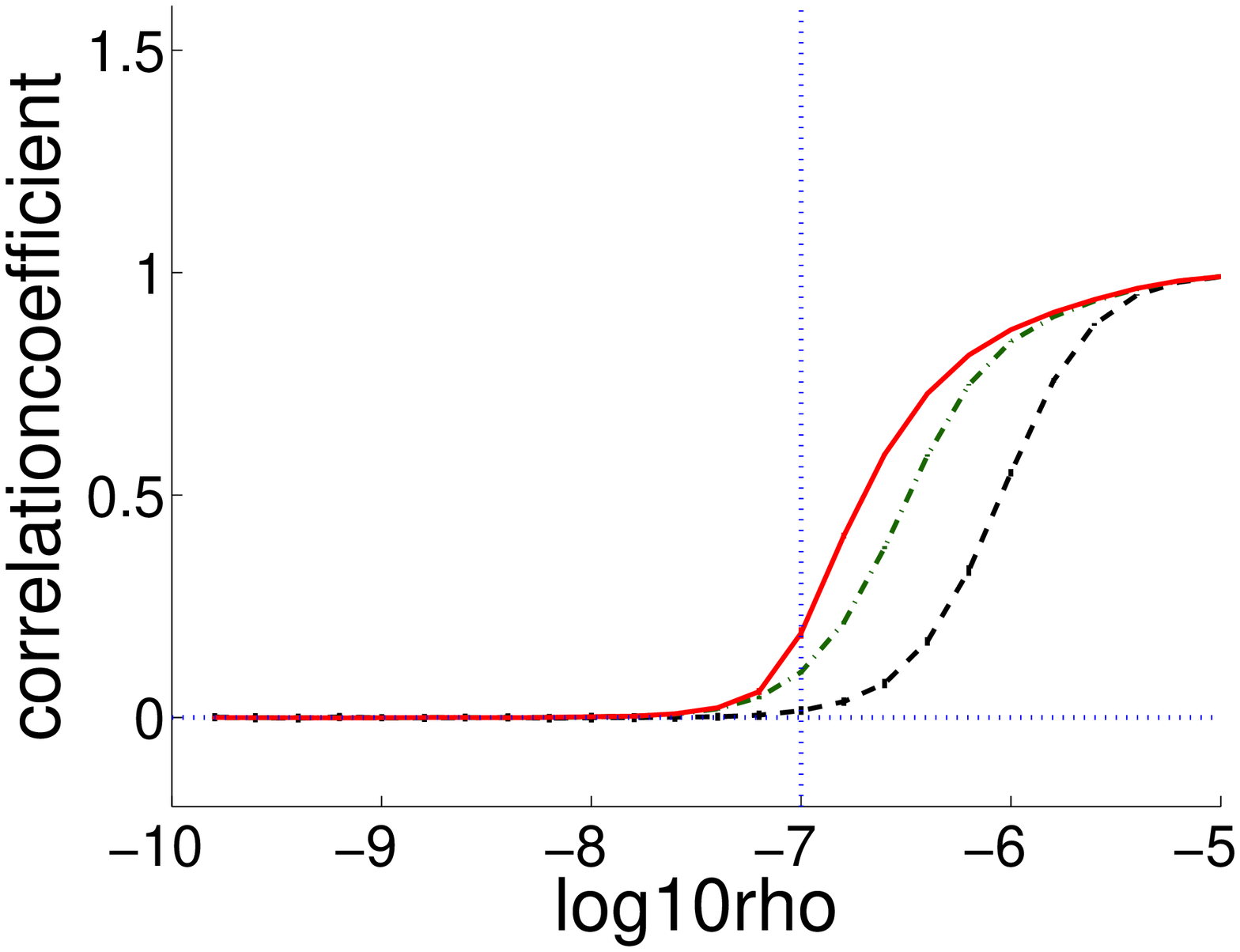}\hspace{1cm}
\includegraphics[width=5cm]{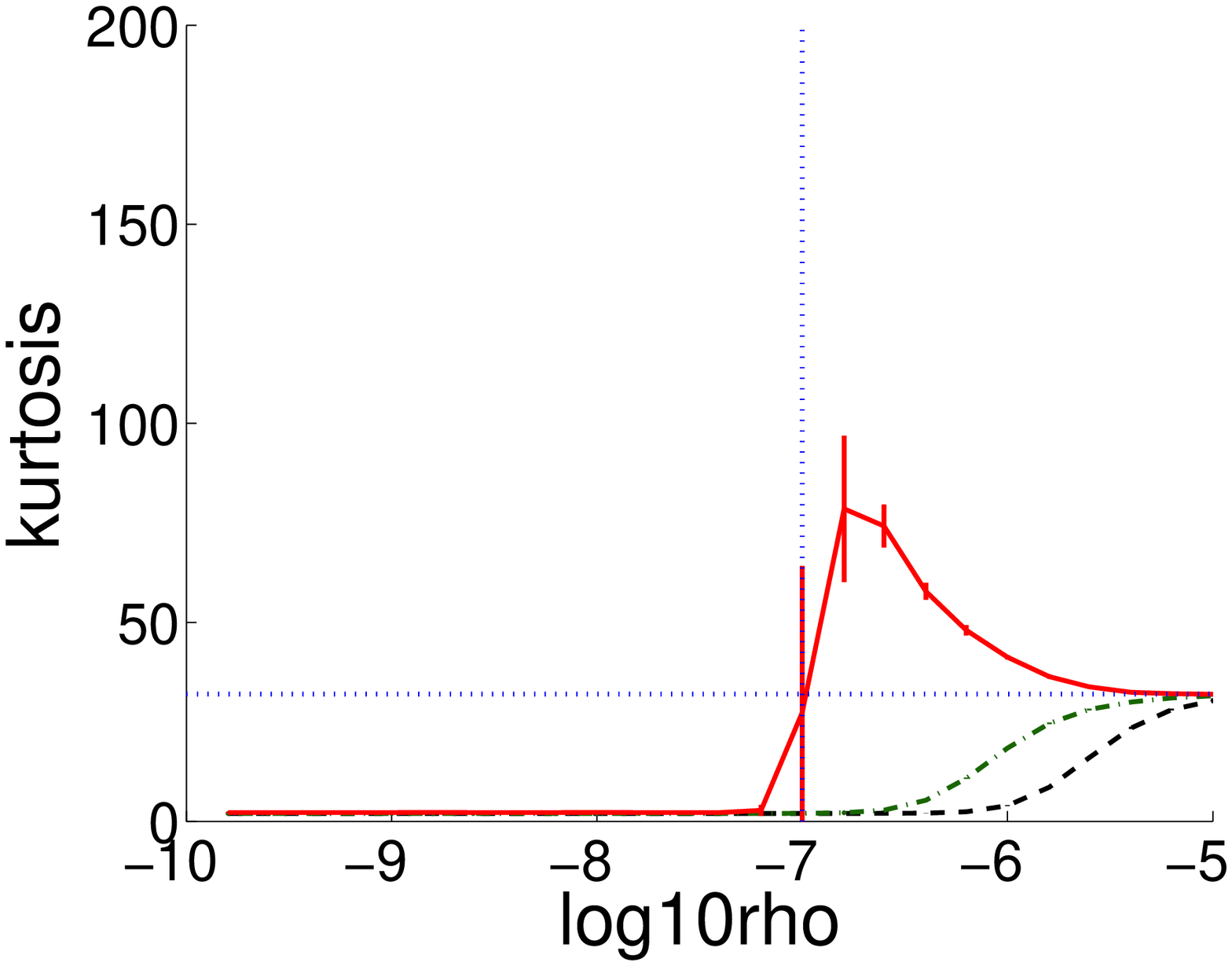}\vspace{1cm}

\includegraphics[width=5cm]{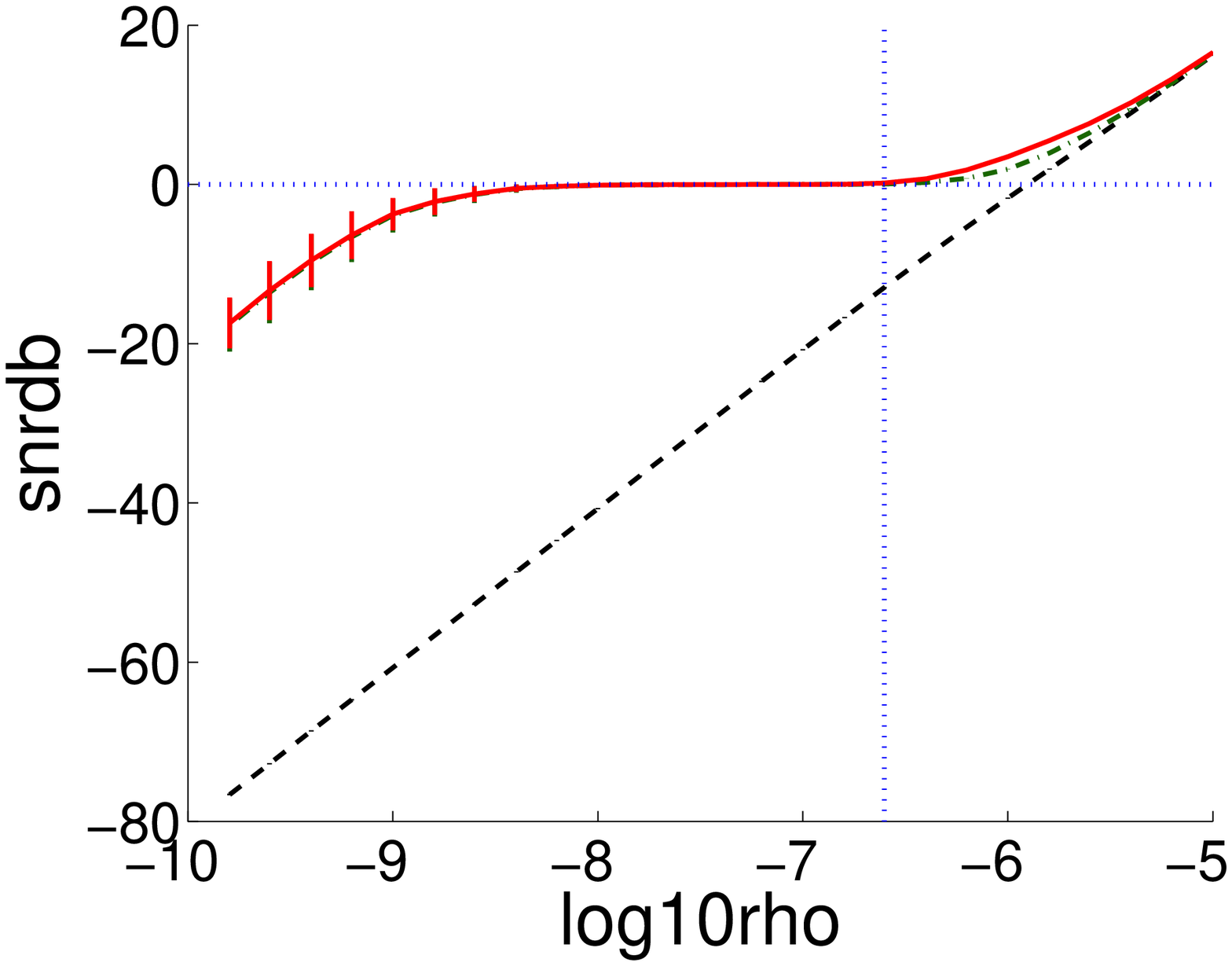}\hspace{1cm}
\includegraphics[width=5cm]{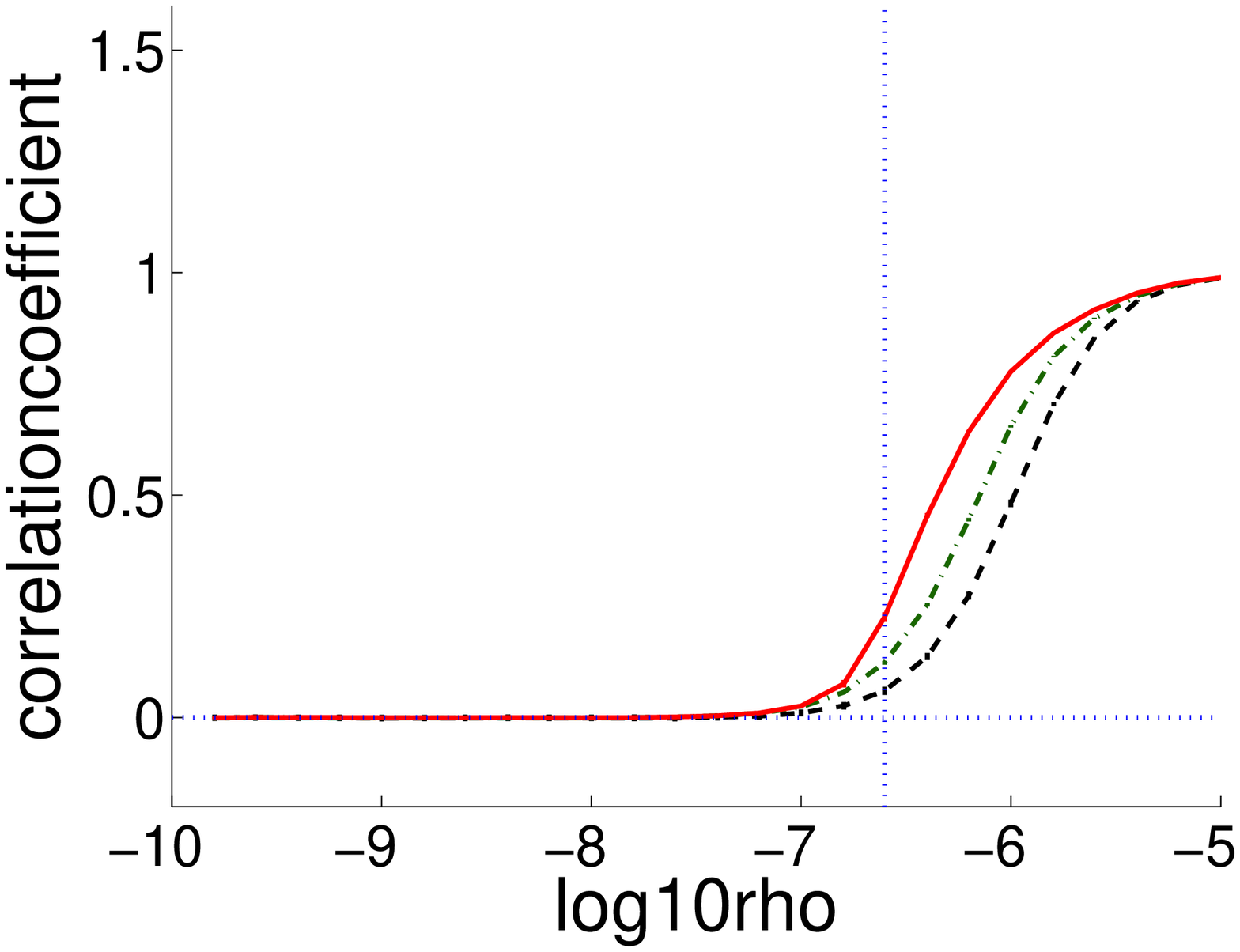}\hspace{1cm}
\includegraphics[width=5cm]{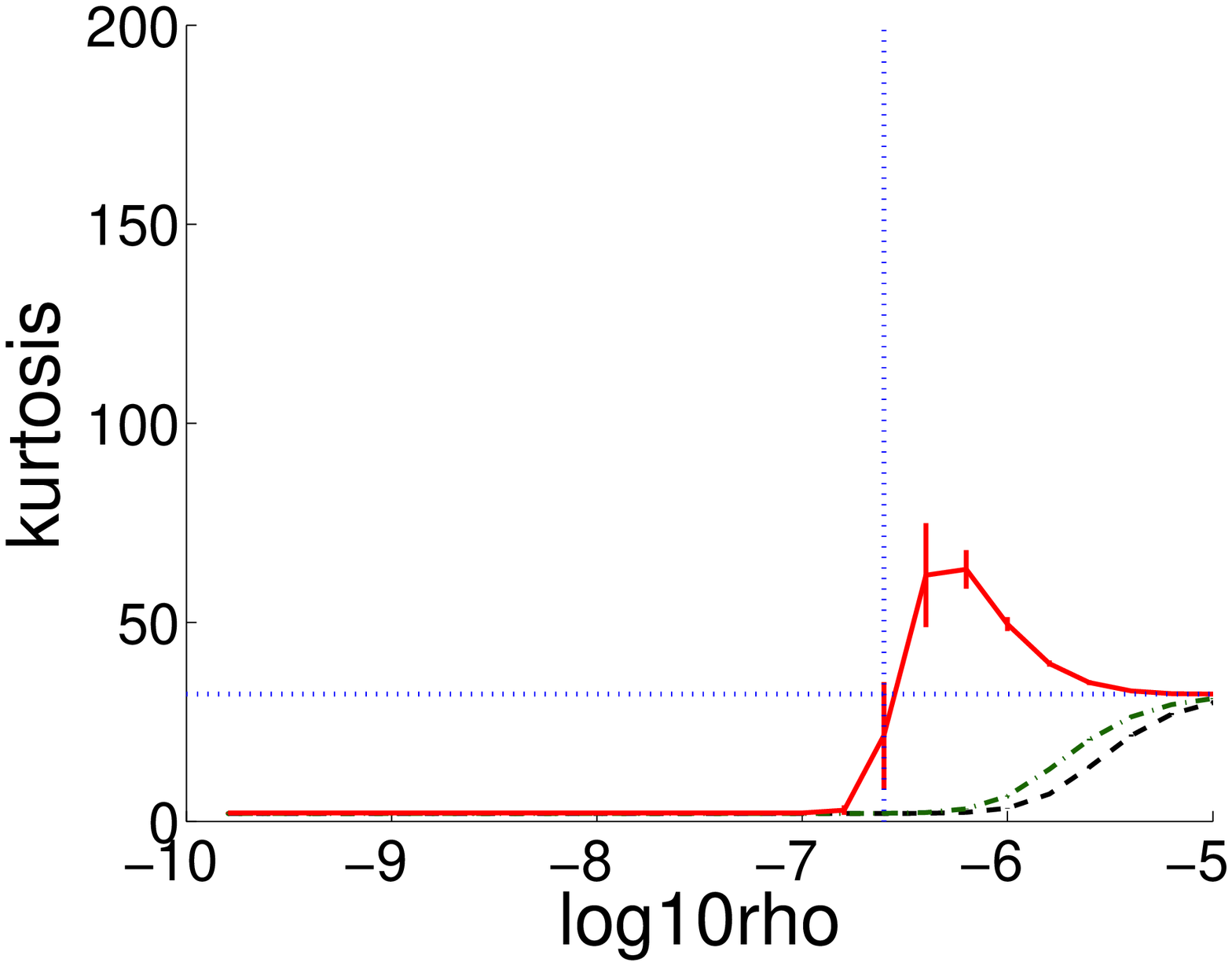}\end{center}

\caption{\label{fig:kurtosis-secondary}
  Signal-to-noise ratio (left panels) in decibels (dB), correlation
  coefficient (middle panels), and kurtosis (right panels) of the
  magnitude of gradient as functions of the string tension in
  logarithmic scaling in the range $\log_{10}\rho\in[-10,-05]$ and at
  $1$ arcminute resolution. The top panels relate to the noise
  conditions SA$-$tSZ, while the bottom panels relate to the noise
  conditions SA$+$tSZ. The black dashed curves represent values before
  denoising, while the red solid curves and green dot-dashed curves
  represent values after WDBD and Wiener filtering respectively. The
  vertical lines on the curves represent the variability at one
  standard deviation of the estimated statistic across the $100$ test
  simulations considered (these lines are not visible by eye where
  smaller than the width of the curves). The blue dotted vertical
  lines represent the eye visibility thresholds
  $\rho=\scinot{1.0}{-7}$ for the top panels and
  $\rho=\scinot{2.5}{-7}$ for the bottom panels.  The blue dotted
  horizontal lines identify either the limit of zero signal-to-ratio,
  zero correlation coefficient, or the kurtosis of the magnitude of
  gradient of a pure string signal:
  $\kappa^{\vert\vecnabla s\vert}\simeq32$.}

\end{figure*}
The magnitude of gradient of the string signal
before and after WDBD and Wiener filtering is represented in Figure
\ref{fig:gradient-beforeafter-secondary} for various string tensions,
from a single simulation.  In the noise conditions SA$-$tSZ and for a
value of the string tension around the experimental upper bound
$\rho=\scinot{4.0}{-7}$, a very reduced number of strings is visible
by eye before denoising.  Part of the network is visible by eye after
WDBD and Wiener filtering, but the resulting map is clearly more noisy
in the second case. The value $\rho=\scinot{1.0}{-7}$ is the lower
bound on the string tension where a very reduced number of strings is
visible by eye through WDBD, while no string is visible by eye before
denoising.  Wiener filtering only provides noise at that level. In the
noise conditions SA$+$tSZ, the value $\rho=\scinot{2.5}{-7}$ is the
lower bound on the string tension where a very reduced number of
strings is visible by eye through WDBD, while no string is visible by
eye before denoising.  Again, Wiener filtering only provides noise at
that level. At the lower bounds for the string tensions in both noise
conditions only string loops are recovered, still with some spurious
point sources.  

The posterior probability distributions for the string tension are
reported in Figure \ref{fig:Posterior-pdf-secondary} as computed from
the signals observed in the three cases of interest in Figure
\ref{fig:gradient-beforeafter-secondary}. The graphs still highlight
the high precision of the localization of $\rho$ by the PSM.

The signal-to-noise ratio, correlation coefficient, and kurtosis of
the magnitude of gradient of the string signal before and after WDBD
and Wiener filtering are represented in Figure
\ref{fig:kurtosis-secondary} as functions of the string tension.  As
for the PA$-$IN and PA$+$IN cases, both WDBD and Wiener filtering
increase the signal-to-noise ratio and correlation coefficient to
strictly positive values for tensions above the eye visibility
threshold.  The correlation coefficient is significantly higher for
WDBD than for Wiener filtering in the whole range of string tensions
of interest. As before, the kurtosis of the magnitude of gradient is
also significantly increased from its value before denoising towards
higher values through WDBD, with a peak at low string tensions due to
the fact that the denoising recovers a thresholded version of the
string signal.  Wiener filtering essentially fails to increase the
kurtosis values towards the expected value in the whole range of
string tensions of interest, once more reflecting its poorer denoising
performance.  As before, we see that for both SA$-$tSZ and SA$+$tSZ,
the eye visibility thresholds are very close to the lowest string
tensions where each of our quantitative measures begin to show
effective denoising performance.

Let us acknowledge the fact that, in the noise conditions SA$-$tSZ and
SA$+$tSZ, the lowest string tensions where denoising is effective are
greatly increased relative to the noise condition PA$-$IN. For
SA$-$tSZ, the eye visibility threshold is slightly below the best
experimental bound, while for SA$+$tSZ it is slightly above.  These
results are absolutely in the line of those obtained in the noise
condition PA$+$IN, as the secondary anisotropies represent even
stronger higher frequency noise.
 
\subsection{Comparison to PSM detectability threshold}

\begin{table}
\begin{center}\begin{tabular}{ccc}
\hline 
\noalign{\vskip\doublerulesep}
Noise condition & PSM detectability  & Eye visibility  \\
\hline
PA$-$IN  & \scinot{2.2}{-10} & \scinot{6.3}{-10}\\ 
PA$+$IN  & \scinot{1.7}{-8}  & \scinot{2.5}{-8}\\
SA$-$tSZ & \scinot{6.1}{-8}  & \scinot{1.0}{-7}\\
SA$+$tSZ & \scinot{1.9}{-7}  & \scinot{2.5}{-7}\\
\hline
\end{tabular}\end{center}

\caption{\label{tab:detectionthresholds} 
  PSM detectability and eye visibility thresholds on the string
  tension  determined on the basis
  of the PSM for each of the noise conditions considered.  All values
  are given with two significant figures.}  
\end{table}
As our WDBD algorithm uses the PSM for preliminary localization of the
string tension, it is a natural question to ask whether the overall
denoising performance at low string tensions is limited by this
preliminary PSM localization. We address this by defining and studying
the detectability threshold for the PSM, which provides a measure of
the minimum string tension where the PSM alone provides robust
detection of strings.

We firstly describe how the PSM detection threshold is computed, based
on a hypothesis test for string detection. We may define an
estimation $\widehat{\rho}$ of the string tension from the observed
signal $f$ as the expectation value of the posterior probability
distribution (\ref{5-11}) computed on the basis of the PSM:
\begin{equation}\label{detection-1}
\widehat{\rho}=E\left[p\left(\rho\vert\widehat{f}\right)\right].
\end{equation}
For any possible string tension, the probability distribution function
for $\widehat{\rho}$ may consequently be obtained from simulations.

We identify the critical value $\rho_{0}$ such that, for null string
tension, one has $p(\widehat{\rho}\geq\rho_{0})=\alpha$, for some
suitable positive value $\alpha$ much smaller than unity. The test for
the hypothesis of null string tension is then defined as follows. For
estimated values $\widehat{\rho}\geq\rho_{0}$, the hypothesis of null
string tension may be rejected with a significance level $\alpha$. On
the contrary for estimated values $\widehat{\rho}<\rho_{0}$, the
hypothesis of null string tension may not be rejected.

We define the detectability threshold $\rho^{\star}$ such that, for a
string tension $\rho^{\star}$, one has
$p(\widehat{\rho}\geq\rho_{0})=1-\beta$, for some other suitable
positive value $\beta$ much smaller than unity. Consequently, for
string tensions larger than $\rho^{\star}$, the probability of
rejecting a null string tension on the basis of the hypothesis test
defined is larger than $1-\beta$.  The value $\rho^{\star}$ is the
smallest string tension that can be discriminated from the hypothesis
of null string tension for given values of $\alpha$ and $\beta$. It
may thus be understood as a detectability threshold determined on the
basis of the PSM.  As our overall denoising method is using the PSM as
a preliminary estimation of the string tension, $\rho^{\star}$
identifies an effective lower bound on the string tension range where
denoising could reasonably be expected to be effective.  

The PSM detectability thresholds in the various noise conditions
considered are reported in Table \ref{tab:detectionthresholds} for
$\alpha\simeq\beta\simeq 0.01$. In all cases except SA$+$tSZ
the PSM detectability thresholds are below the best experimental
bound, while for SA$+$tSZ the PSM detectability threshold is around
the best experimental bound.

Secondly, we compare these thresholds with the eye visibility
thresholds, which as noted previously also indicate the lower limit of
string tensions where our quantitative measures show effective
performance for the WDBD algorithm.  For each of the noise conditions
with significant high frequency content, i.e. PA$+$IN, SA$-$tSZ and
SA$+$tSZ, the PSM detectability threshold $\rho^*$ is slightly below
the eye visibility threshold.  For the PA$-$IN case, this difference
is larger, the PSM detectability threshold being about one third of
the eye visibility threshold.  This indicates that the PSM is able to
more effectively exploit the high spatial frequency ranges where the
string signal dominates the primary anisotropies.

The discrepancy between these two thresholds shows that the detection
problem alone can be solved with the PSM at slightly lower string
tensions than the more difficult denoising problem.  Indeed for values
of the string tension between the two thresholds, denoising does not
produce visible strings even though the PSM posterior probability
distributions for $\rho$ are distinctly peaked away from zero.  It is
one thing to estimate a single global parameter such as the string
tension on the basis of a PSM, but quite another to explicitly
reconstruct the string network itself.

\subsection{Algorithm robustness}

We comment here on the robustness of the WDBD algorithm relative to
both additional noise from foreground point sources and to the
possible improvements in the definition of the denoising procedure
itself.

We have explicitly disregarded the problem of foreground emissions
such as radio and infrared point sources. The discrimination of point
sources from string loops imprinted in the CMB may appear to be a
difficult task. However, the dipolar structure of the string loops
represents an essential difference with point sources
\citep{fraisse08}.  In that context, the odd symmetry of the wavelets
used in the WDBD algorithm (see Figure \ref{fig:wavelet}) to match the
string imprints is adequate both for long strings and string loops,
and might help to discriminate between string loops and point
sources. The algorithm was shown to be effective at detecting string
loops, even at low tensions where long strings are not reconstructed
anymore. However spurious point sources were also reconstructed at
very small string tensions in the noise conditions PA$-$IN, in the
absence of foreground point sources. A thorough analysis should be
conducted in order to assess the real robustness of the algorithm to
discriminate between string loops and point sources, and to discuss
necessary enhancements.

Our approach explicitly assumes the statistical independence of
coefficients of the wavelet decomposition, when conditioned on the
string tension. However, significant correlations are present in the
wavelet coefficients, and exploiting them should lead to improved
denoising performance. Gaussian scale mixture (GSM) models may be
considered which allow one to explicitly account for local
correlations of the wavelet coefficients in the denoising process
\citep{andrews74,portilla03}. An enhanced version of this model called
the orientation-adapted Gaussian scale mixture (OAGSM) model relies on
steerable wavelets in order to integrate directionality information in
the local correlations \citep{hammond08}. A preliminary implementation
of the OAGSM model suggests that an enhancement relative to the WDBD
algorithm may indeed be expected, albeit at significant computational
cost.

An improvement of the similarity of the shape of the filters to better
match the typical string imprints may also be envisaged. Even though
steerable wavelets can be very directional, their spatial support is
not especially narrow.  Filters with a more elongated support such as
curvelets \citep{candes99,starck02} might be expected to provide
better performance for the detection of long strings. Let us notice
however that such filters would not be adequate anymore for string
loops. Moreover a preliminary implementation of this evolution
provides no improvement relative to the WDBD algorithm for the
detection of long strings.

Finally, a discretization of the wavelet scales finer than the dyadic
discretization used might provide an improved statistical model of the
coefficients of the string signal at each spatial scale $b$. We did
not consider this evolution here.

\section{Conclusion}

\label{sec:c}

We have described a Bayesian framework for mapping the CMB signal
induced by cosmic strings, based on a generalized Gaussian model
capturing the sparse behaviour of the string signal in the steerable
wavelet domain. This signal is buried in the standard primary and
secondary CMB anisotropies, which we model as Gaussian noise.  For a
fixed string tension we compute the Bayesian least squares estimator
for each wavelet coefficient of the string signal. Our overall
estimator is then formed as an average of these estimates for
different string tensions, weighted by the posterior probability of
the string tension under a power spectral model.

We have demonstrated the performance of our denoising algorithm
through a series of numerical analyses at $1$ arcminute resolution
consistent with upcoming experiments.  The maps of the magnitude of
the gradient of the denoised string signal produced by our algorithm
were evaluated on the basis of three quantitative measures: the
signal-to-noise ratio and correlation coefficient computed with
respect to the known original string signal, and the kurtosis.  In the
idealized case of primary anisotropies without instrumental noise, the
strings can be identified for tensions down to
$\rho=\scinot{6.3}{-10}$, more than two orders of magnitude below the
current experimental upper bound. With instrumental noise around
$1\,\mu\textnormal{K}$ per pixel this lower bound is increased by more
than one order of magnitude. The inclusion of secondary anisotropies
further raises this bound to $\rho=\scinot{1.0}{-7}$ disregarding the
thermal Sunyaev-Zel'dovich effect and to $\rho=\scinot{2.5}{-7}$
including this effect in the Rayleigh-Jeans limit. These values
nonetheless remain slightly below or near the current experimental
upper bound on the string tension, demonstrating that the proposed
algorithm will be useful for analysis of upcoming high resolution
data.

\section*{Acknowledgments}

The authors wish to thank A. A. Fraisse, C. Ringeval, D. N. Spergel,
and F. R. Bouchet for valuable comments and for kindly providing $2$
simulations of a string signal. The authors also wish to thank M.
P. Hobson, J. D. McEwen, and G. Puy for useful discussions. The work
of Y. W. is funded by the Swiss National Science Foundation (SNF)
under contract No. 200020-113353. Y. W. is also a Postdoctoral Researcher
of the Belgian National Science Foundation (F.R.S.-FNRS).

\label{lastpage}


\begin{thebibliography}{\protect\citeauthoryear{Antoine \& Vandergheynst}{1999}}
\bibitem[\protect\citeauthoryear{Albrecht \& Turok}{1985}]{albrecht85}Albrecht
A., Turok N., 1985, Phys. Rev. Lett., 54, 1868 

\bibitem[\protect\citeauthoryear{Albrecht \& Turok}{1989}]{albrecht89}Albrecht
A., Turok N., 1989, Phys. Rev. Lett., 40, 973 

\bibitem[\protect\citeauthoryear{Allen \& Shellard}{1990}]{allen90}Allen
B., Shellard E. P. S., 1990, Phys. Rev. Lett., 64, 119 

\bibitem[\protect\citeauthoryear{Amsel et al.}{2007}]{amsel07}Amsel
S., Berger J., Brandenberger R. H., 2008, JCAP, 04, 015

\bibitem[\protect\citeauthoryear{Andrews \& Mallows}{1974}]{andrews74}Andrews
D., Mallows C., 1974, J. Royal Stat. Soc. B, 36, 99 

\bibitem[\protect\citeauthoryear{Antonini et al.}{1992}]{antonini92}Antonini
M., Barlaud M., Mathieu P., Daubechies I., 1992, IEEE Trans. Image
Proc., 1, 205 

\bibitem[\protect\citeauthoryear{Barker et al.}{2006}]{barker06}Barker
R. et al., 2006, MNRAS, 369, L1 

\bibitem[\protect\citeauthoryear{Barreiro \& Hobson}{2001}]{barreiro01}Barreiro
R. B., Hobson M. P., 2001, 327, 813 

\bibitem[\protect\citeauthoryear{Belge et al.}{2000}]{belge00}Belge
M., Kilmer M. E., Miller E. L., 2000, IEEE Trans. Image Proc., 9,
597 

\bibitem[\protect\citeauthoryear{Bennet}{1986}]{bennet86}Bennet D.
P., 1986, Phys. Rev. D, 33, 872 

\bibitem[\protect\citeauthoryear{Bennet \& Bouchet}{1989}]{bennet89}Bennet
D. P., Bouchet F. R., 1989, Phys. Rev. Lett., 63, 2776 

\bibitem[\protect\citeauthoryear{Bennet \& Bouchet}{1990}]{bennet90}Bennet
D. P., Bouchet F. R., 1990, Phys. Rev. D, 41, 2408 

\bibitem[\protect\citeauthoryear{Bennett et al.}{2003}]{bennett03}Bennett
C.L. et al., 2003, ApJS, 148, 1 

\bibitem[\protect\citeauthoryear{Bevis et al.}{2004}]{bevis04}Bevis
N., Hindmarsh M., Kunz M., 2004, Phys. Rev. D, 70, 043508 

\bibitem[\protect\citeauthoryear{Bevis et al.}{2007}]{bevis07}Bevis
N., Hindmarsh M., Kunz M., Urrestilla J., 2007, Phys. Rev. D, 75,
065015 

\bibitem[\protect\citeauthoryear{Bobin et al.}{2008}]{bobin08}Bobin
J., Moudden Y., Starck J.-L., Fadili J., Aghanim N., 2008, Stat. Method.,
5, 307 

\bibitem[\protect\citeauthoryear{Bouchet et al.}{1988}]{bouchet88}Bouchet
F. R., Bennet D. P., Stebbins A., 1988, Nature, 335, 410 

\bibitem[\protect\citeauthoryear{Bouchet}{2004}]{bouchet04}Bouchet
F. R., 2004, preprint (arXiv:astro-ph/0401108v1) 

\bibitem[\protect\citeauthoryear{Cand\`{e}s \& Donoho}{1999}]{candes99}
Cand\`es E., Donoho D. L., 1999, in Cohen A., Rabut C., Schumaker L. L., eds,
Curve and Surface fitting: Saint-Malo 1999. Vanderbilt University
Press, Nashville, p. 105 

\bibitem[\protect\citeauthoryear{Davis \& Kibble}{2005}]{davis05}Davis
A. C., Kibble T. W. B., 2005, Contemp. Phys., 46, 313 

\bibitem[\protect\citeauthoryear{Delabrouille et al.}{2003}]{delabrouille03}Delabrouille
J., Cardoso J.-F., Patanchon G., 2003, MNRAS, 346, 1089 

\bibitem[\protect\citeauthoryear{Forni \& Aghanim}{2004}]{forni04}Forni
O., Aghanim N., 2004, A\&A, 420, 49 

\bibitem[\protect\citeauthoryear{Fraisse et al.}{2007}]{fraisse07}Fraisse
A. A., 2007, JCAP, 03, 008

\bibitem[\protect\citeauthoryear{Fraisse et al.}{2008}]{fraisse08}Fraisse
A. A., Ringeval C., Spergel D. N., Bouchet F. R., 2008, Phys. Rev.
D, 78, 043535 

\bibitem[\protect\citeauthoryear{Gott}{1985}]{gott85}Gott J. R.,
1985. ApJ, 288, 422 

\bibitem[\protect\citeauthoryear{Gull \& Skilling}{1999}]{gull99}Gull
S. F., Skilling J., 1999. Quantified Maximum Entropy, MemSys5 Users'
Manual. Maximum Entropy Data Consultants Ltd., Suffolk

\bibitem[\protect\citeauthoryear{Hammond \& Simoncelli}{2008}]{hammond08}Hammond
D. K., Simoncelli E. P., 2008, IEEE Trans. Image Proc., 17, 2089

\bibitem[\protect\citeauthoryear{Hindmarsh}{1995a}]{hindmarsh95a}Hindmarsh
M., 1995, Nucl. Phys. Proc. Suppl., 43, 50 

\bibitem[\protect\citeauthoryear{Hindmarsh \& Kibble}{1995b}]{hindmarsh95b}Hindmarsh
M., Kibble T. W. B., 1995, Rep. Prog. Phys., 58, 477 

\bibitem[\protect\citeauthoryear{Hinshaw et al.}{2007}]{hinshaw07}Hinshaw
G. et al., 2007, ApJS, 170, 288 

\bibitem[\protect\citeauthoryear{Hinshaw et al.}{2009}] 
{hinshaw09}Hinshaw
G. et al., 2009, ApJS, 180, 225

\bibitem[\protect\citeauthoryear{Hobson et al.}{1998}]{hobson98}Hobson
M. P., Jones A. W., Lasenby A. N., Bouchet F. R., 1998, MNRAS, 300,
1 

\bibitem[\protect\citeauthoryear{Hobson et al.}{1999}]{hobson99}Hobson
M. P., Jones A. W., Lasenby A. N., 1999, MNRAS, 309, L125 

\bibitem[\protect\citeauthoryear{Jeong \& Smoot}{2005}]{jeong05}Jeong
E., Smoot G. F., 2005, ApJ, 624, 21 

\bibitem[\protect\citeauthoryear{Jones et al.}{2002}]{jones02}Jones
M. E., 2002, in Chen L.-W., Ma C.-P., Ng K.-W., Pen U.-L., eds, ASP
Conf. Ser. Vol. 257, AMiBA 2001: High-z Clusters, Missing Baryons,
and CMB Polarization. Astron. Soc. Pac., San Francisco, p. 35 

\bibitem[\protect\citeauthoryear{Jones et al.}{2006}]{jones06}Jones
W. C. et al., 2006, ApJ, 647, 823 

\bibitem[\protect\citeauthoryear{Kaiser \& Stebbins}{1984}]{kaiser84}Kaiser
N., Stebbins A., 1984, Nature, 310, 391 

\bibitem[\protect\citeauthoryear{Kibble}{1985}]{kibble85}Kibble T.
W. B., 1985, Nucl. Phys. B, 252, 227 

\bibitem[\protect\citeauthoryear{Komatsu \& Seljak}{2002}]{komatsu02}Komatsu
E., Seljak U., 2002, MNRAS, 336, 1256 

\bibitem[\protect\citeauthoryear{Komatsu et al.}{2009}] 
{komatsu09}Komatsu
E. et al., 2009, ApJS, 180, 330

\bibitem[\protect\citeauthoryear{Kosowsky}{2006}]{kosowsky06}Kosowsky
A., 2006, New Astron. Rev., 50, 969 

\bibitem[\protect\citeauthoryear{Lo \& Wright}{2005}]{lo05}Lo A.
S., Wright E. L., 2005, Bull. of the AAS, 37, 1429

\bibitem[\protect\citeauthoryear{Maisinger et al.}{1999}]{maisinger04}Maisinger
K., Hobson M. P., Lasenby A. N., 2004, MNRAS, 347, 339 

\bibitem[\protect\citeauthoryear{Mallat}{1998}]{mallat98}Mallat S.
G., 1998, A wavelet tour of signal processing. Academic Press, San
Diego 

\bibitem[\protect\citeauthoryear{Moulin \& Liu}{1999}]{moulin99}Moulin
P., Liu J.,1999, IEEE Trans. Information Theo., 45, 909 

\bibitem[\protect\citeauthoryear{Moessner et al.}{1994}]{moessner94}Moessner
R., Perivolaropoulos L., Brandenberger R., 1994, ApJ, 425, 365 

\bibitem[\protect\citeauthoryear{Perivolaropoulos}{1993}]{perivolaropoulos93}Perivolaropoulos
L., 1993, Phys. Lett. B, 1993, 305 

\bibitem[\protect\citeauthoryear{Pires et al.}{2006}]{pires06}Pires
S., Juin J.-B., Yvon D., Moudden Y., Anthoine S., Pierpaoli E., 2006,
A\&A, 455, 741 

\bibitem[\protect\citeauthoryear{Portilla et al.}{2003}]{portilla03}Portilla
J., Strela V., Wainwright M. J., Simoncelli E. P., 2003, IEEE Trans.
Image Proc., 12, 1338 

\bibitem[\protect\citeauthoryear{Readhead et al.}{2004}]{readhead04}Readhead
A. C. S. et al., 2004, ApJ, 609, 498 

\bibitem[\protect\citeauthoryear{Reichardt et al.}{2009}] 
{reichardt09}Reichardt
et al., 2009, ApJ, 694, 1200

\bibitem[\protect\citeauthoryear{Rocha et al.}{2005}]{rocha05}Rocha
G., Hobson M. P., Smith S., Ferreira P., Challinor A., 2005, MNRAS,
357, 1 

\bibitem[\protect\citeauthoryear{Ruhl et al.}{2004}]{ruhl04}Ruhl
J. E. et al., 2004, in Zmuidzinas J., Holland W. S., and Withington
S., eds, Proc. SPIE Conf. Vol. 5498, Millimeter and Submillimeter
Detectors for Astronomy II. SPIE, Bellingham, p. 11 

\bibitem[\protect\citeauthoryear{Simoncelli et al.}{1992}]{simoncelli92}Simoncelli
E. P., Freeman W. T., Adelson E. H., Heeger D. J., 1992, IEEE Trans.
Information Theo., 38, 587 

\bibitem[\protect\citeauthoryear{Simoncelli \& Adelson}{1996}]{simoncelli96}Simoncelli
E. P., Adelson E. H., 1996, Proc. IEEE Conf. Vol. I. IEEE Signal Proc.
Soc., p. 379 

\bibitem[\protect\citeauthoryear{Spergel et al.}{2003}]{spergel03}Spergel
D. N. et al., 2003, ApJS, 148, 175 

\bibitem[\protect\citeauthoryear{Spergel et al.}{2007}]{spergel07}Spergel
D. N. et al., 2007, ApJS, 170, 377 

\bibitem[\protect\citeauthoryear{Starck et al.}{2002}]{starck02}Starck
J.-L., Cand\`es E., Donoho D. L., 2002, IEEE Trans. Image Proc., 11,
670 

\bibitem[\protect\citeauthoryear{Starck et al.}{2006a}]{starck06a}Starck
J.-L., Murtagh F., 2006a, Astronomical Image and Data Analysis (Second
Edition). Springer, Berlin 

\bibitem[\protect\citeauthoryear{Sunyaev \& Zel'dovich}{1980}]{sunyaev80}Sunyaev
R. A., Zel'dovich Y. B., 1980, Ann. Rev. Astron. Astrophys, 18, 537 

\bibitem[\protect\citeauthoryear{Turok \& Spergel}{1990}]{turok90}Turok
N., Spergel D. N., 1990, Phys. Rev. Lett. 64, 2736 

\bibitem[\protect\citeauthoryear{Vachaspati \& Vilenkin}{1984}]{vachaspati84}Vachaspati
T., Vilenkin A., 1984, Phys. Rev. D, 30, 2036 

\bibitem[\protect\citeauthoryear{Vilenkin \& Shellard}{1994}]{vilenkin94}Vilenkin
A., Shellard E. P. S., 1994, Cosmic Strings and Other Topological
Defects. Cambridge Univ. Press, Cambridge 

\bibitem[\protect\citeauthoryear{Wyman et al.}{2005}]{wyman05}Wyman
M., Pogosian L., Wasserman I., 2005, Phys. Rev. D, 72, 023513 

\bibitem[\protect\citeauthoryear{Wyman et al.}{2006}]{wyman06}Wyman
M., Pogosian L., Wasserman I., 2006, Phys. Rev. D, 73, 089905(E) 

\bibitem[\protect\citeauthoryear{Zwart et al.}{2008}]{zwart08}Zwart
J. T. L., 2008, 391, 1545

\end{thebibliography}
\end{document}